\documentclass[fleqn,twoside]{article}
\usepackage{amsmath,amssymb,espcrc2,color,cite}
\usepackage[usenames,dvipsnames]{xcolor}

\usepackage{graphicx,soul}
\usepackage[figuresright]{rotating}
\newcommand{\url}[1]{{\tt #1}}

\newcommand{\gsim}
{\;\raisebox{-.3em}{$\stackrel{\displaystyle >}{\sim}$}\;}

\newcommand{\gmt}{\ensuremath{(g-2)_\mu}}
\newcommand{\br}{{\rm BR}}
\newcommand{\bsg}{\ensuremath{\br(b \to s \gamma)}}

\newcommand{\bsmm}{\ensuremath{\br(B_s \to \mu^+\mu^-)}}

\newcommand{\bsdmm}{\ensuremath{\br(B_{s, d} \to \mu^+\mu^-)}}

\newcommand{\ssi}{\ensuremath{\sigma^{\rm SI}_p}}

\newcommand{\MW}{\ensuremath{M_W}}

\newcommand{\Mh}{\ensuremath{M_h}}

\newcommand{\MA}{\ensuremath{M_A}}
\newcommand{\mgrav}{\ensuremath{m_{3/2}}}

\newcommand{\mt}{m_t}

\newcommand{\mgl}{\ensuremath{m_{\tilde g}}}
\newcommand{\msq}{\ensuremath{m_{\tilde q}}}

\newcommand{\stau}[1]{\ensuremath{\tilde \tau_{#1}}}

\newcommand{\mstop}[1]{\ensuremath{m_{\tilde t_{#1}}}}

\newcommand{\msqr}{\ensuremath{m_{\tilde q_R}}}

\newcommand{\cha}[1]{\tilde \chi^\pm_{#1}}
\newcommand{\chap}[1]{\tilde \chi^+_{#1}}
\newcommand{\cham}[1]{\tilde \chi^-_{#1}}

\newcommand{\mcha}[1]{\ensuremath{m_{\tilde \chi^\pm_{#1}}}}
\newcommand{\neu}[1]{\tilde \chi^0_{#1}}
\newcommand{\mneu}[1]{\ensuremath{m_{\tilde \chi^0_{#1}}}}

\newcommand{\mst}[1]{m_{\tilde t_{#1}}}

\newcommand{\mstau}[1]{\ensuremath{m_{\tilde \tau_{#1}}}}

\newcommand{\tb}{\ensuremath{\tan\beta}}

\newcommand{\tev}{\ensuremath{\,\, \mathrm{TeV}}}
\newcommand{\gev}{\ensuremath{\,\, \mathrm{GeV}}}
\newcommand{\mev}{\ensuremath{\,\, \mathrm{MeV}}}

\newcommand{\ifb}{\ensuremath{{\rm fb}^{-1}}}

\definecolor{orange}{rgb}{1,0.5,0}

\definecolor{Gray}{named}{Gray}

\newcommand{\ETslash}{\ensuremath{/ \hspace{-.7em} E_T}}

\graphicspath{{figs/}}

\title{\vspace{-5cm}
\bf \LARGE Likelihood Analysis of the Minimal AMSB Model} 

\author{\vspace{-0.1cm}
{\bf E.~Bagnaschi}\address[DESY]{DESY, Notkestra{\ss}e 85, D--22607 Hamburg, Germany},
{\bf M.~Borsato}\address[USdC]{Universidade de Santiago de Compostela, 
E-15706 Santiago de Compostela, Spain},
{\bf K.~Sakurai}\address[Durham]{Institute for Particle Physics Phenomenology, Department of Physics, 
University of Durham, Science Laboratories, South Road, Durham, DH1 3LE, UK}\hbox{$^{\rm ,}$}\address[Warsaw]{Institute of Theoretical Physics, Faculty of Physics, University of Warsaw, ul.~Pasteura 5, PL--02--093 Warsaw, Poland},
{\bf O.~Buchmueller}{\address[Imperial]
   {High\,Energy\,Physics\,Group,\,Blackett\,Laboratory,\,Imperial\,College,\,Prince\,Consort\,Road,\,London\,SW7\,2AZ,\,UK},
\bf R.~Cavanaugh}\address[FNAL]
   {Fermi National Accelerator Laboratory, P.O. Box 500, 
    Batavia, Illinois 60510, USA}\hbox{$^{\rm ,}$}\address[UIC]
   {Physics Department, University of Illinois at Chicago, Chicago, 
    Illinois 60607-7059, USA},
{\bf V.~Chobanova}\addressmark[USdC],
{\bf M.~Citron}\addressmark[Imperial],
{\bf J.C.~Costa}\addressmark[Imperial],
{\bf A.~De~Roeck}\address[CERNEP]
   {Experimental Physics Department, CERN, CH--1211 Geneva 23, Switzerland; \\  Antwerp University, B--2610 Wilrijk, Belgium},
 {\bf M.J.~Dolan}\address[SLAC]
{ARC Centre of Excellence for Particle Physics at the Terascale, School of Physics, University of Melbourne, 3010, Australia},
{\bf J.R.~Ellis}\address[KCL]{Theoretical Particle Physics
  and Cosmology Group, Department of Physics, King's College London, London~WC2R~2LS, UK; Theoretical Physics Department, CERN, CH--1211 Geneva 23, Switzerland},
{\bf H.~Fl\"acher}\address[Bristol]
   {H.H.~Wills Physics Laboratory, University of Bristol, Tyndall Avenue, Bristol BS8 1TL, UK},
{\bf S.~Heinemeyer}\address[Madrid]
{Campus of International Excellence UAM+CSIC, Cantoblanco, E--28049 Madrid, Spain;\\
  Instituto de F\'{\i}sica Te{\'o}rica UAM-CSIC, C/ Nicolas Cabrera 13-15, E--28049 Madrid, Spain; \\
   Instituto de F\'{\i}sica de Cantabria (CSIC-UC), Avda. de Los Castros s/n, 
    E--39005 Santander, Spain},
{\bf G.~Isidori}\address[Zurich]
{Physik-Institut, Universit\"at Z\"urich, CH-8057 Z\"urich, Switzerland},
{\bf M.~Lucio}\addressmark[USdC],
{\bf F.~Luo}\address[KIPMU]{Kavli IPMU (WPI), UTIAS, The University of Tokyo, Kashiwa, Chiba 277-8583, Japan},
{\bf D.~Mart\'inez~Santos}\addressmark[USdC],
{\bf K.A.~Olive}\address[Minnesota] 
{William I.\ Fine Theoretical Physics Institute, School of Physics and
 Astronomy, University of Minnesota, Minneapolis, Minnesota 55455, USA}, 
{\bf A.~Richards}\addressmark[Imperial] and
{\bf G.~Weiglein}\addressmark[DESY]
}

\begin{document}
\begin{abstract}
\vspace{-0.3cm}

{\small We perform a likelihood analysis of the minimal Anomaly-Mediated Supersymmetry Breaking (mAMSB) 
model using constraints from cosmology and accelerator experiments. We find that 
{a wino-like or a Higgsino-like neutralino LSP, $\neu1$, may provide the cold dark matter (DM) with similar likelihood}. 
The upper limit on the DM density from Planck and other experiments enforces {$\mneu1 \lesssim 3 \tev$ }
after the inclusion of Sommerfeld enhancement in its annihilations.
If most of the cold DM density is provided by the $\neu1$, the measured value of the  Higgs mass 
favours a limited range of $\tan \beta \sim 5$ (or for $\mu > 0$, $\tb \sim 45$) but the scalar mass $m_0$ is poorly constrained. In the wino-LSP case, \mgrav\ is constrained to about $900\tev$ and \mneu1 to $2.9\pm0.1\tev$, whereas in the Higgsino-LSP case \mgrav\ has just a lower limit {$\gtrsim 650\tev$ ($\gtrsim 480\tev$)} and $\mneu1$ is constrained to $1.12 ~(1.13) \pm0.02\tev$ in the $\mu>0$ ($\mu<0$) scenario. 
In neither case can the anomalous magnetic moment of the muon, $\gmt$, be improved 
{significantly} relative to its Standard Model (SM) value, nor do flavour measurements constrain the 
model significantly, and there are poor prospects for discovering supersymmetric  particles at the LHC, 
{though there} are some prospects for direct DM detection. 
On the other hand, if  the $\neu1$ contributes only a fraction of the cold DM density, 
{future LHC $\ETslash$-based searches for gluinos, squarks and heavier chargino and neutralino states
as well as disappearing track searches in the wino-like LSP region will be relevant}, and
{interference effects enable \bsdmm\ to agree with the data
better than in the SM in the case of wino-like DM with $\mu > 0$}.}

\vspace{-0.2cm}
\begin{center}
{\tt KCL-PH-TH/2016-58, CERN-PH-TH/2016-220, DESY 16-155, IFT-UAM/CSIC-16-112\\
IPMU16-0157, FTPI-MINN-16/30, UMN-TH-3610/16, FERMILAB-PUB-16-502-CMS, IPPP/16/104}
\end{center}

\end{abstract}

\thispagestyle{empty}
\newpage

\maketitle


\section{Introduction}
\label{sec:intro}

In previous papers~\cite{oldmc,mc9,mc10,mc11,mc12, mc14-SU5, mcweb} we have presented likelihood analyses of the parameter spaces of various scenarios for supersymmetry (SUSY) breaking, including the CMSSM~\cite{CMSSM},  in which soft SUSY breaking parameters are constrained to be universal at the grand unification scale, models in which Higgs masses are allowed to be non-universal (NUHM1,2)~\cite{NUHM1,NUHM2}, a model in which 10 soft SUSY-breaking parameters  were treated as free phenomenological parameters (the pMSSM10)~\cite{pMSSM10} and one with SU(5) GUT boundary conditions on soft supersymmetry-breaking parameters~\cite{SU5}. These analyses took into account the strengthening direct constraints from sparticle searches at the LHC, as well as indirect constraints based on electroweak precision observables (EWPOs), flavour observables and the contribution to the
density of cold dark matter (CDM) in the Universe from the lightest supersymmetric particle (LSP), assuming that it is a neutralino and that $R$-parity is conserved~\cite{EHNOS}. In particular, we analysed the prospects within these scenarios for discovering SUSY at the LHC and/or in future direct dark matter searches~\cite{mc12}.

In this paper we extend our previous analyses of GUT-based models~\cite{oldmc,mc9,mc10,mc11,mc12,mc14-SU5} by   presenting a likelihood analysis of the parameter space of the minimal scenario for anomaly-mediated SUSY breaking (the mAMSB)~\cite{anom,mAMSB}.  The spectrum of this model is {quite} different from those of the CMSSM, NUHM1 and NUHM2, with a different composition of the LSP. Consequently,
{different} issues arise in the application of the experimental constraints, as we discuss below. In the mAMSB there are 3 relevant continuous parameters, the gravitino mass, \mgrav, which sets the scale of SUSY breaking, the supposedly universal soft SUSY-breaking scalar mass\footnote{In pure gravity-mediated models \cite{pgm}, $m_0$ is constrained to be equal to the gravitino mass, resulting in a two-parameter model in which $\tan \beta$ is strongly constrained to a value near 2.}, $m_0$, and the ratio of Higgs vacuum expectation values, $\tan \beta$, to which may be added the sign of the Higgsino mixing parameter, $\mu$. 
The LSP is either a Higgsino-like or a wino-like neutralino $\neu1$.
{In both cases the} $\neu1$ is almost degenerate with its chargino partner, $\cha1$.
It is well known that, within  this mAMSB framework, if one requires that a wino-like $\neu1$ is the dominant source of the CDM density indicated by Planck measurements of the cosmic microwave background radiation, namely $\Omega_{\rm CDM} h^2 = 0.1186 \pm 0.0020$~\cite{Planck15}, $\mneu1 \simeq 3 \tev$~\cite{winomass1,winomass} after inclusion of Sommerfeld enhancement effects~\cite{Sommerfeld1931}. If instead the CDM density is to be explained by a Higgsino-like $\neu1$, $\mneu1$ takes a value of $1.1 \tev$. 
In both cases, sparticles are probably too heavy to be discovered at the LHC, and supersymmetric contributions to EWPOs, flavour observables and \gmt\ are small.

In the first part of our likelihood analysis of the mAMSB parameter space, we combine the assumption that the LSP is the dominant source of CDM with other measurements, notably of the mass of the Higgs boson, $\Mh = 125.09 \pm 0.24 \gev$~\cite{Aad:2015zhl} (including the relevant theory uncertainties~{\cite{DeltaMhTH}}) and its
{production and decay rates}~\cite{HiggsSignals}. 
{In addition to solutions in which the $\neu1$ is wino- or Higgsino-like, we also find less-favoured
solutions in which the $\neu1$ is a mixed wino-Higgsino state}. 
In the wino case, whereas \mgrav\ and hence $\mneu1$ are relatively well determined, 
as is the value of $\tan \beta$, 
the value of $m_0$ is quite poorly determined, and there is little difference between the
values of the global likelihood functions for the two signs of $\mu$. 
On the other hand, in the case of a Higgsino-like $\neu1$, while $\tan\beta$ has values around 5, $m_0$ and \mgrav\ are only constrained to be larger than $20 \tev$ and  $600\tev$, respectively, in the positive $\mu$ case. For negative $\mu$, the $m_0$ and \mgrav\ constraints are lowered to $18 \tev$ and $500 \tev$, respectively.

If there is some other contribution to the CDM, so that $\Omega_{\neu1} < \Omega_{\rm CDM}$, the SUSY-breaking mass scale \mgrav\ can be reduced, and hence also $\mneu1$, although the value of $\Mh$ still imposes a significant lower limit. In this case, some direct searches for sparticles at the LHC also become relevant, notably 
{$\ETslash$-based searches for gluinos, squarks and heavier chargino and neutralino states
as well as disappearing track searches for the next-to-LSP charged wino.}
We discuss the prospects for sparticle searches at the LHC in this case
{and at the 100 TeV FCC-hh collider}, and also find that some deviations from Standard Model (SM) predictions for flavour observables may become important, notably \bsg\ and \bsdmm.

Using the minimum value of the $\chi^2$ likelihood function and the number of effective degrees of freedom
(excluding the constraint from {\tt HiggsSignals}, as was done in~\cite{mc9,mc10,mc11}) leads to an {estimate of
{$\sim 11\%$} for the $\chi^2$ probability of the mAMSB model if most of the CDM is due to the $\neu1$,
for both signs of $\mu$ in both the wino- and Higgsino-like cases}. {When this CDM condition is relaxed, the $\chi^2$ probability is unchanged if $\mu < 0$, but increases to $18\%$ in the wino-like LSP case
if $\mu >0$ thanks to improved consistency with the experimental
measurement of \bsdmm.}
These $\chi^2$ probabilities for the mAMSB model cannot be compared directly with those found previously for the 
CMSSM~\cite{mc9}, the NUHM1~\cite{mc9}, the NUHM2~\cite{mc10} and the pMSSM10~\cite{mc11}, since
those models were studied with a different dataset that included an older set of LHC data.

The outline of this paper is as follows. In Section~\ref{sec:specification} we review briefly the specification of the mAMSB model. In Section~\ref{sec:Constraints}  we review our implementations of the relevant theoretical, phenomenological, experimental, astrophysical and cosmological constraints, including those from the flavour and Higgs sectors, and from LHC and dark matter searches
(see~\cite{mc11,mc14-SU5} for details of our other LHC search implementations). In the case of dark matter we describe in detail our implementation of Sommerfeld enhancement in the calculation of the relic CDM density. Section~\ref{sec:Framework} reviews the {\tt MasterCode} framework. 
Section~\ref{sec:Results} then presents our results, first under the assumption that the lightest neutralino $\neu1$ is the dominant form of CDM, and then in the more general case when other forms of CDM may dominate. This Section is concluded by the presentation and discussion of the $\chi^2$ likelihood functions for observables of interest. Finally, we present our conclusions in Section~\ref{sec:conx}.

\section{Specification of the mAMSB Model}
\label{sec:specification}

In AMSB, SUSY breaking arises via a loop-induced super-Weyl anomaly~\cite{anom}.  
Since the gaugino masses $M_{1, 2, 3}$ are suppressed by loop factors relative to the gravitino mass, $\mgrav$, the latter is fairly heavy in this scenario ($\mgrav \gtrsim 20 \tev$) and the wino-like states are lighter than the bino-like ones, with the following ratios of gaugino masses at NLO: $|M_1| : |M_2| : |M_3| \approx 2.8:1:7.1$. 
Pure AMSB is, however, an unrealistic model, because renormalization leads to negative squared masses for sleptons and, in order to avoid tachyonic sleptons, the minimal AMSB scenario (mAMSB) adds a constant $m_0^2$ to all  squared scalar masses \cite{mAMSB}. Thus the mAMSB model has three continuous free parameters: $m_{3/2}$, $m_0$ and the ratio of Higgs vevs, $\tb$. In addition, the sign of the Higgsino mixing parameter, $\mu$, is also free. 
The trilinear soft SUSY-breaking mass terms, $A_i$, are determined by anomalies, like the gaugino masses, and are thus proportional to $m_{3/2}$. The $\mu$ term and the Higgs bilinear, $B$, are determined phenomenologically via the minimization of the Higgs potential, as in the CMSSM.

The following are some characteristic features of mAMSB: near mass-degeneracy of the left and right sleptons: $m_{\tilde{l}_R} \approx m_{\tilde{l}_L}$, and of the lightest chargino and neutralino, $\mcha1 \approx \mneu1$. The mass hierarchy between sleptons and gauginos is dependent on the numerical values of the input parameters, and the squark masses are typically very heavy, because they contain a term proportional to $g_3^4 m^2_{3/2}$.
In addition, the measured Higgs mass and the relatively low values of the trilinears $A_i$
together imply that the stop masses must also be relatively high.
The LSP {composition may be wino-, Higgsino-like or mixed}, as we discuss in more detail below.


\section{Implementations of Constraints}
\label{sec:Constraints}

Our treatments in this paper of many of the relevant constraints follow very closely the implementations in our previous analyses which were recently summarized in ~\cite{mc14-SU5}.
In the following subsections we review the implementations, highlighting new constraints
and instances where we implement constraints differently from our previous work. 


\subsection{Flavour, Electroweak and Higgs Constraints}
\label{sec:flav-ew-higgs}

Constraints from $B$-physics and $K$-physics observables are {the same as in~\cite{mc14-SU5}.
In particular, we include} the recent ATLAS result in our global combination of measurements
of \bsdmm~~\cite{LHCBsmm}. 
{In contrast to our previous studies~\cite{mc9,mc10,mc11,mc12,mc14-SU5},  in this study we do not
evaluate independently the constraints from EWPOs, since for SUSY-breaking
parameters in the multi-TeV range
they are indistinguishable from the Standard Model values within the current
experimental uncertainties, as we have checked using {\tt FeynWZ}~\cite{Svenetal}. 
The only exception is the mass of the $W$~boson, $\MW$, 
which is evaluated using {\tt FeynHiggs}}\footnote{{We imposed
SU(2) symmetry on the soft SUSY-breaking terms in the $\overline{\rm DR}$-on-shell conversion of the 
parameters in the scalar top/bottom sector, leading to a small shift in the values of the scalar bottom
masses.}}.
For the other EWPOs
we use the theoretical and experimental values given in the review~\cite{PDG2016}.
We use the combination of ATLAS and CMS measurements of the mass of the
Higgs boson: $\Mh = 125.09 \pm 0.24 \gev$~\cite{Aad:2015zhl}. We use 
{a beta-version of the {\tt FeynHiggs 2.12.1} code~{\cite{FH,DeltaMhTH}} to evaluate the
constraint this imposes on the mAMSB parameter space. It improves on
the {\tt FeynHiggs} versions used for previous analyses~\cite{mc9,mc10,mc11,mc12}
by including two-loop QCD corrections in the
evaluation of the $\overline{\rm {DR}}$ running top mass and an improved evaluation
of the top mass in the $\overline{\rm{DR}}$-on-shell conversion for the scalar tops.
At low values of $\mstop1$, we use, as previously, a one-$\sigma$
theoretical uncertainty of $1.5 \gev$. In view of the larger theoretical uncertainty
at large input parameter values, this uncertainty is smoothly inflated up to $3.0 \gev$ at
$\mstop1  > 7.5 \tev$, as a conservative estimate.} 
The $\chi^2$ contributions of Higgs search channels from LHC and
Tevatron are evaluated using {\tt HiggsSignals}~\cite{HiggsSignals} and
{\tt HiggsBounds}~\cite{HiggsBounds,HBtautau} as detailed in our
previous paper~\cite{mc14-SU5}.  


\subsection{LHC Constraints}

If the entire CDM relic density is provided by the lightest neutralino, all sparticles are heavy, 
and the current results of the direct sparticle searches at the LHC have no impact on our global fit,
{though there is some impact from $H/A$ searches~\cite{HA8,ICHEP}}.
On the other hand, if $\neu1$ accounts only for a fraction of the relic CDM density, some sparticles can be light enough to be produced at the LHC. However, as we discuss in more detail later, even for this case we find that the sleptons, the first two generations of squarks  and the third-generation squarks are heavier than 0.7, 3.5 and 2.5 \tev\ at the 2\,$\sigma$ level, respectively, well beyond the current LHC sensitivities~\cite{CMS:2015fih,Khachatryan:2014qwa,Aad:2014vma}. On the other hand, gluinos and winos can be as light as 2.5 and $0.5 \tev$, respectively, 
at the 2\,$\sigma$ level, so we have considered in more detail the constraints from searches at the LHC.
{Currently they do not impact the 68 and 95\% CL ranges we find for the mAMSB,
but some impact can be expected for future LHC runs, as we discuss in Section~5.4.}  

\subsection{Dark Matter Constraints}

\subsubsection{{Density Calculations Implementing Sommerfeld Enhancement}}

{For a wino-like dark matter particle, the non-perturbative Sommerfeld 
effect~\cite{Sommerfeld1931} needs to be taken into 
account in the calculation of the thermal relic abundance. Dedicated studies have been 
performed in the literature~\cite{winomass1,winomass}, with the result that
the correct relic abundance is obtained for $\mneu1 \simeq 3.1 \tev$ after inclusion of Sommerfeld enhancement in the thermally-averaged coannihilation cross sections, compared to $\mneu1 \simeq 2.3 \tev$ at tree level. 

Because of the large number of points in our mAMSB sample, 
we seek a computationally-efficient implementation of the
Sommerfeld enhancement. 
We discuss this now, and consider its implications
in the following subsections.
}
 
 It is sufficient for our $\chi^2$ likelihood analysis to use a phenomenological
fit for the Sommerfeld enhancement that is applicable near $3.1 \tev$. One reason is that,
away from $\sim 3.1 \tev$, the $\chi^2$ price rises rapidly due to the very small
uncertainty in the Planck result for $\Omega_{\rm CDM} h^2$. Another reason is that the enhancement factor 
depends very little on the particle spectrum and mostly on $\mneu1$. Therefore, we extract the Sommerfeld factor 
by using a function to fit the `non-perturbative' curve in the right panel of Fig.~2 in~\cite{winomass1}. 
One can see that the curve has a dip at $\sim 2.4 \tev$, due to the appearance of a loosely-bound state. 
The calculated relic abundance near the dip is much smaller than the Planck value, so it gives a very large $\chi^2$, 
and therefore we do not bother to fit the dip. Considering that the Yukawa potential approaches the 
Coulomb limit for $\mneu1 \gg \MW$, and that only the electromagnetic force is relevant for $\mneu1 \ll \MW$,  we fit the annihilation cross section using~\footnote{We 
emphasize that one can choose a different fitting function, as long as the fit is good near $3.1 \tev$.},    
\begin{eqnarray}
{a_{\rm eff}}  &\equiv&  {a_{\rm eff}}_{\rm SE=0} \big[ \left(c_{\rm pm} S_{\alpha_{\rm em}} + 1 - c_{\rm pm}\right) 
\nonumber \\ 
&& \left( 1 - \exp({- \kappa \, \MW / \mneu1}) \right)  \nonumber \\
&&+  S_{\alpha_2} \, \exp({- \kappa \, \MW / \mneu1}) \big] \, , 
\label{somapp}
\end{eqnarray}
where $a_{\rm eff}$ is the effective s-wave coannihilation cross section (including the Sommerfeld enhancement) for the wino system including the wino-like LSP, 
$\neu1$, and the corresponding chargino, $\cha1$, and
${a_{\rm eff}}_{\rm SE=0}$
is the effective s-wave coannihilation cross section calculated ignoring the enhancement.
The latter is defined as 
\begin{equation}
{a_{\rm eff}}_{\rm SE=0} \equiv \sum_{i, j} a_{ij} r_i r_j \, ,
\end{equation}
where $r_i \equiv  g_i \left(1+ \Delta_i\right)^{3/2} \exp(-\Delta_i \mneu1/T) / g_{\rm eff}$, 
and $g_{\rm eff} \equiv \sum_{k} g_k \left(1+ \Delta_k\right)^{3/2} \exp(-\Delta_k \mneu1/T)$ expressed
as functions of the temperature, $T$, {at which the coannihilations take place}.
The indices refer to $\neu1$, $\tilde{\chi}^+_1$ and $\tilde{\chi}^-_1$, and $g_i$ is the number of
degrees of freedom, which is 2 for each of the three particles, $\Delta_i \equiv (m_i /  \mneu1 - 1)$,
$a_{ij}$ is the total s-wave (co)annihilation cross section for the processes with incoming particles $i$ and $j$,
and $c_{\rm pm}$ is the fraction of the contribution of the $\chap1 \cham1$ s-wave cross section in ${a_{\rm eff}}_{\rm SE=0}$, namely, 
\begin{equation}
c_{\rm pm} \equiv \frac{2 a_{\chap1 \cham1}}{{a_{\rm eff}}_{\rm SE=0}}  r_{\chap1} r_{\cham1} \,.
\end{equation} 
In practice, since $m_{\chap1} - \mneu1 \simeq 0.16 \gev$, which is much smaller than the 
typical temperature of interest in the calculation of the relic abundance for $\mneu1$ near $3.1 \tev$, we have
${a_{\rm eff}}_{\rm SE=0} \simeq (a_{\neu1 \neu1} + 4 a_{\neu1 \chap1} + 2 a_{\chap1 
\cham1} + 2 a_{\chap1 \chap1})/9$, and $c_{\rm pm} \simeq \frac{2}{9} a_{\chap1 
\cham1} / {a_{\rm eff}}_{\rm SE=0}$. In Eq.~(\ref{somapp}), $S_{\alpha_{\rm em}}$ and $S_{\alpha_2}$ are the thermally-averaged 
s-wave Sommerfeld enhancement factors for attractive Coulomb potentials with couplings $\alpha_{\rm em}$ and $\alpha_2$, 
respectively. We use the function given in Eq. (11) of~\cite{winomass2} for these quantities, namely
\begin{equation}
{S_{\alpha_x}} \equiv \frac{1 + 7 y/4 + 3y^2/2 + (3/2-\pi/3)y^3}{1+3y/4+(3/4-\pi/6)y^2} \, ,
\end{equation}
where $y \equiv {\alpha_x} \sqrt{\pi \mneu1 / T}$. 

Because the curve in~\cite{winomass1} is obtained by taking the massless limit of the SM particles in $a_{ij}$, 
we do the same for our fit to obtain the fitting parameter $\kappa$. We find that a $\kappa = {\cal O}(1)$ 
can give a good fit for the curve, and that the fit is not sensitive to the exact value of $\kappa$. 
We choose $\kappa = 6$ in our calculation, which gives a good fit around $\mneu1 \simeq 3.1 \tev$, in particular. 

Eq.~(\ref{somapp}) is used in our calculation of the relic abundance {$\Omega_{\neu1} h^2$}
for mAMSB models, 
for which we evaluate ${a_{\rm eff}}_{\rm SE=0}$ and $c_{\rm pm}$ for any parameter point using {\tt SSARD} \cite{SSARD}. 
The perturbative p-wave contribution is also included. We note that, whereas the Sommerfeld enhancement 
depends almost entirely on $\mneu1$, the values of $a_{\rm eff}$ and $c_{\rm pm}$ depend on the details of the 
supersymmetric particle spectrum. In particular, due to a cancellation between $s$- and $t$-channel contributions
in processes with SM fermion anti-fermion pairs in the final states, ${a_{\rm eff}}_{\rm SE=0}$ becomes smaller 
when the sfermion masses are closer to $\mneu1$.

\begin{figure*}[htb!]
\vspace{0.5cm}
\begin{center}
{\resizebox{7cm}{!}{\includegraphics{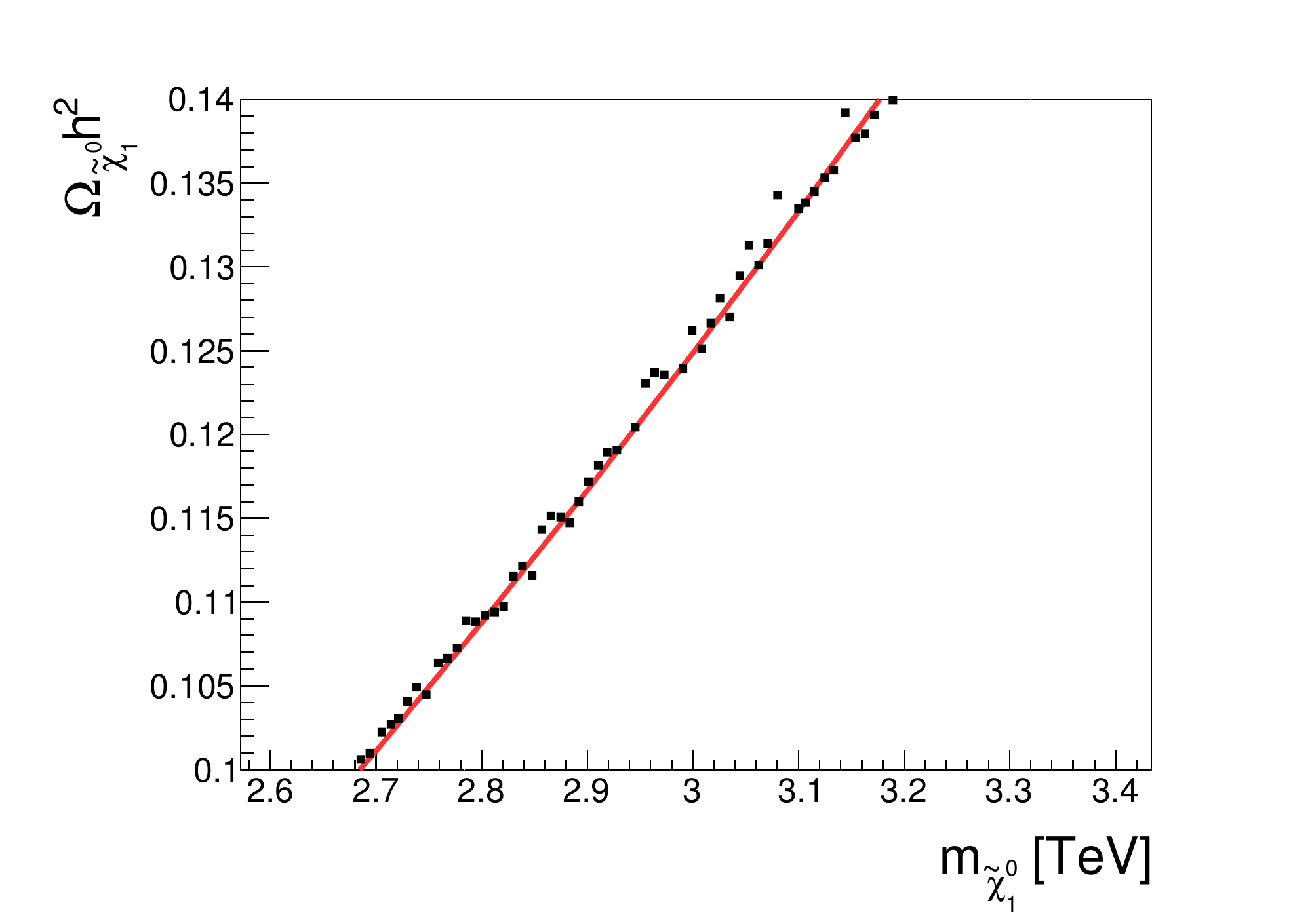}}
\resizebox{7cm}{!}{\includegraphics{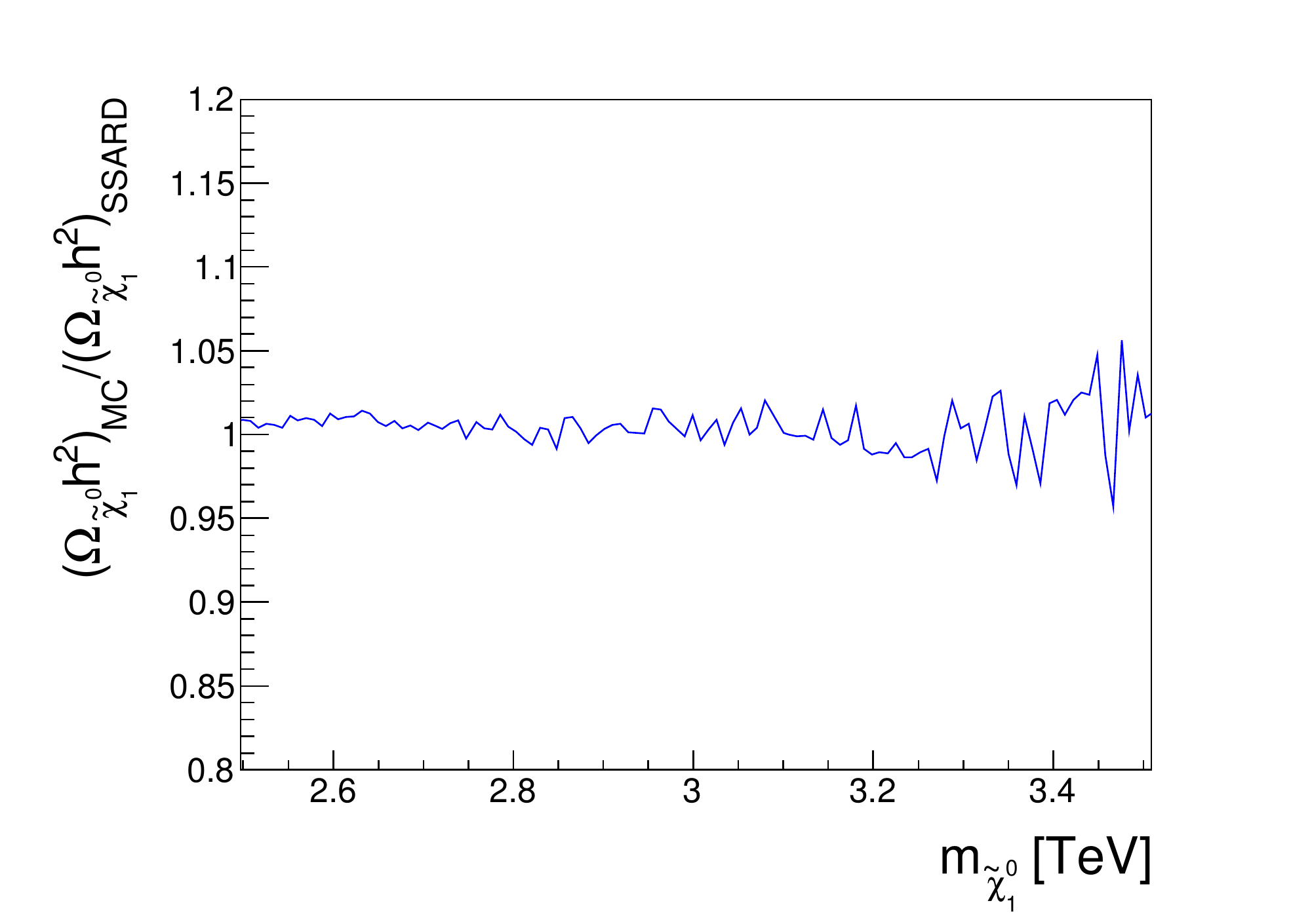}}} \\
\vskip .2in
\end{center}
\caption{\it Calculations of  $\Omega_{\neu1} h^2$ comparing 
{results from {\tt SSARD}~\protect\cite{SSARD}
and our simplified treatment of the Sommerfeld enhancement in the case of wino dark matter.} The
left panel compares the {\tt SSARD} calculations (black dots) with our Sommerfeld implementation (red line), and the
right panel shows the ratio of the calculated relic densities, connecting the points in the left panel by a continuous blue line.}
\label{fig:Sommerfeld1}
\end{figure*}

For a small subset of our mAMSB parameter sample, we have compared results obtained from our
approximate implementation of the Sommerfeld enhancement in the case of wino dark matter with more precise
results obtained with {\tt SSARD}.
As seen in the left panel of Fig.~\ref{fig:Sommerfeld1}, our implementation (red line) yields results for the
relic density that are very similar to those of complete calculations (black dots). In the right panel we plot
the ratio of the relic density calculated using our
simplified Sommerfeld implementation for the sub-sample of mAMSB points to {\tt SSARD} results,
connecting the points at different $\mneu1$ by a continuous blue line. We see that our Sommerfeld
implementation agrees with the exact results at the $\lesssim 2$\% level (in particular when $\mneu1 \sim 3 \tev$),
an accuracy that is comparable to the current experimental uncertainty from the Planck data.
We conclude that our simplified Sommerfeld implementation is adequate for our general study of the mAMSB parameter 
space~\footnote{As stated above, a 
full point-by-point calculation of the relic density would be impractical for our large sample of mAMSB parameters.}.

\begin{figure*}[htb!]
\vspace{0.5cm}
\begin{center}
\vskip .2in
{\resizebox{7.0cm}{!}{\includegraphics{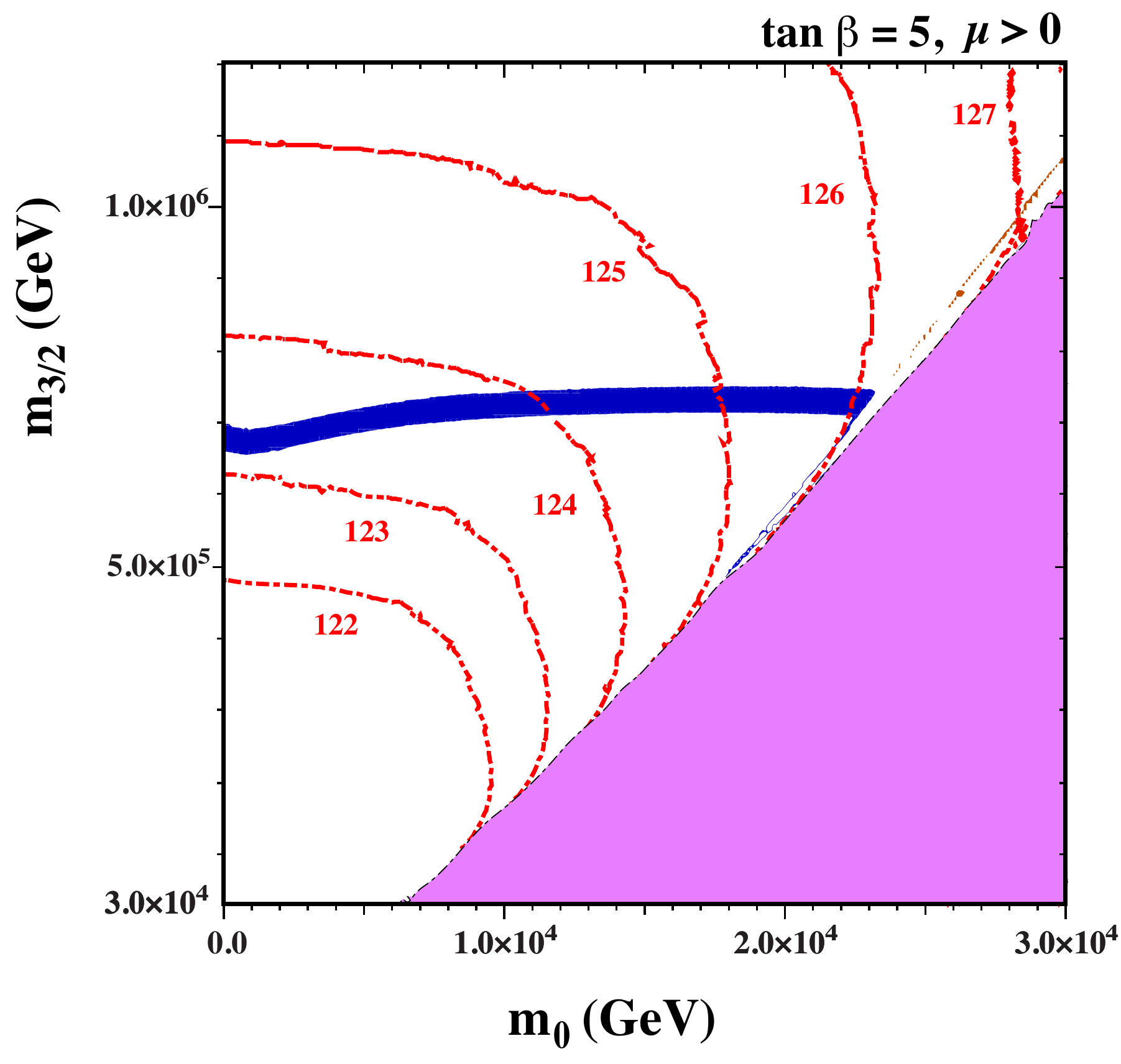}}
\resizebox{7.0cm}{!}{\includegraphics{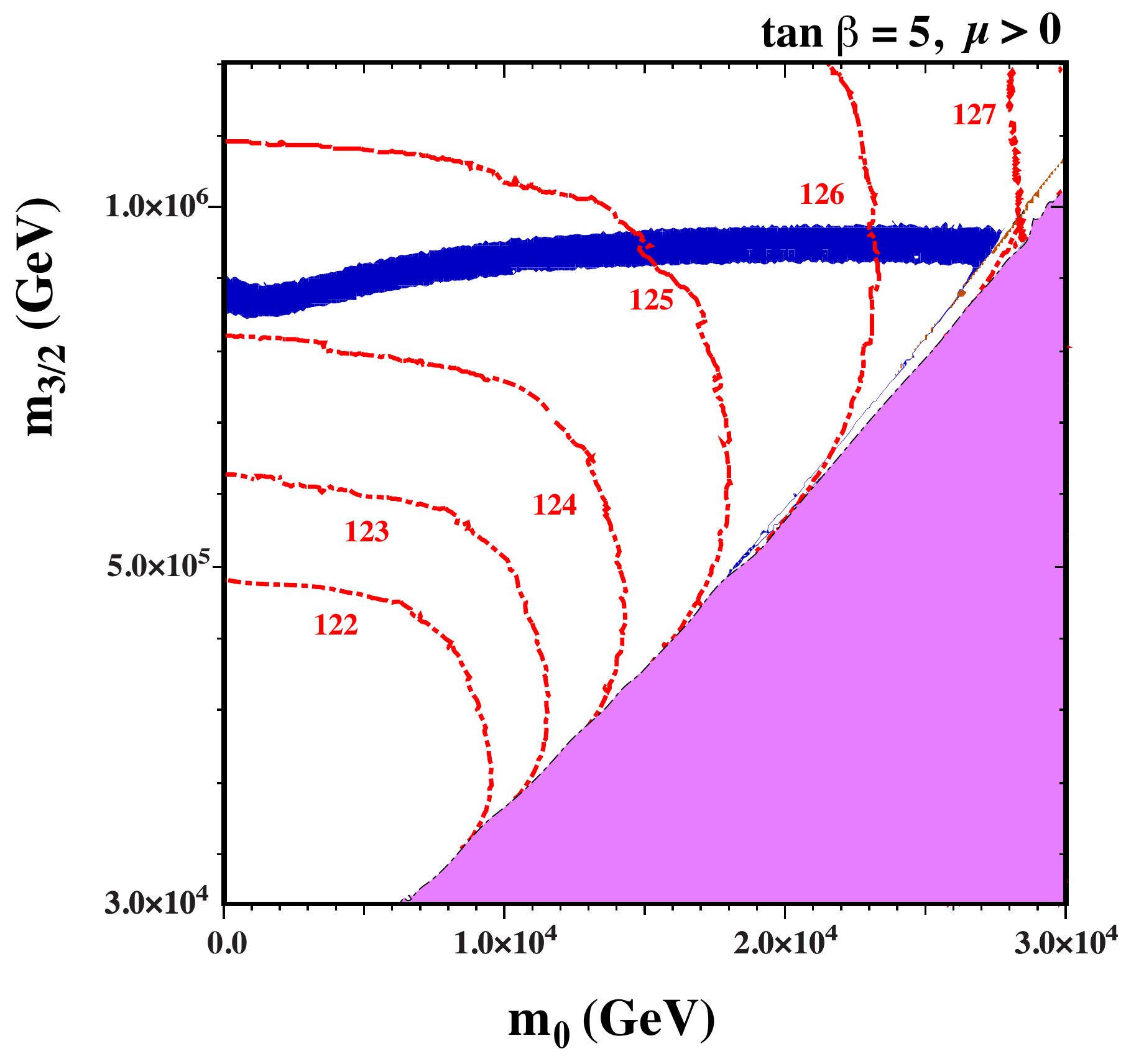}}}
\vspace{1cm}
\end{center}
\vspace{-1.5cm}
\caption{\it The $(m_0, \mgrav)$ plane for $\tan \beta = 5$ without (left panel) and with (right panel) the  Sommerfeld enhancement, as calculated using {\tt SSARD}~\protect\cite{SSARD}. There are no consistent solutions of the electroweak vacuum conditions in the pink shaded triangular regions {at lower right}. The $\neu1$ LSP density falls within the range of the CDM density indicated by Planck and other experiments in the dark blue shaded bands. Contours of $\Mh$ calculated using  {\tt FeynHiggs 2.11.3} 
{(see text)} are shown as red dashed lines.}
\label{fig:Sommerfeld2}
\end{figure*}

\begin{figure*}[htb]
\vspace{0.5cm}
\begin{center}
{\resizebox{10cm}{!}{\includegraphics{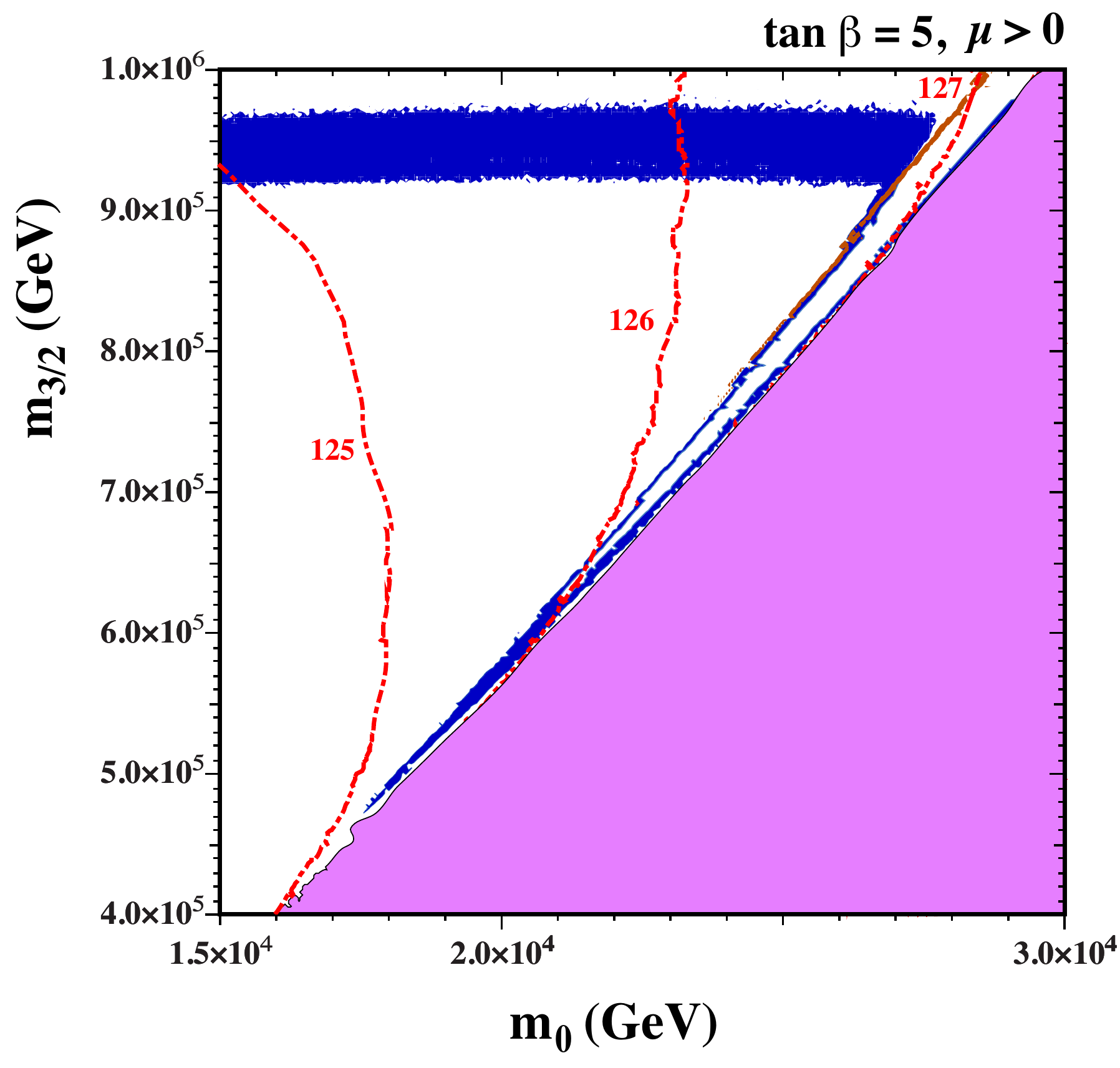}}}
\end{center}
\vspace{-1cm}
\caption{\it Blowup of the right panel in Fig.~\ref{fig:Sommerfeld2}.
When $\mgrav \le 9.1 \times 10^5 \gev$, we shade dark blue regions with $0.1126 \le  \Omega_{\tilde{\chi}_1^0} h^2 \le 0.2$ so as to
thicken the slanted V-shaped Higgsino LSP strip. Towards the upper part of the Higgsino strip, 
there is a thin brown shaded strip that is excluded because the LSP is a chargino.
Contours of $\Mh$ calculated (labelled in GeV) using  {\tt FeynHiggs 2.11.3} {(see text)} are shown as red dashed lines.}
\label{fig:bu}
\end{figure*}


Figure~\ref{fig:Sommerfeld2} illustrates the significance of the Sommerfeld enhancement via a dedicated scan of the $(m_0, \mgrav)$ plane for $\tan \beta = 5$ using {\tt SSARD}.
The pink triangular region at large $m_0$ and relatively small $\mgrav$ is excluded because there are no consistent solutions to the electroweak vacuum conditions in that region. The border of that region corresponds to the line where $\mu^2 = 0$, like that often encountered in the CMSSM at low $m_{1/2}$ and large $m_0$ near the so-called focus-point region \cite{fp}. 
The dark blue strips indicate where the calculated $\neu1$ density falls within the 3-$\sigma$ CDM density range preferred by the Planck data~\cite{Planck15}, and the red dashed lines are
contours of $\Mh$ (labelled in GeV) calculated using  {\tt FeynHiggs 2.11.3}~\footnote{{This
version is different from that used for our
$\chi^2$ evaluation, and is used here for illustration only. The
numerical differences do not change the picture in a
significant way.}}.
The Sommerfeld enhancement is omitted in the left panel and included in the right panel of Fig.~\ref{fig:Sommerfeld2}.
We see that the Sommerfeld enhancement increases the values of $\mgrav$  along the prominent near-horizontal band (where the LSP is predominantly wino) by $\sim 200 \tev$, which is much larger than the uncertainties associated with the CDM density range and our approximate implementation of the Sommerfeld enhancement. We stress that any value of $\mgrav$ below this band would also be allowed if the $\neu1$ provides only a fraction of the total CDM density. 

\subsubsection{Higgsino Region}

We note also the presence in both panels of a very narrow 
V-shaped diagonal strip running close to the electroweak vacuum boundary, 
where the $\neu1$ LSP has a large Higgsino component as mentioned previously. 
As this Higgsino strip is rather difficult to see,
we show in Fig. \ref{fig:bu} a blowup of the Higgsino region {for $\mu > 0$
(the corresponding region for $\mu < 0$ is similar)}, where we have thickened
artificially the Higgsino strips by shading dark blue regions 
with $\mgrav \le 9.1 \times 10^5 \gev$ where $0.1126 \le  \Omega_{\tilde{\chi}_1^0} h^2 \le 0.2$.
As the nearly horizontal wino strip approaches the electroweak symmetry breaking
boundary, the blue strip deviates downward to a point, 
and then tracks the boundary back up to higher $m_0$ and $\mgrav$, forming a slanted V shape.

The origin of these two strips can be understood as follows. 
In most of the triangular region beneath
the relatively thick horizontal strip, the LSP is a wino with mass below 3 TeV,
and the relic density is below the value preferred by the Planck data.
For fixed $\mgrav$, as $m_0$ is increased, $\mu$ drops so that, eventually, 
the Higgsino mass becomes comparable to the wino mass. When {$\mu > 1$~TeV}, 
the crossover to a Higgsino LSP (which occurs when {$\mu \lesssim M_2$}) 
yields a relic density that reaches and then exceeds the Planck relic density, 
producing the left arm of the slanted V-shape strip near the focus-point boundary 
where coannihilations between the wino and Higgsino are important.
As one approaches closer to the focus point, {$\mu$} continues to fall and, 
when {$\mu \simeq 1 \tev$}, the LSP becomes mainly a Higgsino and its relic density
returns to the Planck range, thus producing right arm of the slanted V-shape strip 
corresponding to the focus-point strip in the CMSSM. In the right panel of 
Fig.~\ref{fig:Sommerfeld2} and in Fig.~\ref{fig:bu} the tip of the V where these narrow dark matter strips 
merge occurs when $m_0 \sim 1.8 \times 10^4 \gev$. 

{In the analysis below, we model the transition region by using {\tt Micromegas~3.2}~\cite{MicroMegas} to calculate the relic density, with a correction
in the form of an analytic approximation to the Sommerfeld enhancement given by {\tt SSARD} 
that takes into account the varying wino and Higgsino fractions
in the composition of the LSP. In this way we interpolate between the wino approximation based on {\tt SSARD}
discussed above for 
winos, and {\tt Micromegas~3.2} for Higgsinos. }

Comparing the narrowness of the strips  in Figs. \ref{fig:Sommerfeld2} and \ref{fig:bu}
with the thickness of the near-horizontal wino strip, it is clear that they are
relatively finely tuned. We also note in Fig.~\ref{fig:bu} a thin brown shaded region
towards the upper part of the V-shaped Higgsino strip
that is excluded because the LSP is a chargino.

We also display in these $(m_0, \mgrav)$ planes contours of $\Mh$
(labelled in GeV) as calculated using {\tt FeynHiggs 2.11.3} {(see above)}. Bearing in mind the estimated uncertainty in the theoretical calculation of $\Mh$~{\cite{DeltaMhTH}}, all the broad near-horizontal band and the narrow diagonal strips are compatible with the measured value of $\Mh$, both with and without the inclusion of the Sommerfeld enhancement. 


\subsubsection{{Dark Matter Detection}}

{We implement direct constraints on the spin-independent dark matter proton
scattering cross section, \ssi, using the {\tt SSARD} code~\cite{SSARD}, as reviewed previously~\cite{mc9,mc10,mc11,mc12,mc14-SU5}. As discussed there and in Section~5.5,
\ssi\ inherits considerable uncertainty from the poorly-constrained $\langle p | {\bar s} s | p \rangle$
matrix element and other hadronic uncertainties, which are larger than those associated with
the uncertainty in the local CDM halo density.}

{We note also that the relatively large annihilation cross section
of wino dark matter is in tension with gamma-ray observations of the Galactic center,
dwarf spheroidals and satellites of the Milky Way made by
the Fermi-LAT and H.E.S.S. telescopes \cite{wino}.
However, there are still considerable ambiguities in the dark matter profiles near the Galactic center
and in these other objects. Including these indirect constraints on dark matter annihilation
in our likelihood analysis would require estimates of these underlying astrophysical
uncertainties~\cite{Benito:2016kyp}, which are beyond the scope of the present work.}

\section{Analysis Procedure}
\label{sec:Framework}

\subsection{MasterCode Framework}

We define a global $\chi^2$ likelihood function that combines the theoretical predictions with experimental constraints, as done in our previous analyses~{\cite{mc9,mc10,mc11,mc12,mc14-SU5}}.

We calculate the observables that go into the likelihood using the  {\tt
  MasterCode} framework~\cite{oldmc,mc9,mc10,mc11,mc12,mc14-SU5,mcweb},
which interfaces various public and private codes: {\tt
  SoftSusy~3.7.2}~\cite{Allanach:2001kg}  for the spectrum,
{\tt FeynHiggs~2.12.1}~{\cite{DeltaMhTH,FH}} {(see Section~\ref{sec:flav-ew-higgs})} for the Higgs sector, the $W$~boson mass
and \gmt, {\tt SuFla}~\cite{SuFla} for the $B$-physics observables, {\tt Micromegas~3.2}~\cite{MicroMegas} 
{(modified as discussed above)} for the dark matter relic density, {\tt SSARD}~\cite{SSARD} for the spin-independent cross-section \ssi\ and the wino dark matter relic density, {\tt SDECAY~1.3b}~\cite{Muhlleitner:2003vg} for calculating sparticle branching ratios, and {\tt HiggsSignals~{1.4.0}}~\cite{HiggsSignals}  and {\tt HiggsBounds~{4.3.1}}~\cite{HiggsBounds,HBtautau} for calculating constraints on the Higgs sector. 
The codes are linked using the SUSY Les Houches Accord (SLHA)~\cite{SLHA}.

We use {\tt SuperIso}~\cite{SuperIso} and {\tt Susy\_Flavor}~\cite{Susy_Flavor} to check our evaluations
of flavour observables, and we have used {\tt Matplotlib}~\cite{matplotlib} and {\tt PySLHA}~\cite{pyslha}
to plot the results of our analysis.

\subsection{Parameter Ranges}

The ranges of the mAMSB parameters that we sample are shown in Table~\ref{tab:ranges}. We also indicate in the right column of this Table how we divide the ranges of these parameters into segments, as we did previously for our analyses of the CMSSM, NUHM1, NUHM2, pMSSM10 and SU(5) ~\cite{mc9,mc10,mc11, mc12, mc14-SU5}. The combinations of these segments constitute boxes, in which we sample the parameter space using the {\tt MultiNest} package~\cite{multinest}. 
For each box, we choose a prior for which 80\% of the sample has a flat distribution within the nominal range, and 20\% of the sample is outside the box in normally-distributed tails in each variable. In this way, our total sample exhibits a smooth overlap between boxes, eliminating spurious features associated with box boundaries.
Since it is relatively fine-tuned, we made a dedicated supplementary 36-box scan of the Higgsino-LSP 
region of the mAMSB parameter space, requiring the lightest neutralino to be Higgsino-like. 
We have sampled a total of { $11 (13) \times 10^6$ points for $\mu > 0$ ($\mu < 0$).}

\begin{table*}[htb!]
\begin{center}
\begin{tabular}{|c|c|c|c|} \hline
Parameter   &  \; \, Range      	& Generic  	& Higgsino 	\\ 
            &             			& Segments	& Segments 	\\ 
\hline         
$m_0$       &  ( 0.1  , $50 \tev$)    & 4 		& 6  		\\
$m_{3/2}$   &  ( 10  , $1500 \tev$) & 3 		& 3  		\\
\tb         &  ( 1  , 50)      		& 4 		& 2  		\\
\hline \hline
\multicolumn{2}{|c|}{Total number of boxes}    	& 48 		& 36 		\\
\hline
\end{tabular}
\caption{\it Ranges of the mAMSB parameters sampled, together with the numbers of
segments into which each range was divided, and the corresponding number of sample boxes. The numbers of segments and boxes are shown both for the generic scan and for the 
{supplementary scan} where we constrain the neutralino to be Higgsino like.} 
\label{tab:ranges}
\end{center}
\end{table*}

\section{Results}
\label{sec:Results}

\subsection{{Case I:} CDM is mainly the lightest neutralino}

\begin{figure*}[htb!]
\vspace{0.5cm}
\begin{center}
\resizebox{7.5cm}{!}{\includegraphics{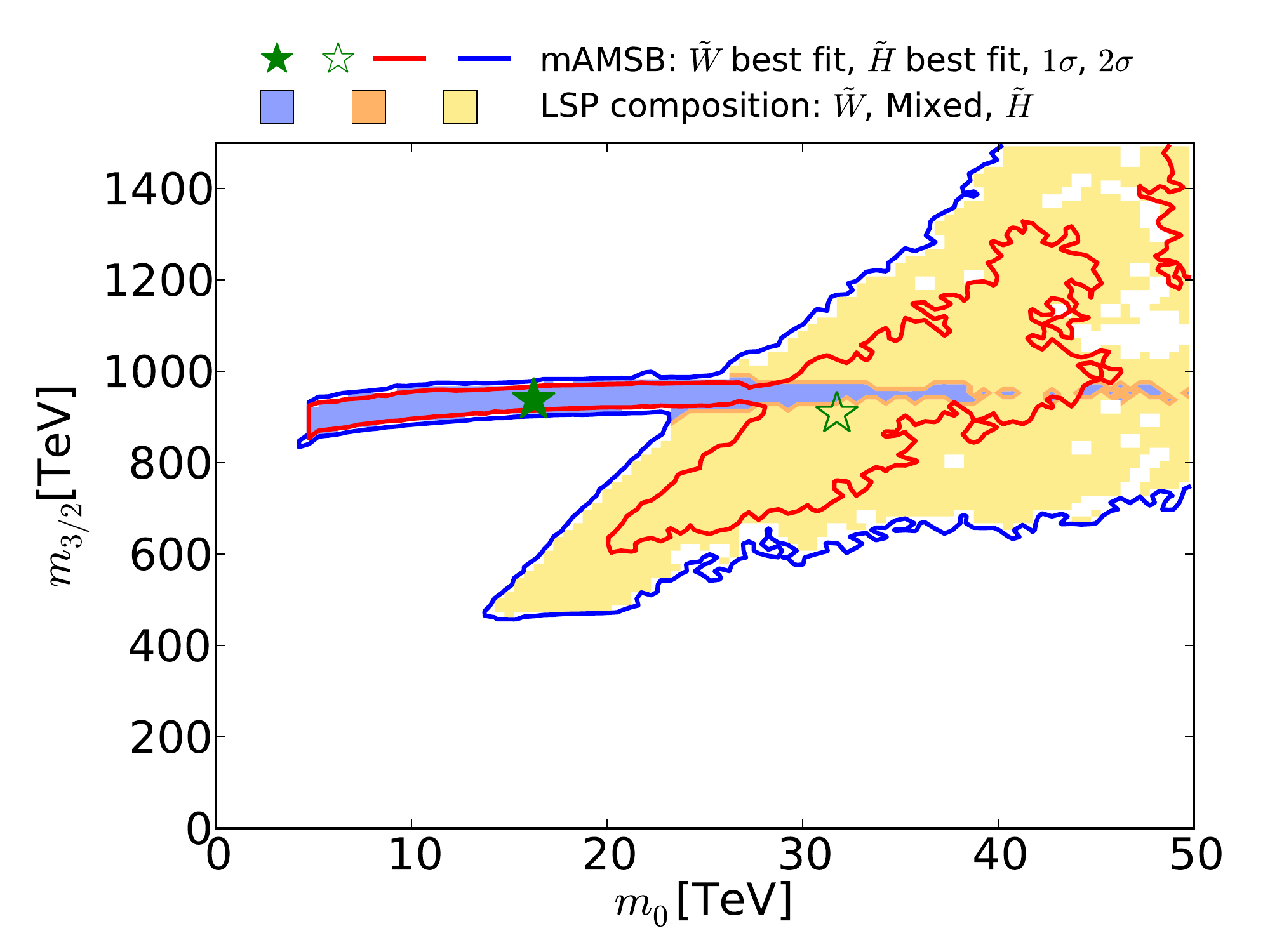}}\put(-169, +123){\footnotesize $\mu>0$, $\Omega_{\neu1}=\Omega_{\rm CDM}$}
\resizebox{7.5cm}{!}{\includegraphics{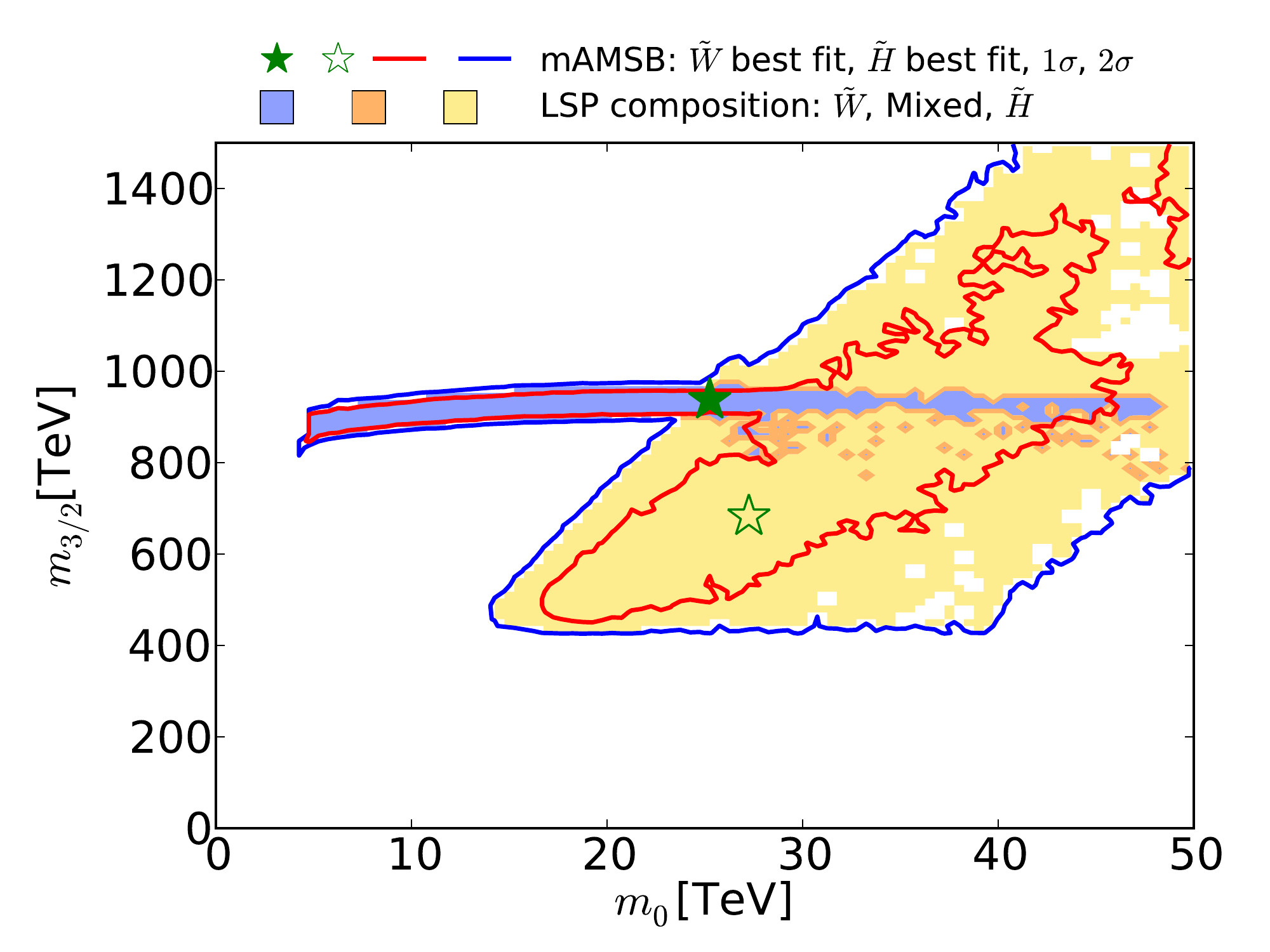}}\put(-169, +123){\footnotesize $\mu<0$, $\Omega_{\neu1}=\Omega_{\rm CDM}$}
\end{center}
\vspace{-1.0cm}
\caption{\it The $(m_0, \mgrav)$ planes for $\mu > 0$ (left panel) and $\mu < 0$ (right panel). The 
red and blue coloured contours surround regions that are allowed at the 68 and 95\% confidence levels (CLs),
corresponding {approximately} to one and two standard deviations, respectively,
{assuming that all the CDM is provided by the $\neu1$}. {The wino- (Higgsino-)like DM regions are shaded blue (yellow),
and mixed wino-Higgsino regions are shaded orange.}
The best-fit points for the two signs of $\mu$ are indicated by green stars, {closed in the wino-like region and open in the Higgsino-like region}. }
\label{fig:m0m32}
\end{figure*}

We display in Fig.~\ref{fig:m0m32} the $(m_0, m_{3/2})$ planes for our
sampling of mAMSB parameters with $\mu > 0$ (left panel) and $\mu < 0$
(right panel). The coloured contours bound regions of parameter space
{with $\Delta\chi^2 = 2.30$ and 
$\Delta\chi^2 = 5.99$ contours, which we use as proxies for the 
boundaries of the 68\% (red) and 95\% (dark blue) CL regions.}
The best-fit points for the two signs of $\mu$ are indicated by green stars, {closed in the case of
wino-like DM, open in the case of Higgsino-like DM}. {The shadings in this and subsequent planes
indicate the composition of the sample point with the lowest $\chi^2$ in this projection: in general, there will also be
sample points with a different composition and (possibly only slightly) larger $\chi^2$.}
{Different shading colours represent the composition of the $\neu1$ LSP: a region with Higgsino fraction exceeding
90\% is shaded {yellow}, one with wino fraction exceeding 90\% is shaded light blue, while other cases are shaded 
orange}~\footnote{{The uncoloured patches and the irregularities in the contours are due to the limitations of our sampling.}}. Most of blue shading corresponds to a  wino-like LSP, and in only a small fraction of cases to a mixed wino-Higgsino state.
We see that in the case of a wino-like LSP, the regions favoured at the 2-$\sigma$ level are bands with $900 \tev \lesssim \mgrav \lesssim 1000 \tev$ corresponding to the envelope of the  near-horizontal band in the right panel of Fig.~\ref{fig:Sommerfeld2} and in Fig.~\ref{fig:bu} that is obtained when {profiling} over $\tan \beta$. For both
signs of $\mu$, the lower limit $m_0 \gtrsim 5\tev$ is due to the \stau1 {becoming the LSP}.

\begin{figure*}[htb!]
\vspace{0.5cm}
\begin{center}
\resizebox{7.5cm}{!}{\includegraphics{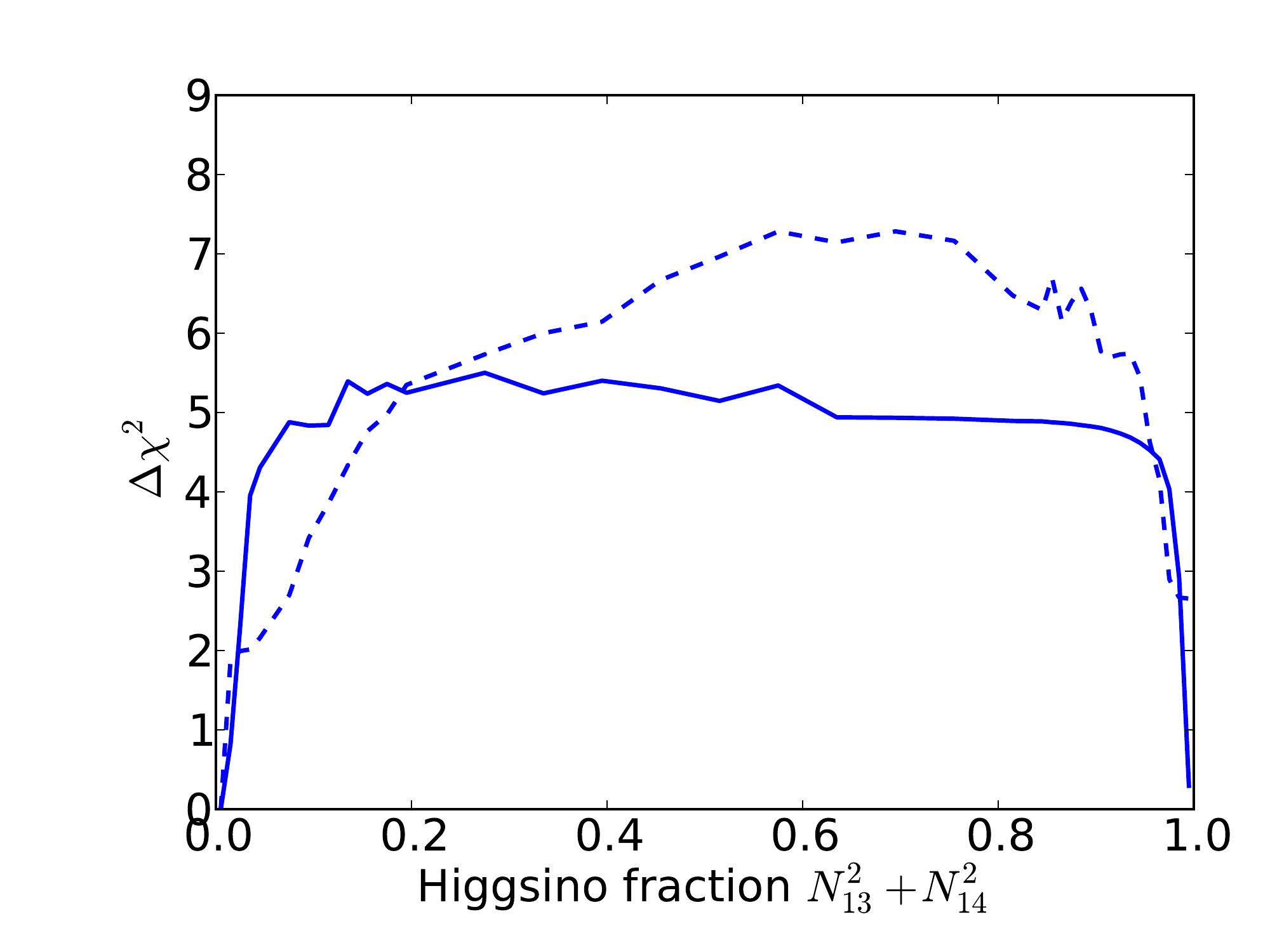}}\put(-169, +123){\footnotesize $\mu>0$}
\resizebox{7.5cm}{!}{\includegraphics{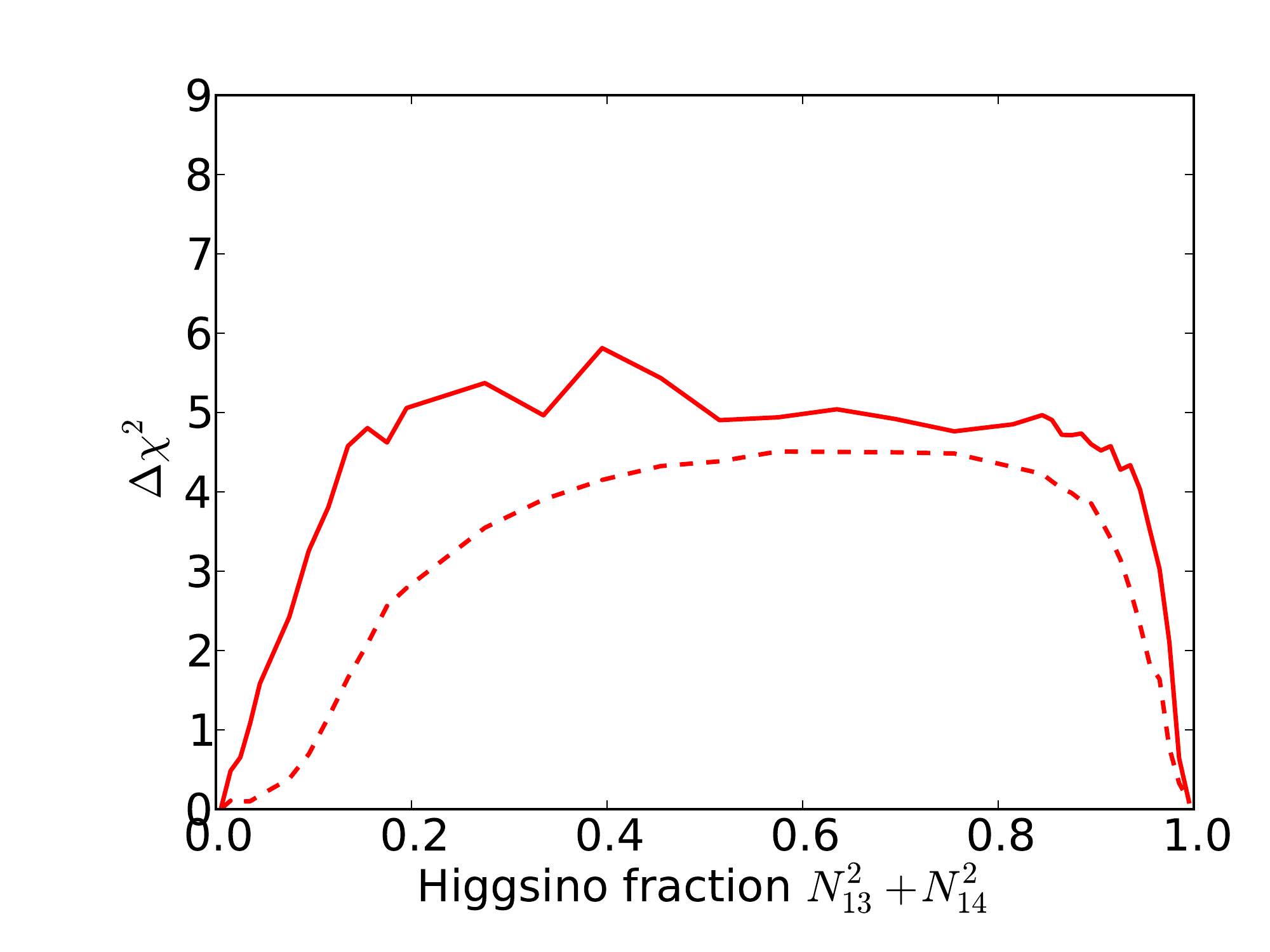}}\put(-169, +123){\footnotesize $\mu<0$}
\end{center}
\vspace{-1.0cm}
\caption{\it Profiled $\Delta\chi^2$ of the $\neu1$ Higgsino fraction for $\mu > 0$ (left panel) and for $\mu < 0$ (right panel). {The profiles for the $\Omega_{\neu1} = \Omega_{\rm CDM}$ case and for the $\Omega_{\neu1} \le \Omega_{\rm CDM}$ case are shown as solid and dashed lines respectively.  The lowest-$\chi^2$ point in the Higgsino-LSP region (${N_{13}^2 + N_{14}^2} \simeq1$) has very similar $\chi^2$ to the wino-LSP best-fit point (${N_{13}^2 + N_{14}^2}\simeq 0$), except in the $\mu>0$
$\Omega_{\neu1} \le \Omega_{\rm CDM}$ case.} }
\label{fig:Higgsino_frac}
\end{figure*}

The yellow Higgsino-LSP regions correspond to the envelope of the V-shaped
diagonal strips seen in Fig.~\ref{fig:Sommerfeld2} and in Fig.~\ref{fig:bu}. 
The locations of these diagonal strips vary significantly with $\tan \beta$ and $\mt$, and their extents
are limited at small and large gravitino mass mainly by the Higgs mass constraint. The best-fit point for the Higgsino-LSP scenario has a total $\chi^2$ very similar to the wino-LSP case, as is shown in Fig.~\ref{fig:Higgsino_frac}.
The $\chi^2$ values at the best-fit points in the wino- and Higgsino-like regions
for both signs of $\mu$ are given in Table~\ref{tab:parameters_DM},
together with more details of the fit results {(see below)}.

\begin{table*}[htb!]
\begin{center}
\begin{tabular}{|c|c|c|c|c|} 
\hline
 					   & \multicolumn{2}{|c|}{Wino-LSP}  					& \multicolumn{2}{|c|}{Higgsino-LSP}       \\
Parameter   	            &      $\mu>0$      			& $\mu<0$ 	            & $\mu>0$            & $\mu<0$            \\ 
\hline       
$m_0$: best-fit value     &      $16 \tev$     	&  $25 \tev$ 	    & $ 32 \tev$     & $27 \tev$    \\
  	      ~~68\% range &      $(4, 40) \tev$     	&  $(4, 43) \tev$ 	    & $(23,50) \tev$     & $(18, 50) \tev$    \\
	      \hline
$\mgrav$: best-fit value  &      $940 \tev$    	&  $940 \tev$ 	& $920 \tev$ & $650 \tev$ \\
~~68\% range              &   $(860, 970) \tev$  &  $(870, 950) \tev$ 	& $(650, 1500) \tev$ & $(480, 1500) \tev$ \\
\hline
$\tb$: best-fit value     &      $5.0 $ 		&  $4.0$             & $4.4$            & $4.2$            \\
  ~~68\% range        &      $(3,8)$ and $(42,48)$ 	&  $(3, 7)$             & $(3,7)$            & $(3,7)$            \\
\hline \hline
$\chi^2 / {\rm d.o.f}$ & 36.4 / 27	       		& 36.4 / 27             & 36.6 / 27          & 36.4 / 27          \\
$\chi^2$ probability        & $10.7\%$               	& $10.7\%$                  & $10.2\%$               & $10.7\%$               \\
\hline
\end{tabular}
\caption{\it Fit results for the mAMSB assuming that the LSP makes the dominant contribution to the cold dark matter density. The 68\% CL ranges correspond to $\Delta\chi^2 = 1$. We also display the values of the global $\chi^2$ function {omitting the contributions from {\tt HiggsSignals}}, and the {corresponding $\chi^2$ probability} values. Each mass range is shown for
both the wino- and higgsino-LSP scenarios as well as for both signs of $\mu$.
} 
\label{tab:parameters_DM}
\end{center}
\end{table*}

\begin{figure*}[htb!]
\vspace{0.5cm}
\begin{center}
\resizebox{7.5cm}{!}{\includegraphics{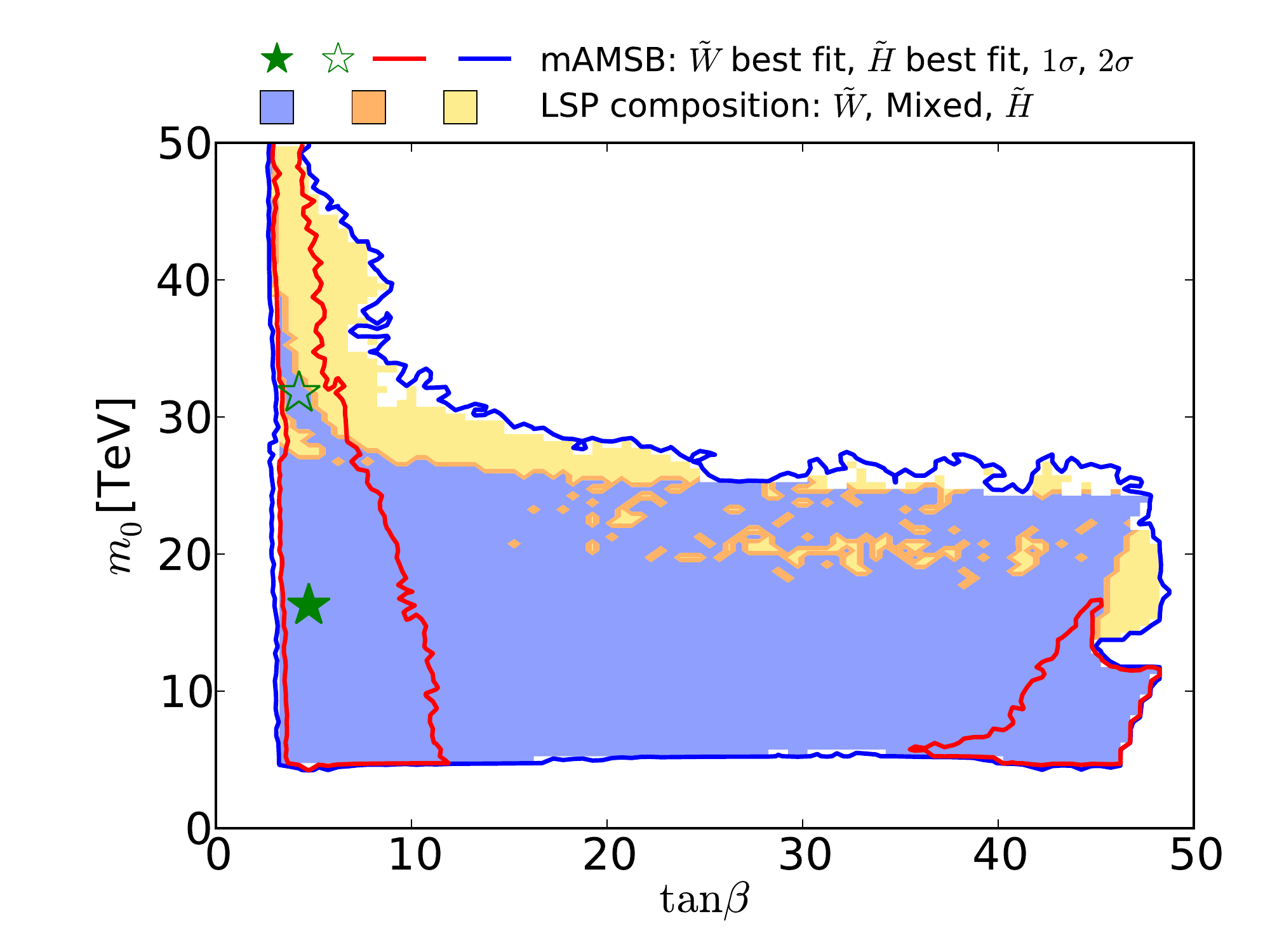}}\put(-95, +123){\footnotesize $\mu>0$, $\Omega_{\neu1}=\Omega_{\rm CDM}$}
\resizebox{7.5cm}{!}{\includegraphics{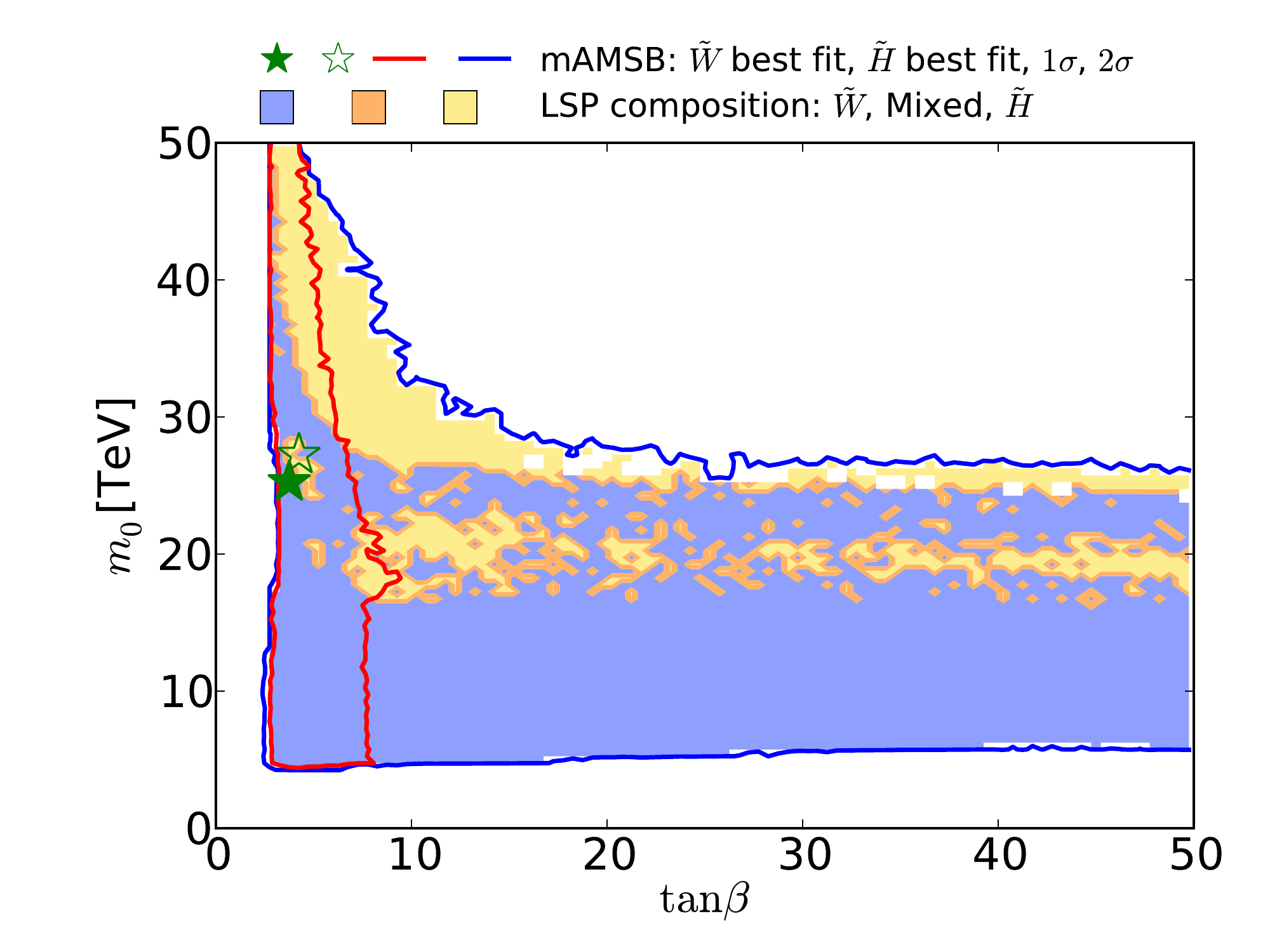}}\put(-95, +123){\footnotesize $\mu<0$, $\Omega_{\neu1}=\Omega_{\rm CDM}$}
\end{center}
\vspace{-1.0cm}
\caption{\it The $(\tb, m_0)$ planes for $\mu > 0$ (left panel) and for $\mu < 0$ (right panel),
{assuming that the $\neu1$ provides all the CDM density}.
The colouring convention for the
{shadings and} contours is the same as in Fig.~\protect\ref{fig:m0m32}, and the best-fit points for the two signs of $\mu$ are again indicated by green stars.}
\label{fig:tbm0}
\end{figure*}

\begin{figure*}[htb!]
\vspace{0.5cm}
\begin{center}
\resizebox{7.5cm}{!}{\includegraphics{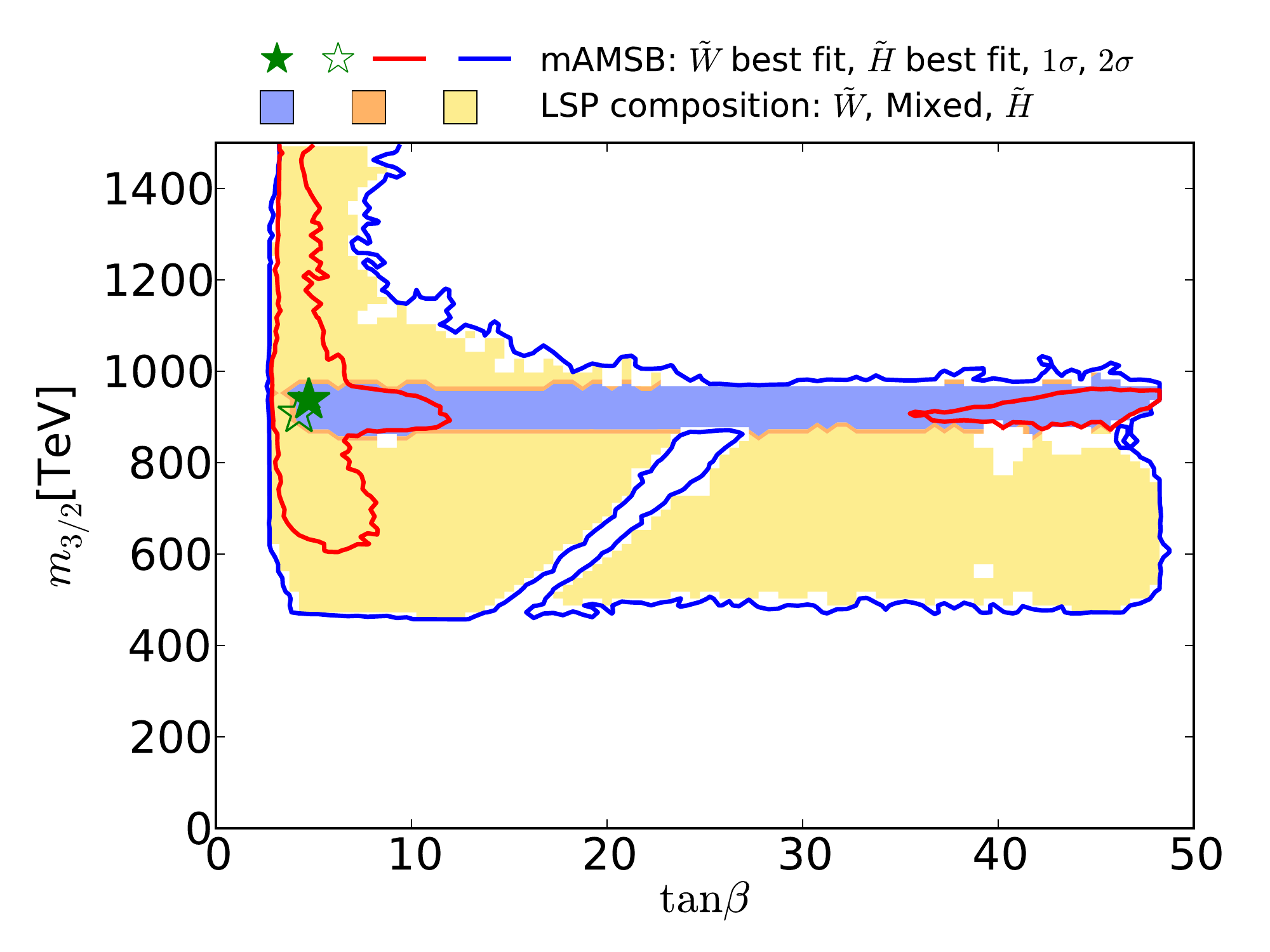}}\put(-95, +123){\footnotesize $\mu>0$, $\Omega_{\neu1}=\Omega_{\rm CDM}$}
\resizebox{7.5cm}{!}{\includegraphics{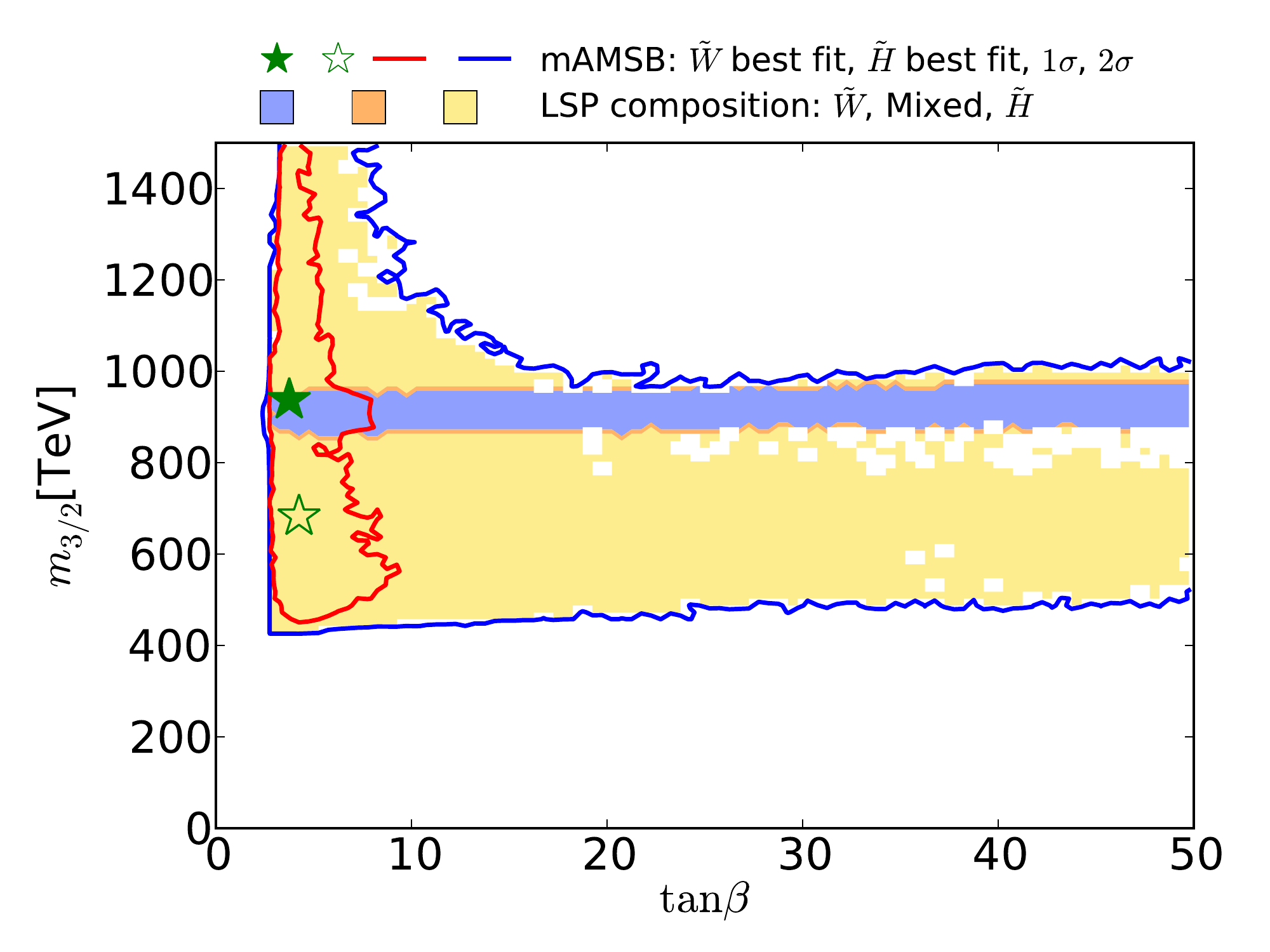}}\put(-95, +123){\footnotesize $\mu<0$, $\Omega_{\neu1}=\Omega_{\rm CDM}$}
\end{center}
\vspace{-1.0cm}
\caption{\it The $(\tb, \mgrav)$ planes for $\mu > 0$ (left panel) and $\mu < 0$ (right panel),
{assuming that the $\neu1$ provides all the CDM density}.
The shadings and colouring convention for the contours are the same as in Fig.~\protect\ref{fig:m0m32},
and the best-fit points for the two signs of $\mu$ are again indicated by green stars.}
\label{fig:tbm32}
\end{figure*}

Figs.~\ref{fig:tbm0} and~\ref{fig:tbm32}  display the $(\tb, m_0)$ and $(\tb, \mgrav)$ planes respectively. Both the $\mu > 0$ case (left panel) and the $\mu < 0$ case (right panel) are shown,
{and are qualitatively similar}. The best-fit points for the two signs of $\mu$ are again indicated by green stars. Larger $m_0$ and \mgrav~values are allowed in the Higgsino-LSP case, provided that \tb\ is small.
Values of $\tan \beta \gtrsim 3$ are allowed at the 95\% CL with an upper limit at 48 only in the $\mu>0$ case.
There are regions favoured at the 68\% CL with small values of $\tb \lesssim 10$ 
in both the wino- and Higgsino-like cases
for both signs of $\mu$. In addition, for $\mu > 0$ there is another 68\% CL preferred region in the wino case
at $\tb \gtrsim 35$, where supersymmetric contributions improve the consistency with the measurements of
\bsdmm, as discussed in more detail in Section~5.3~
\footnote{{The diagonal gap in the left panel of Fig.~\ref{fig:tbm32} for $\mu >0$ is in a region where our
numerical calculations encounter instabilities.}}.

The parameters of the best-fit points for $\mu >0$ and $\mu < 0$ are listed in Table~\ref{tab:parameters_DM}, together with their 68\% CL ranges corresponding to $\Delta\chi^2 = 1$.
We see that at the 68\% CL the range of $\tb$ is restricted to low values for both LSP compositions, with the exception of the $\mu>0$ wino-LSP case, where also larger $\tb$ values around 45 are allowed. In the wino-LSP scenario, \mgrav\ is restricted to a narrow region around $940\tev$ and $m_0$ is required to be larger than $4 \tev$.
The precise location of the Higgsino-LSP region depends on the spectrum calculator employed, 
and also on the version used. These variations can be as large as tens of TeV for $m_0$ 
or a couple of units for $\tan\beta$, and can {change the
$\chi^2$ penalty} coming from the Higgs mass.
In our implementations, we find that \mgrav\ can take masses as low as $650\tev$ ($480\tev$) while $m_0$ is required to be at least $23\tev$ ($18\tev$) at the 68\% CL in the $\mu>0$ ($\mu<0$) case. 
{This variability is related to the uncertainty in the exact location of the electroweak symmetry-breaking boundary,
which is very sensitive 
to numerous corrections, in particular those related to the top quark Yukawa coupling.}

The minimum values of the global $\chi^2$ function for the two signs of $\mu$ are also shown in Table~\ref{tab:parameters_DM}, as are the {$\chi^2$ probability values obtained by {combining} these with the numbers of effective degrees of freedom.  
We see that all the cases studied (wino- and Higgsino-like LSP, $\mu > 0$ and $\mu < 0$) have similar $\chi^2$ {probabilities, around $11\%$.}

{We show in Fig.~\ref{fig:breakdown} the contributions to the total $\chi^2$ of the best-fit point in the 
scenarios with different hypotheses on the sign of $\mu$ and the composition of CDM. In addition, we report the main $\chi^2$ penalties in Table~\ref{tab:mainchi2}.}

\begin{figure}[htb!]
\begin{center}
  \resizebox{5.05cm}{!}{\includegraphics{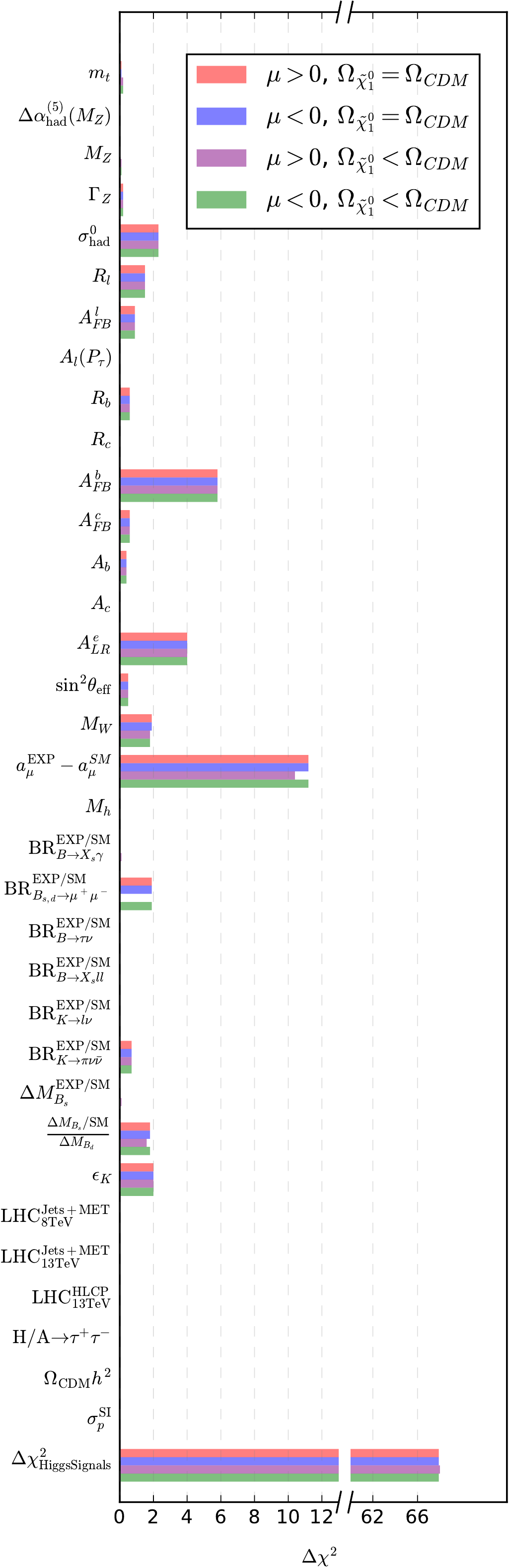}} 
\end{center}
\vspace{-0.5cm}
\caption{\it All the contributions to the total $\chi^2$ for the best-fit points for mAMSB 
assuming different hypotheses on the composition of the dark matter relic density and on the sign of $\mu$
as indicated in the legend.}
\label{fig:breakdown}
\end{figure}

\begin{table*}[htb!]
\begin{center}
\begin{tabular}{|c|c|c|c|c|c|c|c|c|} 
\hline
& \multicolumn{4}{|c|}{$\Omega_{\neu1} = \Omega_{\rm CDM}$ }	& \multicolumn{4}{|c|}{$\Omega_{\neu1} < \Omega_{\rm CDM}$}       \\
& \multicolumn{2}{|c|}{$\tilde{W}$-LSP }	& \multicolumn{2}{|c|}{$\tilde{H}$-LSP}   & \multicolumn{2}{|c|}{$\tilde{W}$-LSP }	& \multicolumn{2}{|c|}{$\tilde{H}$-LSP}    \\
Constraint   	       &  $\mu>0$    & $\mu<0$     & $\mu>0$     & $\mu<0$  &  $\mu>0$    & $\mu<0$     & $\mu>0$     & $\mu<0$   \\ 
\hline         
$\sigma^0_{\mathrm{had}}$ & 2.3 & 2.3 & 2.3 & 2.3 & 2.3 & 2.3 & 2.3 & 2.3\\
$R_l$                & 1.5 & 1.5 & 1.5 & 1.5  & 1.5 & 1.5 & 1.5 & 1.5\\
$A_{\rm FB}^b$           & 5.8 & 5.8 & 5.8 & 5.8  & 5.8 & 5.8 & 5.8 & 5.8\\
$A_{\rm LR}^e$           & 4.0 & 4.0 & 4.0 & 4.0 & 4.0 & 4.0 & 4.0 & 4.0\\
$M_W$                & 1.9 & 1.9 & 2.1 & 1.9 & 1.8 & 1.8 & 1.8 & 1.9 \\
\gmt                 & 11.2       & 11.2       & 11.2   & 11.2 & {10.4} & 11.2 & 11.2 & 11.2 \\
\bsmm                & 1.9        & 1.9        & 1.9    & 1.9 & {0.0} & 1.9 & 1.9 & 1.9   \\
$\frac{\Delta M_{B_s}/{\mathrm{SM}}}{\Delta M_{B_d}}$ & 1.8 & 1.8 & 1.8 & 1.8 & 1.6 & 1.8 & 1.8 & 1.8 \\
$\epsilon_K$ & 2.0 & 2.0 & 2.0 & 2.0 & 2.0 & 2.0 & 2.0 & 2.0 \\
$\Delta\chi^2_{\mathrm{HiggsSignals}}$ & 67.9 & 67.9 & 67.9 & 68.0 & 68.0 & 67.9 & 67.9 & 68.0 \\\hline
\end{tabular}
\caption{\it The most important contributions to the total $\chi^2$ of the best fit points for mAMSB assuming different hypotheses on the composition of the dark matter relic density and on the sign of $\mu$. In the $\mu>0$ scenario with $\Omega_{\neu1} < \Omega_{\rm CDM}$ and $\tilde{W}$-LSP, the experimental constraints from \gmt and \bsmm ~can be accommodated and get a lower $\chi^2$ penalty.} 
\label{tab:mainchi2}
\end{center}
\end{table*}

Figure~\ref{fig:mspectrum_wDM} shows the best fit values (blue lines) of the particle masses
and the 68\% and 95\% CL ranges
 allowed in both the
wino- and Higgsino-like LSP cases
for both signs of $\mu$. 
More complete spectra at the best-fit points for the two signs of $\mu$ are shown in Fig.~\ref{fig:spectrum} in both the wino- and Higgsino-LSP cases, where branching ratios exceeding 20\% are
  indicated by dashed lines. As was apparent from the previous Figures and Tables, a relatively heavy spectrum is favoured in our global fits. {The difference between the best-fit spectra in the Higgsino LSP case 
for $\mu > 0$ and $< 0$ reflects the fact that the likelihood function is quite
flat in the preferred region of {the}
parameter space.} 
In the Higgsino-LSP case, the spectra are even heavier than the other one with a wino LSP, apart from the gauginos, which are lighter. 
Overall, these high mass scales, together with the minimal flavor violation assumption, 
implies that there are, in general,
no significant departures from the SM predictions in the flavour sector or for \gmt.

\begin{figure*}[htb!]
\begin{center}
\resizebox{14cm}{!}{\includegraphics{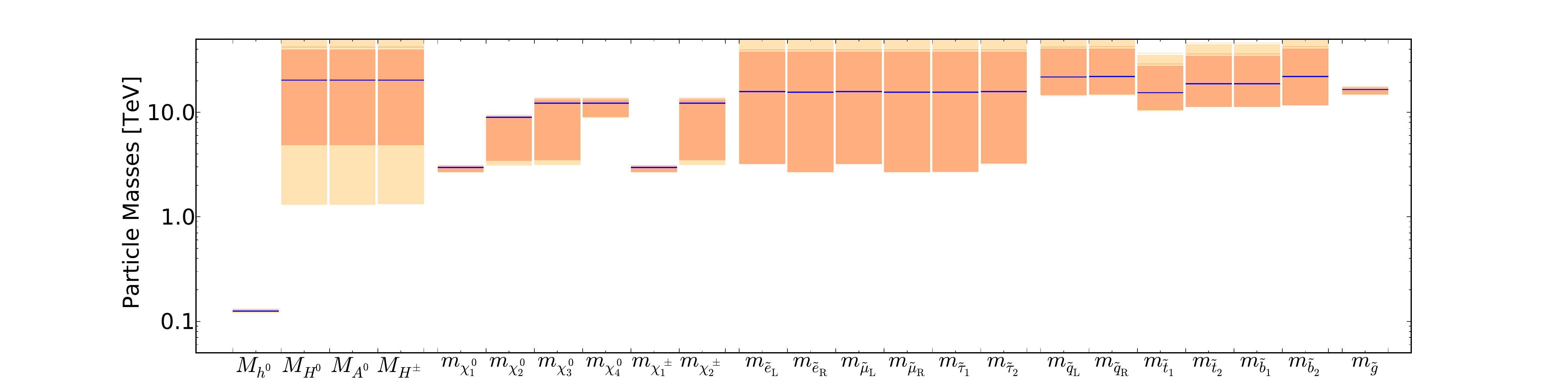}}
\put(-283, +115){\small  $\tilde{W}$-LSP  for $\mu>0$, $\Omega_{\neu1}=\Omega_{\rm CDM}$}
\\\vspace{0.3cm}
\resizebox{14cm}{!}{\includegraphics{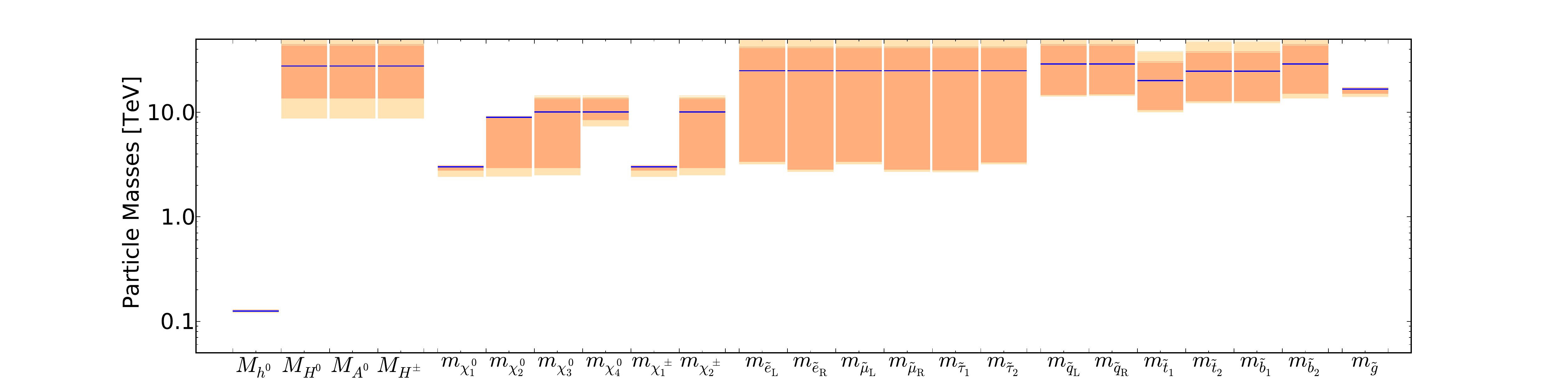}}
\put(-283, +111){\small  $\tilde{W}$-LSP  for $\mu<0$, $\Omega_{\neu1}=\Omega_{\rm CDM}$}
\\\vspace{0.3cm}
\resizebox{14cm}{!}{\includegraphics{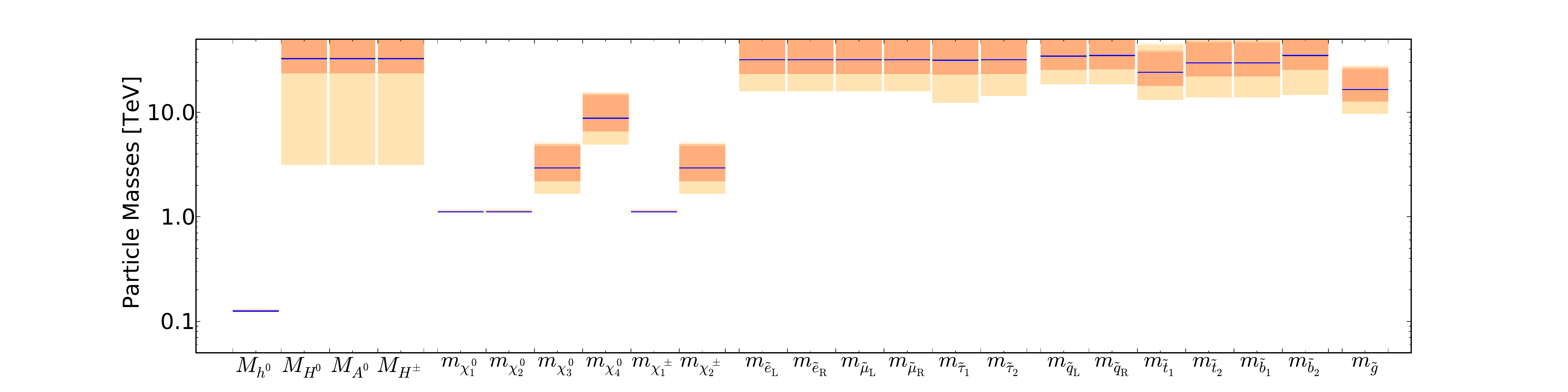}}
\put(-283, +111){\small  $\tilde{H}$-LSP  for $\mu>0$, $\Omega_{\neu1}=\Omega_{\rm CDM}$}
\\\vspace{0.3cm}
\resizebox{14cm}{!}{\includegraphics{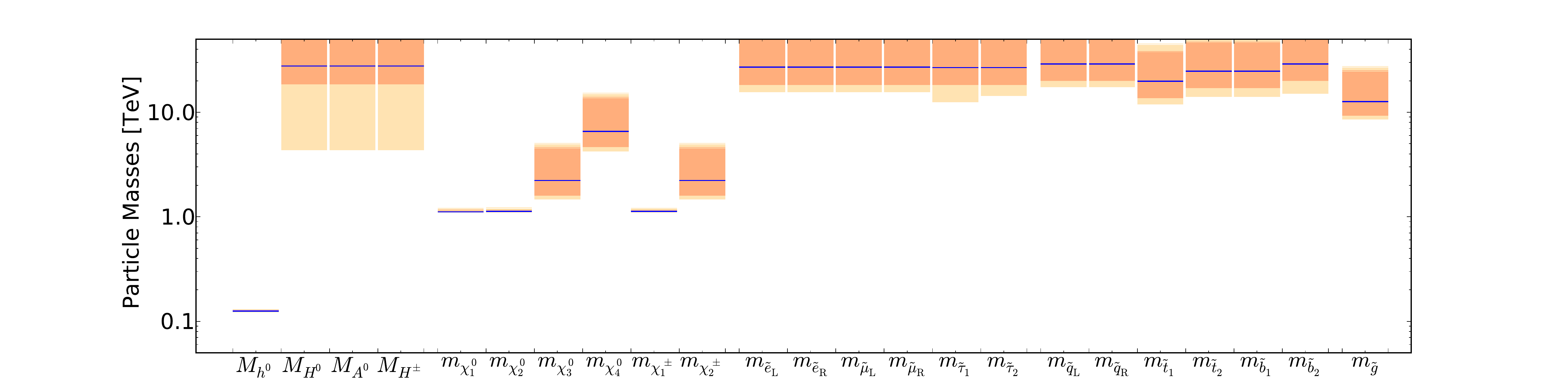}}
\put(-283, +111){\small  $\tilde{H}$-LSP  for $\mu<0$, $\Omega_{\neu1}=\Omega_{\rm CDM}$}
\end{center}
\vspace{-0.5cm}
\caption{\it  The ranges of masses obtained for the wino-like LSP case with $\mu > 0$ (top panel) and 
$\mu < 0$ (second panel), and also for the Higgsino-like LSP case for $\mu > 0$ (third panel)
and $\mu < 0$ (bottom panel), assuming that the LSP makes the dominant contribution to the cold dark matter density.}
\label{fig:mspectrum_wDM}
\end{figure*}

\begin{figure*}[htb!]
\begin{center}
  \resizebox{7.5cm}{!}{\includegraphics{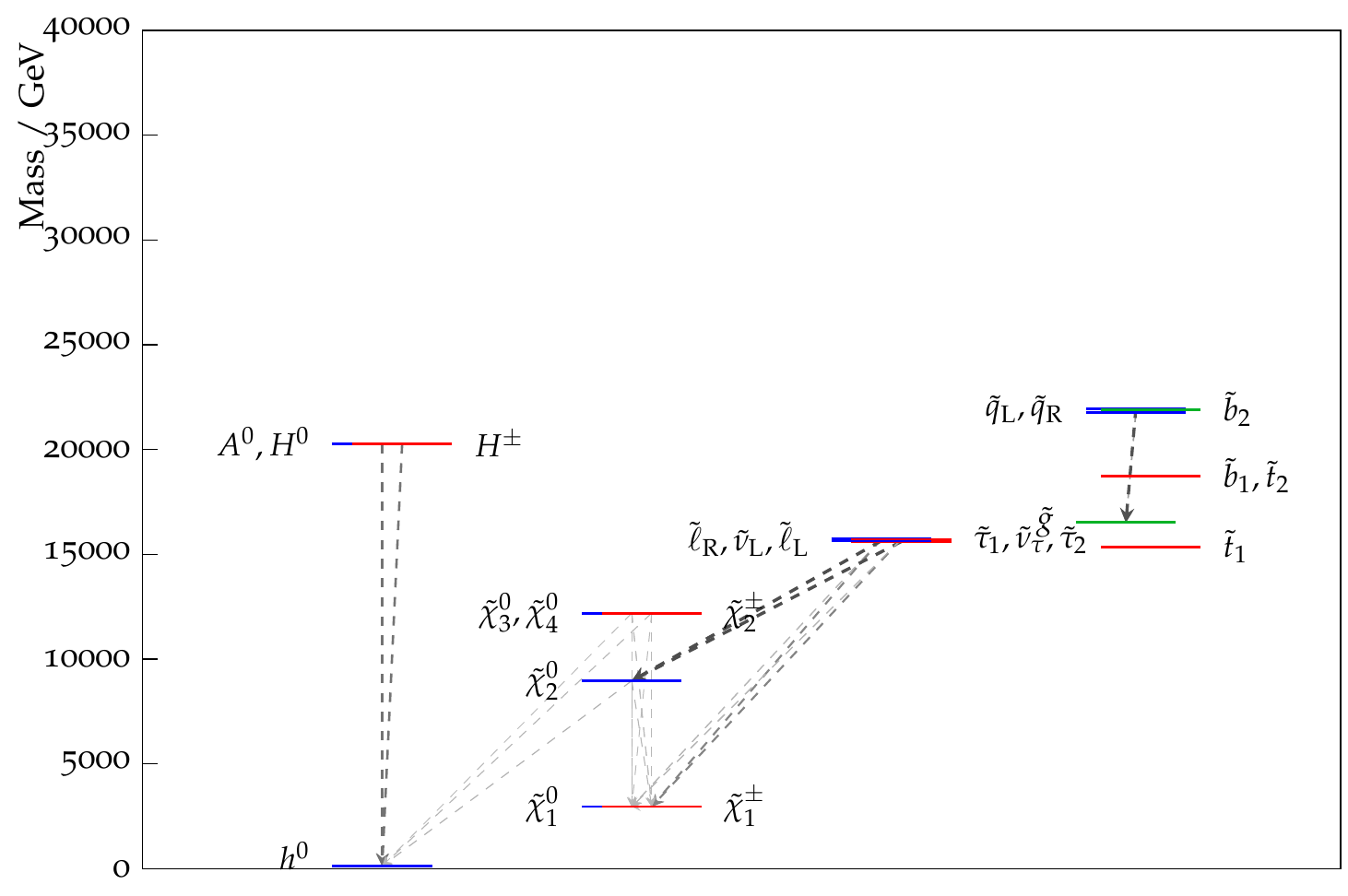}}\put(-174, +144){\small $\tilde{W}$-LSP  for $\mu>0$, $\Omega_{\neu1}=\Omega_{\rm CDM}$}
  \resizebox{7.5cm}{!}{\includegraphics{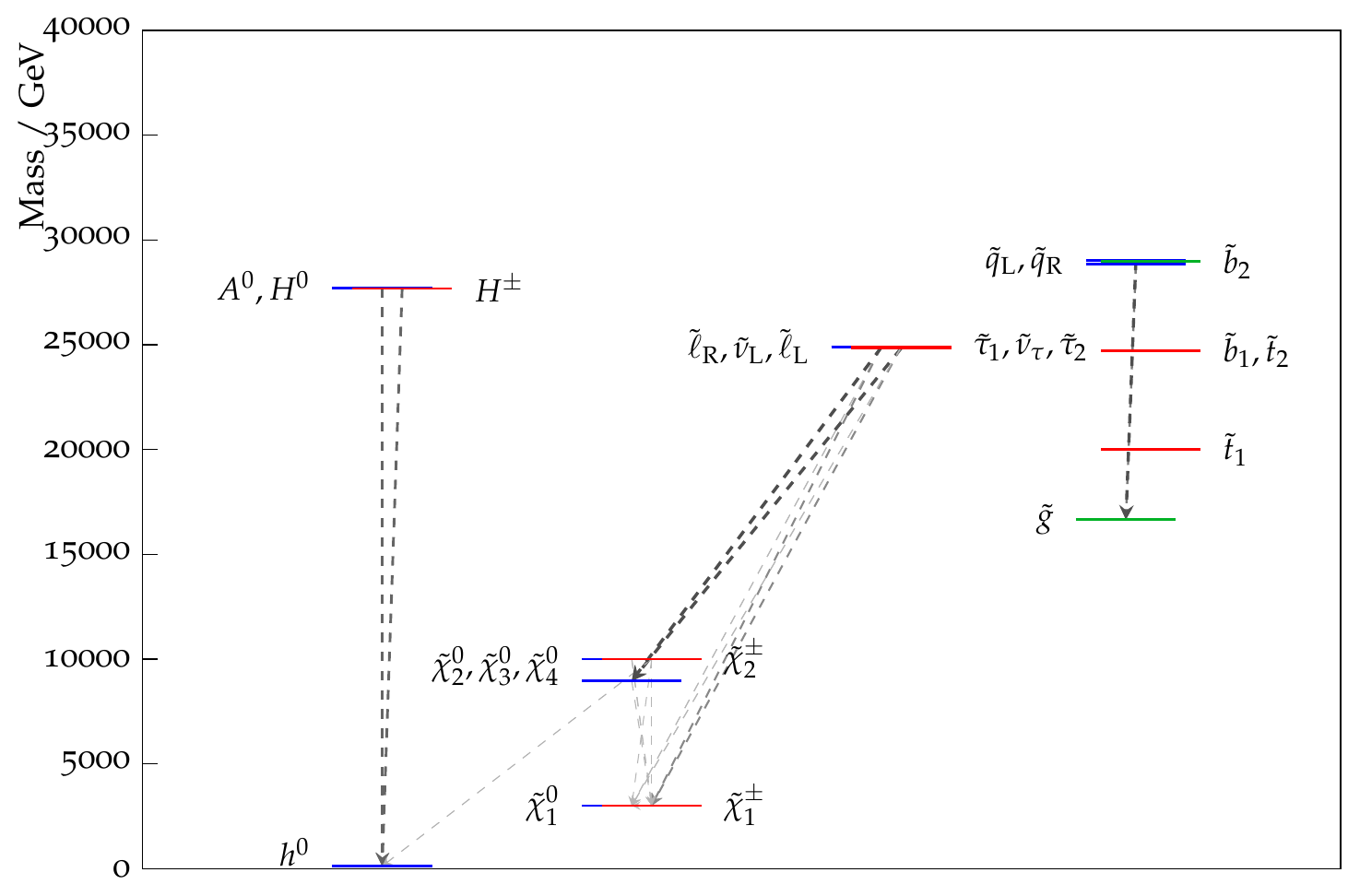}}\put(-174, +144){\small  $\tilde{W}$-LSP  for $\mu<0$, $\Omega_{\neu1}=\Omega_{\rm CDM}$}
\vspace{0.5cm}
\resizebox{7.5cm}{!}{\includegraphics{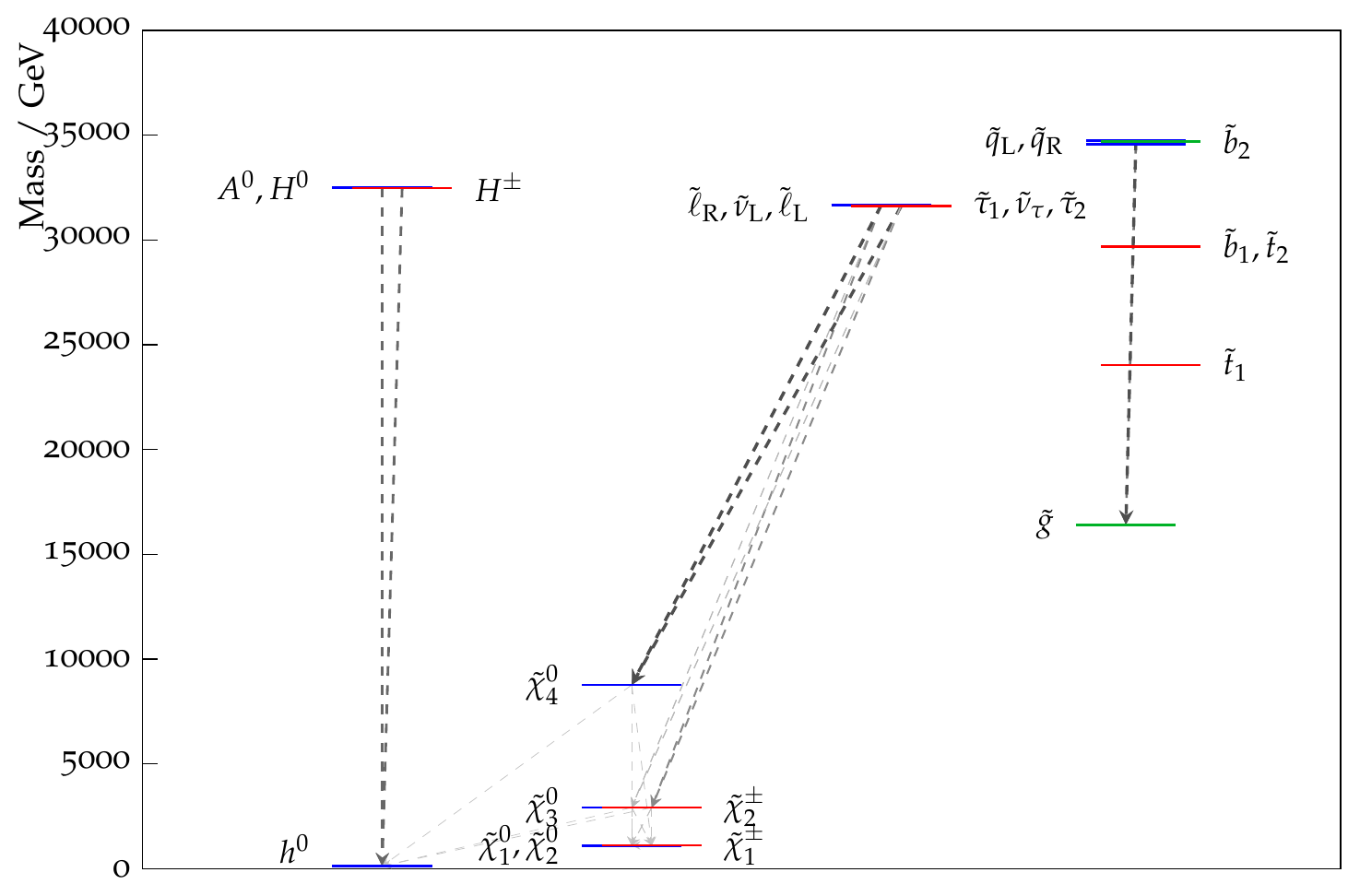}}\put(-174, +144){\small  $\tilde{H}$-LSP  for $\mu>0$, $\Omega_{\neu1}=\Omega_{\rm CDM}$}
\resizebox{7.5cm}{!}{\includegraphics{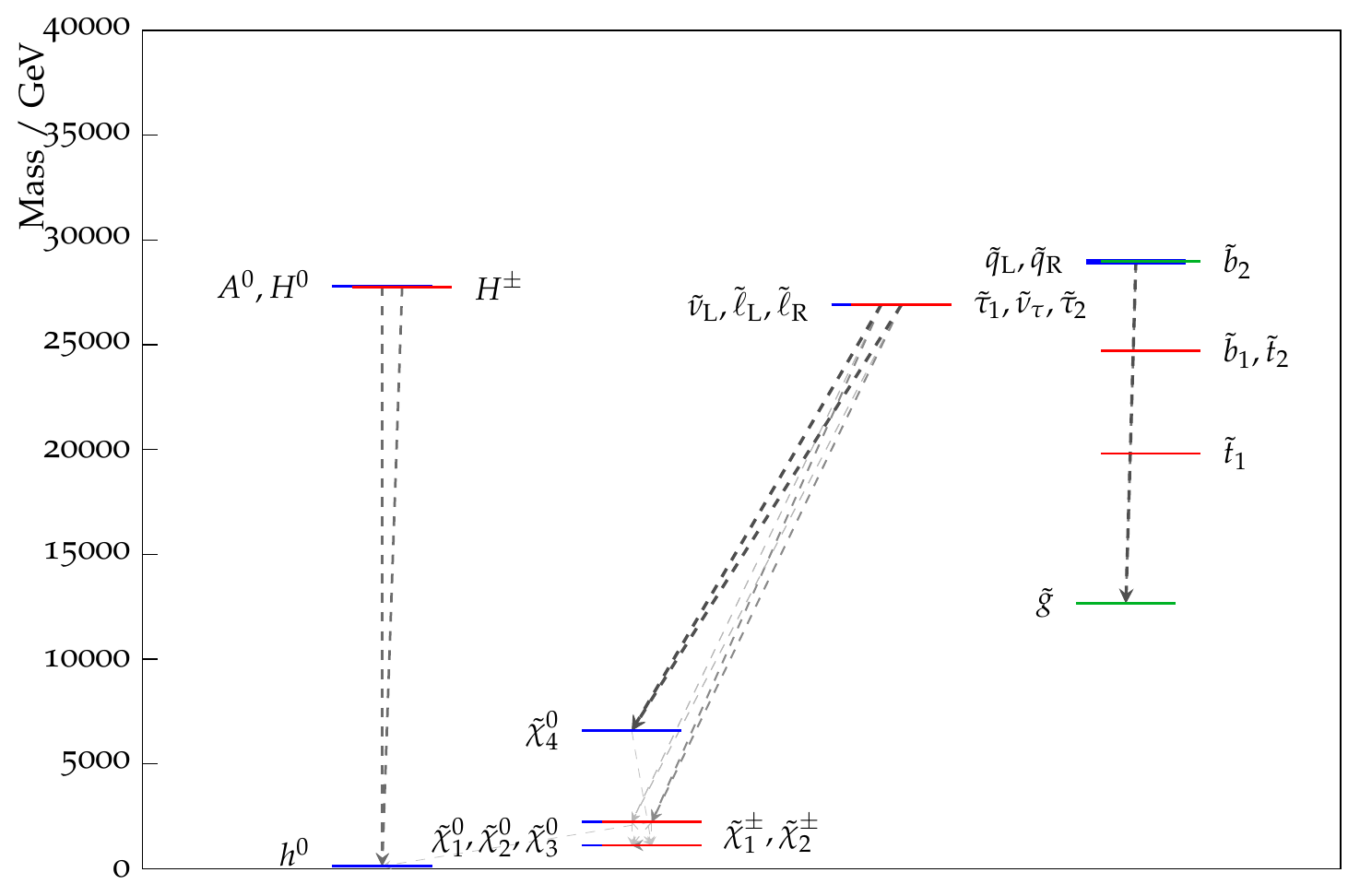}}\put(-174, +144){\small  $\tilde{H}$-LSP  for $\mu<0$, $\Omega_{\neu1}=\Omega_{\rm CDM}$}
\end{center}
\vspace{-0.5cm}
\caption{\it The spectra of our best-fit points for $\mu > 0$ (left panel) and $\mu < 0$ (right panel), assuming that the LSP makes the dominant contribution to the cold dark matter density. Both the wino- (upper) and the Higgsino-like LSP (lower) best-fit points are shown. In each case, we also
indicate all the decay modes with branching ratios above 20\%. }
\label{fig:spectrum}
\end{figure*}

{Figure~\ref{fig:MAtball} shows the $(M_A, \tb)$ planes for $\mu > 0$ (left panel) and for $\mu < 0$ (right panel),
assuming that the $\neu1$ contributes all the CDM density.  As previously, the red (blue) contours represent 
the $68\%$ ($95\%$) CL contours, and the wino- (Higgsino-)like DM regions are shaded blue (yellow),
{and mixed wino-Higgsino regions are shaded orange. 
{We find that the impact of the recent LHC 13-TeV constraints on the $(M_A, \tb)$ plane is small in these plots.}}
{We see here that the large-$\tb$ 68\% CL region mentioned above corresponds to $\MA \lesssim 6 \tev$.}

\begin{figure*}[htb!]
\vspace{0.5cm}
\begin{center}
\resizebox{7.5cm}{!}{\includegraphics{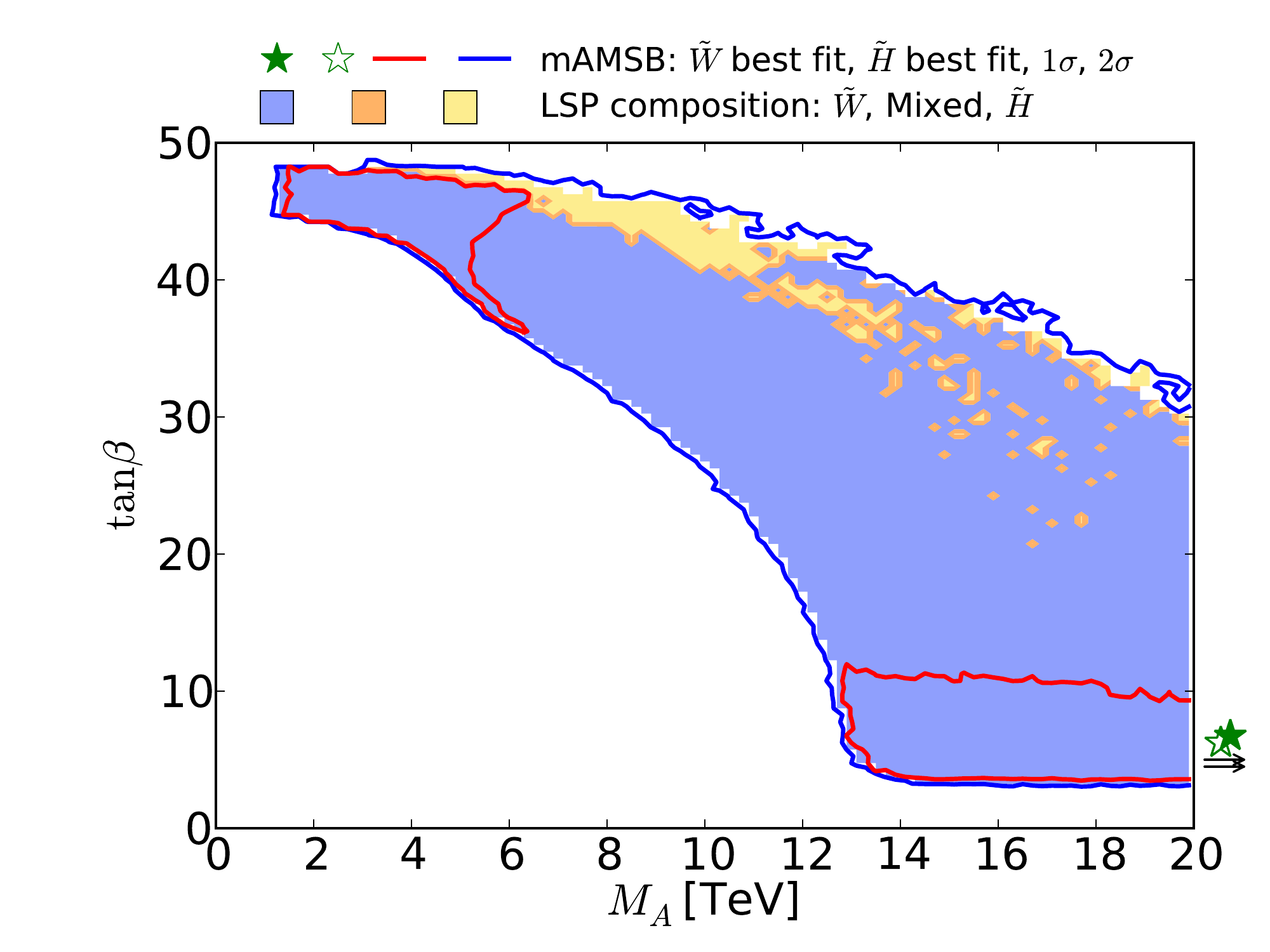}}\put(-169, +33){\footnotesize $\mu>0$, $\Omega_{\neu1}=\Omega_{\rm CDM}$}
\resizebox{7.5cm}{!}{\includegraphics{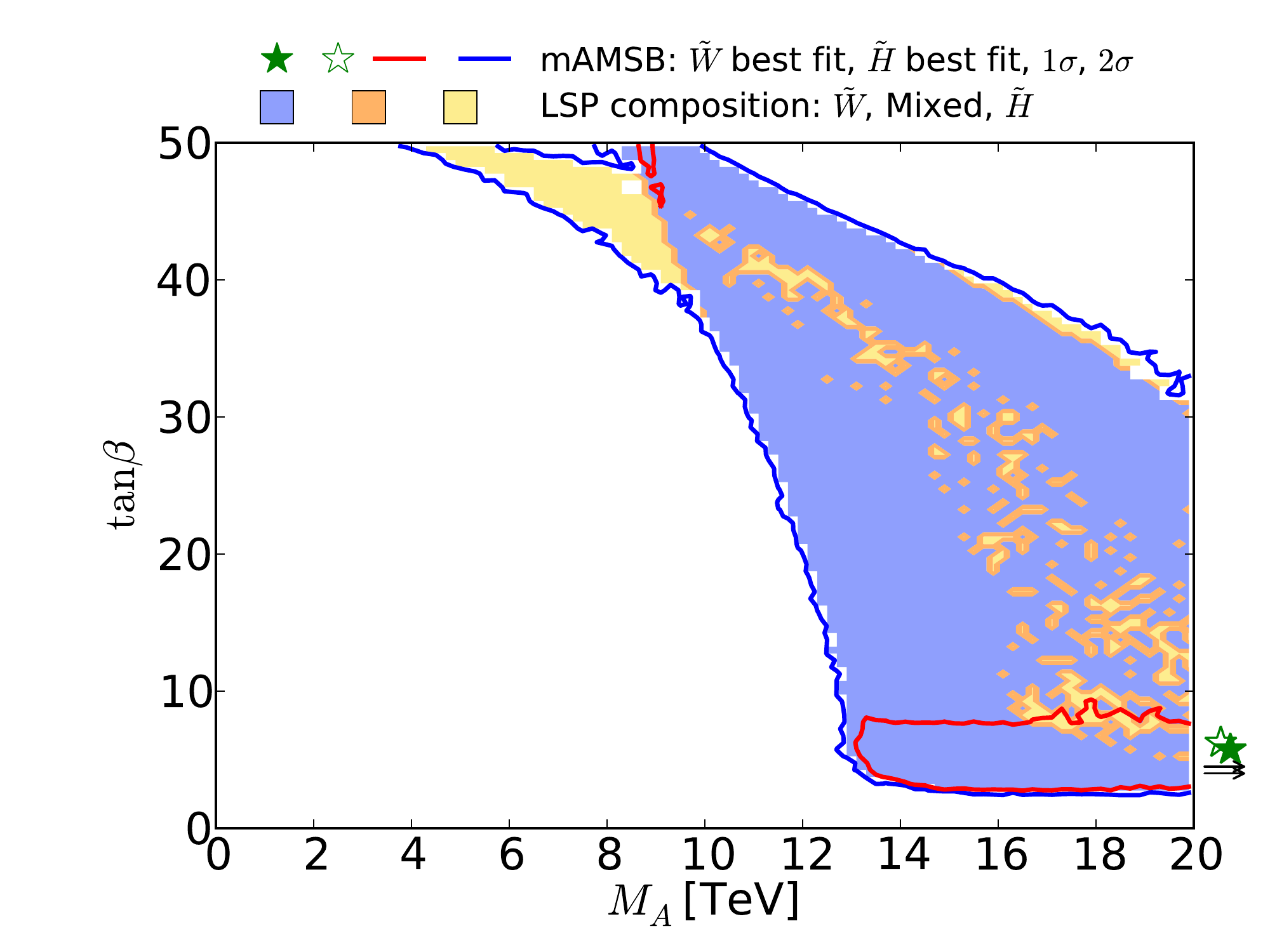}}\put(-169, +33){\footnotesize $\mu<0$, $\Omega_{\neu1}=\Omega_{\rm CDM}$}
\end{center}
\vspace{-0.5cm}
\caption{\it {The $(M_A, \tb)$ planes for $\mu > 0$ (left panel) and for $\mu < 0$ (right panel),
  assuming that the $\neu1$ contributes all the CDM density. As previously,
  the red (blue) contours represent the $68\%$ ($95\%$) CL contours, and the wino- (Higgsino-)like {(mixed wino-Higgsino) DM regions are shaded blue (yellow) (orange)}.}}
\label{fig:MAtball}
\end{figure*}

{As anticipated in Section~\ref{sec:intro}, the wino-LSP is almost degenerate with the lightest chargino, which
acquires a mass about $170 \mev$ larger through radiative corrections.
Therefore,  
because of phase-space suppression the chargino acquires a lifetime around $0.15 \, {\rm ns}$,
and therefore may decay inside the ATLAS tracker. However, the ATLAS search for disappearing tracks~\cite{disappearing-ATLAS} is insensitive to the large mass $\sim 2.9\tev$
expected for the mAMSB chargino if the LSP makes up all the dark matter.
In Section~\ref{sec:noDM} we estimate the LHC sensitivity to the lower chargino masses 
that are possible if the $\neu1$ contributes only a fraction of the cold dark matter density.  
In the Higgsino-LSP case, the chargino has a mass $\sim 1.1\tev$
in the all-DM case, but its lifetime is very short, of the order of few ps. 

{The 68\% CL ranges of the neutralino masses, the gluino mass, the $\cha1 - \neu1$
mass splitting and the $\cha1$ lifetime are reported in Table~\ref{tab:fit_DM},
assuming that the $\neu1$ accounts for all the CDM density. Each parameter is shown for
both the wino- and Higgsino-like LSP scenarios and for the two signs of $\mu$.}

\begin{table*}[htb!]
\begin{center}
\begin{tabular}{|c|c|c|c|c|} 
\hline
 & \multicolumn{2}{|c|}{wino-LSP}  & \multicolumn{2}{|c|}{Higgsino-LSP} \\
Parameter   			&  $\mu>0$      & $\mu<0$ & $\mu>0$      & $\mu<0$ \\ 
\hline                 
$\mneu1$        	       & $2.9 \pm 0.1 \tev$        & $2.9 \pm 0.1 \tev$  & $1.12 \pm 0.02 \tev$ & $1.13 \pm 0.02 \tev$\\
$\mneu2$        		& $(3.4, 9.2) \tev$ 		& $(2.9, 9.1) \tev$   & $1.13 \pm 0.02 \tev$ & $1.14 \pm 0.02 \tev$\\
$\mneu3$        		& $(3.5, 13.5) \tev$ 	& $(2.9, 13.5) \tev$  & $(2.2, 4.9)\tev$     & $(1.7, 4.6)\tev$ \\
$\mneu4$        		& $(9.0, 13.5) \tev$ 	& $(8.4, 13.5) \tev$  & $(6.5, 15.0)\tev$    & $(4.6, 14.0)\tev$\\
$m_{\tilde{g}}$     	       & $16 \pm 1 \tev$ 		& $16 \pm 1 \tev$,    & $(13, 26)\tev$       & $(9, 25) \tev$ \\
$\mcha1-\mneu1$        	& $0.17 \pm 0.01 \gev$ 	& $0.17 \pm 0.01 \gev$    & $(0.7, 1.3) \gev$ & $(1.3, 2.2) \gev$\\
$\tau_{\cha1}$        	& $0.15 \pm 0.02 ~{\rm ns} $       & $0.15 \pm 0.02 ~{\rm ns} $     & $ < 5.0\times 10^{-3} ~{\rm ns}$ & $< 1.0 \times 10^{-3} ~{\rm ns}$\\
\hline
\end{tabular}
\caption{\it The {68\% CL} ranges for the masses of the LSP $\neu1$ and of the {heavier} neutralino{s} 
$\neu2$, $\neu3$ and $\neu4$,
as well as the mass splitting between the lighter chargino $\cha1$ and
the LSP and the corresponding lifetime of the $\cha1$, for the case in which the $\neu1$ accounts for all the CDM density. Each parameter is shown for both the wino- and Higgsino-LSP scenarios,
as well as for both signs of $\mu$.} 
\label{tab:fit_DM}
\end{center}
\end{table*}

 Figure~\ref{fig:Omegastrips} shows our results in the $(\mneu1, \Omega_{\neu1} h^2)$ plane in the case
 when the $\neu1$ is required to provide all the CDM density, within the uncertainties from the Planck and other measurements.
 The left panel is for $\mu > 0$ and the right panel is for $\mu < 0$: they are quite similar, with each
 featuring two distinct strips. 
{The strip where $\mneu1 \sim 1 \tev$ corresponds to a Higgsino LSP near
  the focus-point region, and  the strip where $\mneu1 \sim 3 \tev$ is in the
  wino LSP region of the parameter space. In between these strips, the make-up
  of the LSP changes as the wino- and Higgsino-like neutralino states mix,}
and coannihilations between the three lightest neutralinos and both charginos become important. The Sommerfeld
enhancement varies rapidly (we recall that it is not important in the Higgsino LSP region),
causing the relic density to rise rapidly as well. We expect the gap seen in Fig.~\ref{fig:Omegastrips} to be
populated by points with very specific values of $m_0$.
 

\begin{figure*}[htb!]
\begin{center}
\resizebox{7.5cm}{!}{\includegraphics{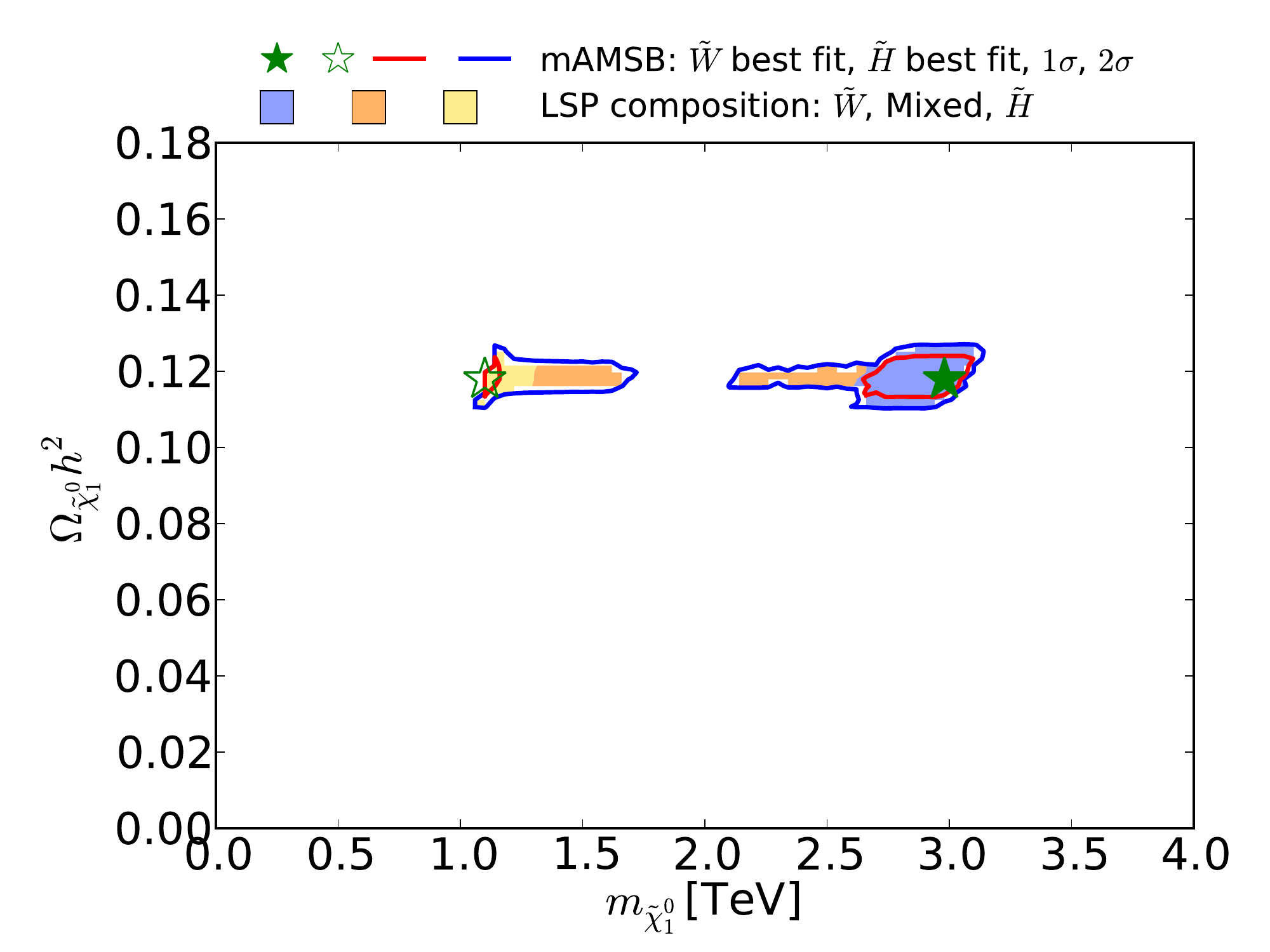}}\put(-169, +123){\footnotesize $\mu>0$, $\Omega_{\neu1}=\Omega_{\rm CDM}$}
\resizebox{7.5cm}{!}{\includegraphics{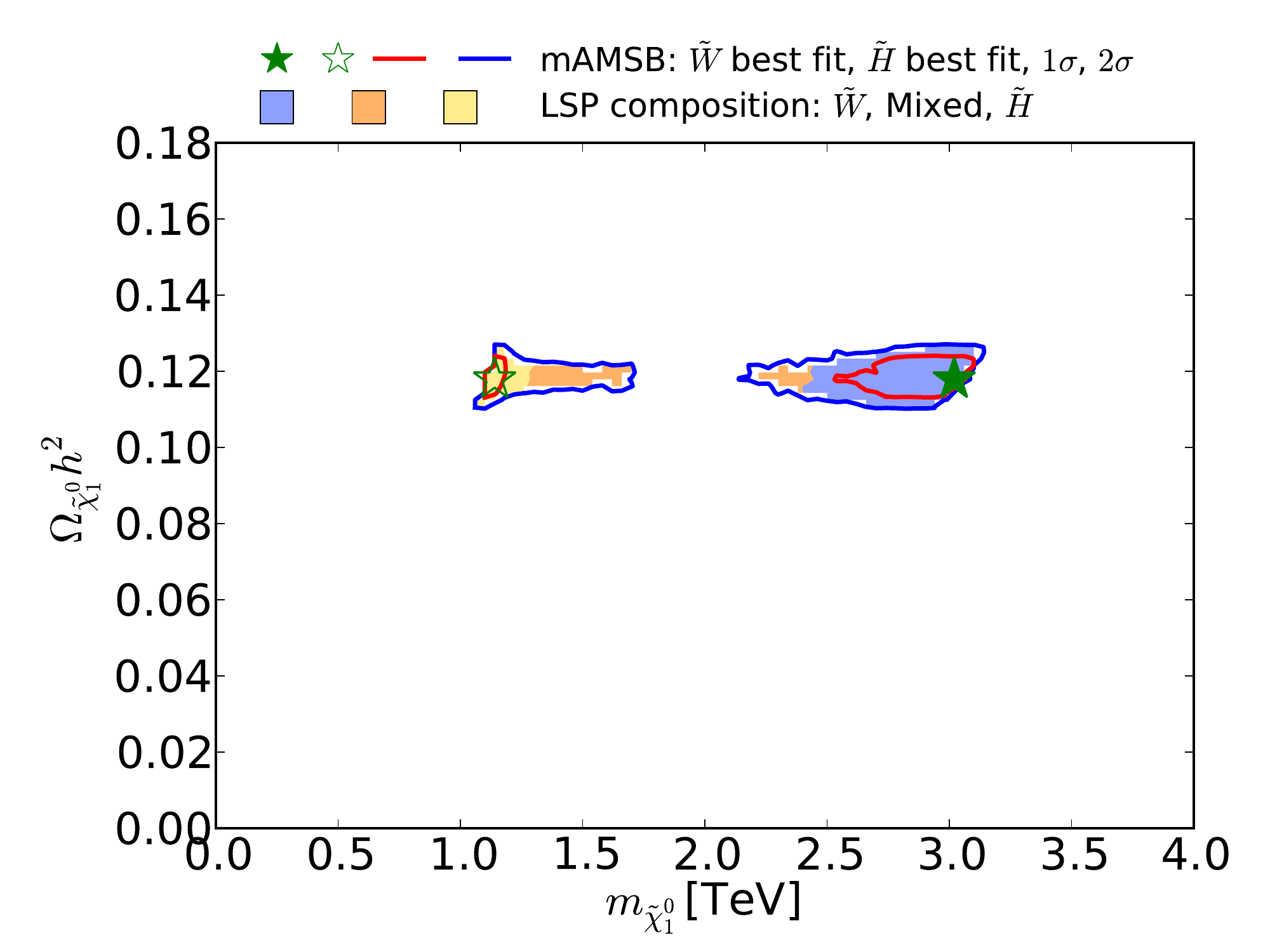}}\put(-169, +123){\footnotesize $\mu<0$, $\Omega_{\neu1}=\Omega_{\rm CDM}$}
\end{center}
\vspace{-1.0cm}
\caption{\it The $(\mneu1, \Omega_{\neu1} h^2)$ planes for $\mu > 0$ (left panel) and $\mu < 0$ (right panel)
{assuming that all the CDM density is provided by the $\neu1$. 
{The shadings are the same as in Fig.~\protect\ref{fig:m0m32}.}}}
\label{fig:Omegastrips}
\end{figure*}


\subsection{{Case II:} the LSP does not provide all the cold dark matter}
\label{sec:noDM}

If the LSP is not the only component of the cold dark matter, \mneu1\ may be smaller,
$m_{3/2}$ may also be lowered substantially, and some sparticles may be within reach of the LHC. 
The preferred regions of the
$(m_0, m_{3/2})$ planes for $\mu > 0$ (left panel) and $\mu < 0$ (right panel)
in this case are shown in  the upper panels of Fig \ref{fig:m0m32UL}~\footnote{{The sharp
boundaries at low $m_0$ in the upper panels of Fig \ref{fig:m0m32UL} are due to the stau
becoming {the LSP}, and the narrow separation between the near-horizontal portions of the 68 and 95\% CL contours
in the upper right panel of Fig. \ref{fig:m0m32UL} is due to the sharp upper limit on the CDM density.}}. 
We see that the wino region allowed at the 95\% CL
extends to smaller $\mgrav$ for both signs of $\mu$, and also to larger $m_0$ at $\mgrav \gtrsim 300 \tev$
when $\mu < 0$. We also see that the 68\% CL region extends to much larger $m_0$ and $\mgrav$
when $\mu < 0$, and the best-fit point also moves to larger masses than for $\mu > 0$,
{though with smaller $\tb$}.

The best-fit points and mass ranges for the case where the LSP relic density falls below the Planck preferred density are given in
Table~\ref{tab:parameters_noDM}. As one can see, the best fit for $\mu > 0$ has a somewhat lower value of $\chi^2$
and a significantly higher value of $\tb$.
{This is because in the case of positive $\mu$ there is negative interference between the mAMSB and SM
contributions to the decay amplitude
in this parameter-space region, reducing \bsdmm\ and allowing a better fit to the latest experimental combination of ATLAS, CMS and LHCb measurements (see Fig.~\ref{fig:indirect_searches}). On the other hand, in the 
negative-$\mu$ case, the interference is constructive and thus the best fit to the experimental measurement is for
{a} SM{-like} branching ratio, which is predicted in a much wider region of the parameter space}.
}

\begin{figure*}[htb!]
\vspace{0.5cm}
\begin{center}
\resizebox{7.5cm}{!}{\includegraphics{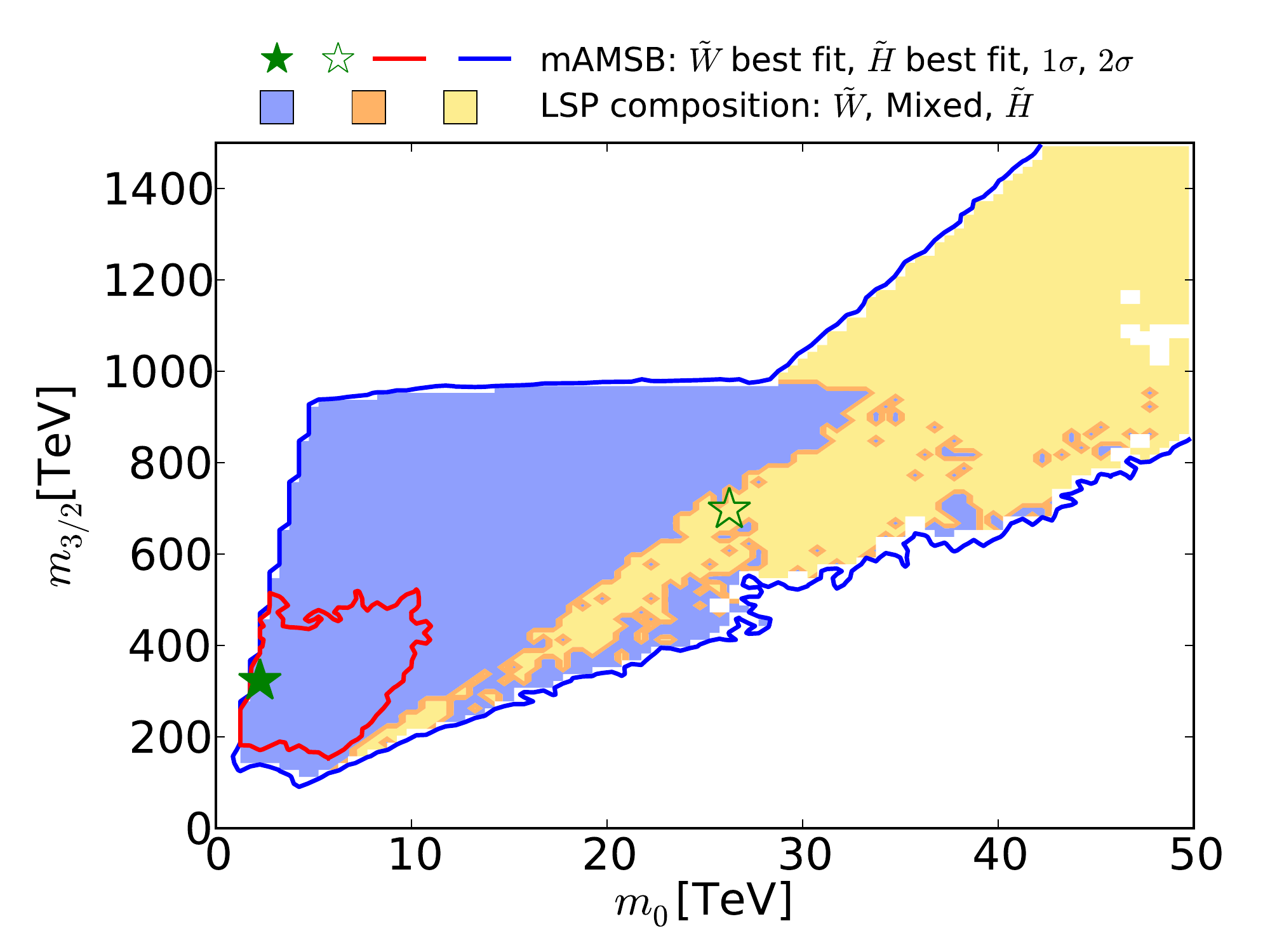}}\put(-169, +123){\footnotesize $\mu>0$, $\Omega_{\neu1}<\Omega_{\rm CDM}$}
\resizebox{7.5cm}{!}{\includegraphics{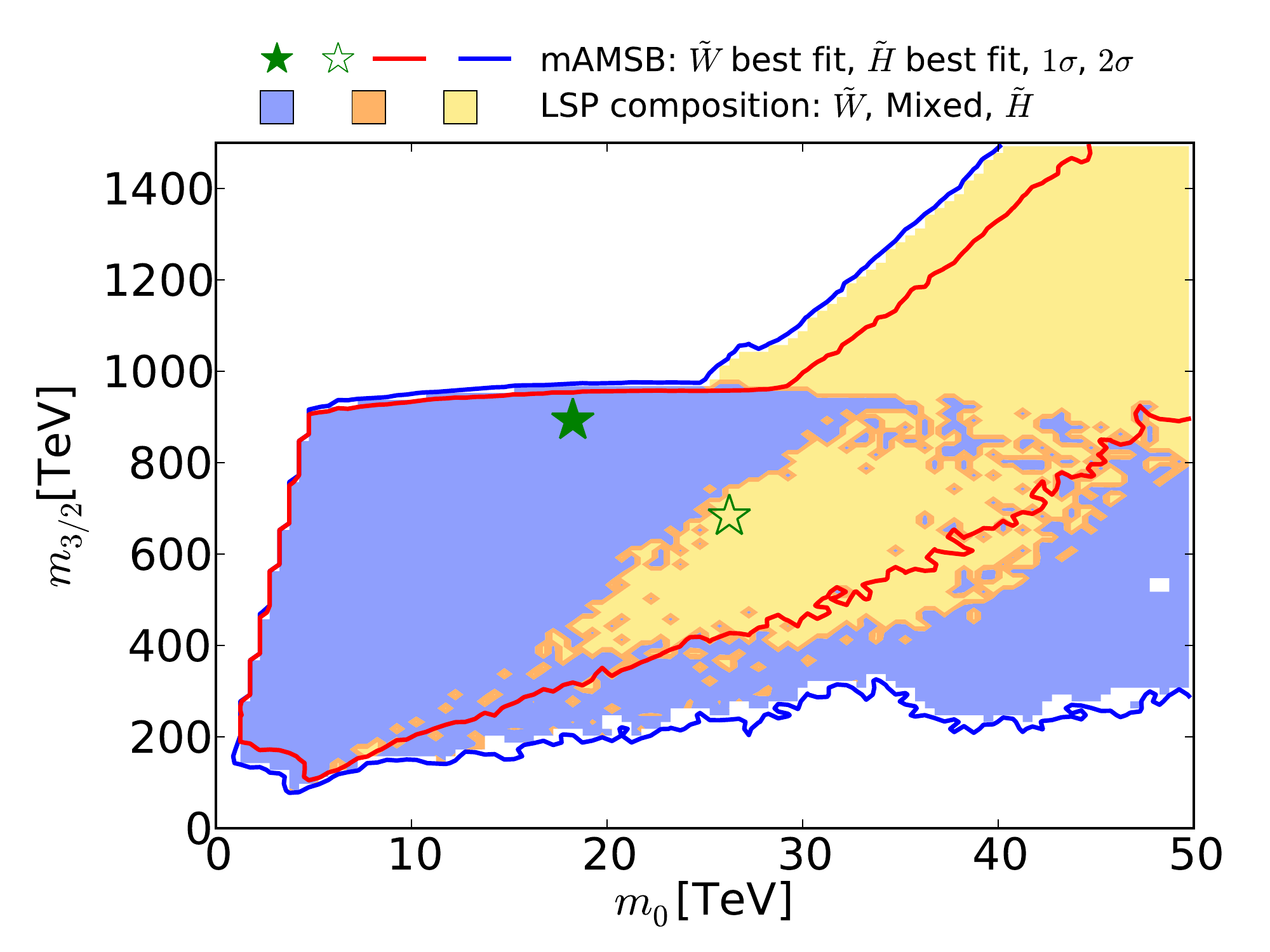}}\put(-169, +123){\footnotesize$\mu<0$, $\Omega_{\neu1}<\Omega_{\rm CDM}$}
\vspace{-3mm}
\resizebox{7.5cm}{!}{\includegraphics{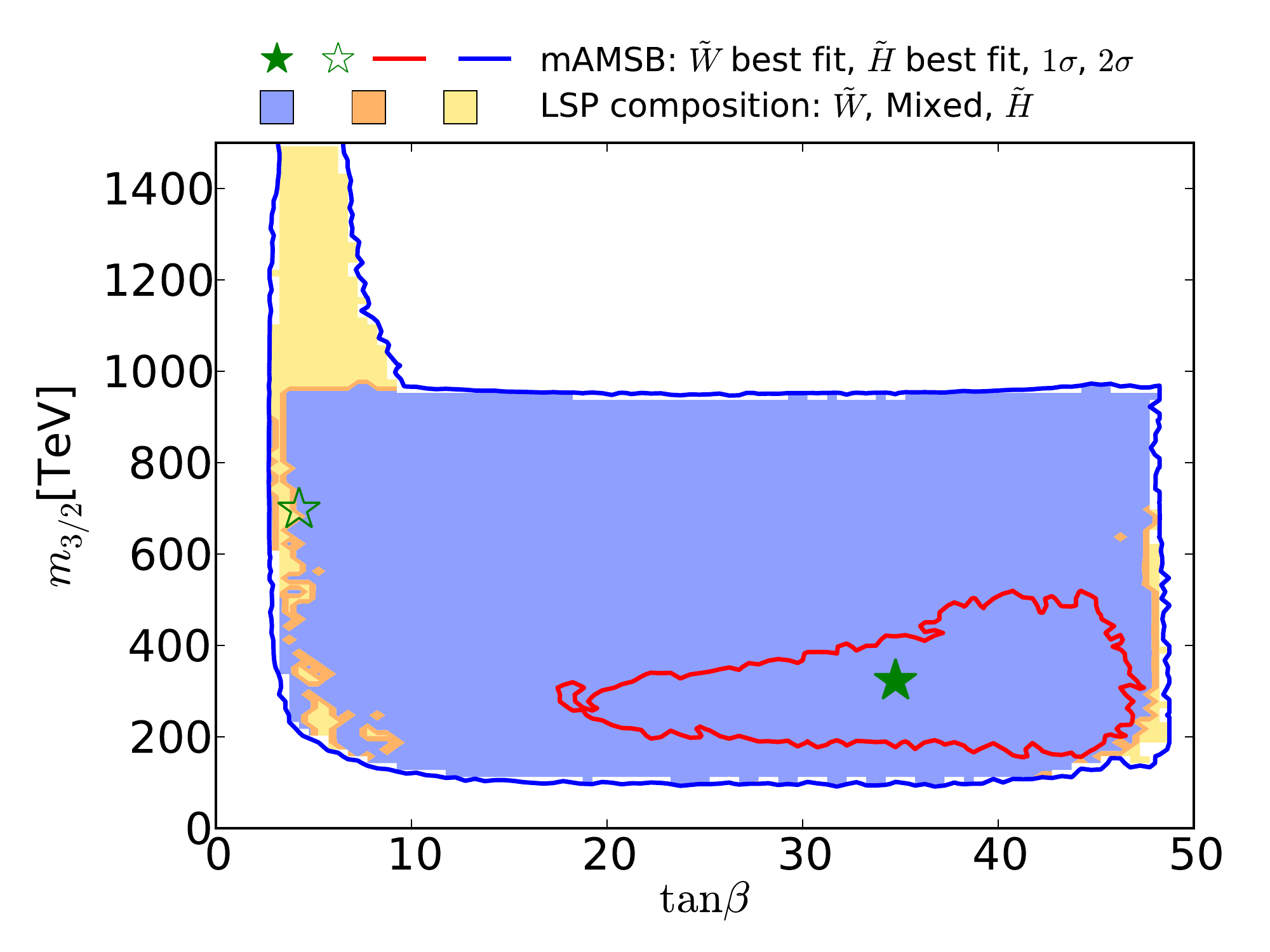}}\put(-95, +123){\footnotesize$\mu>0$, $\Omega_{\neu1}<\Omega_{\rm CDM}$}
\resizebox{7.5cm}{!}{\includegraphics{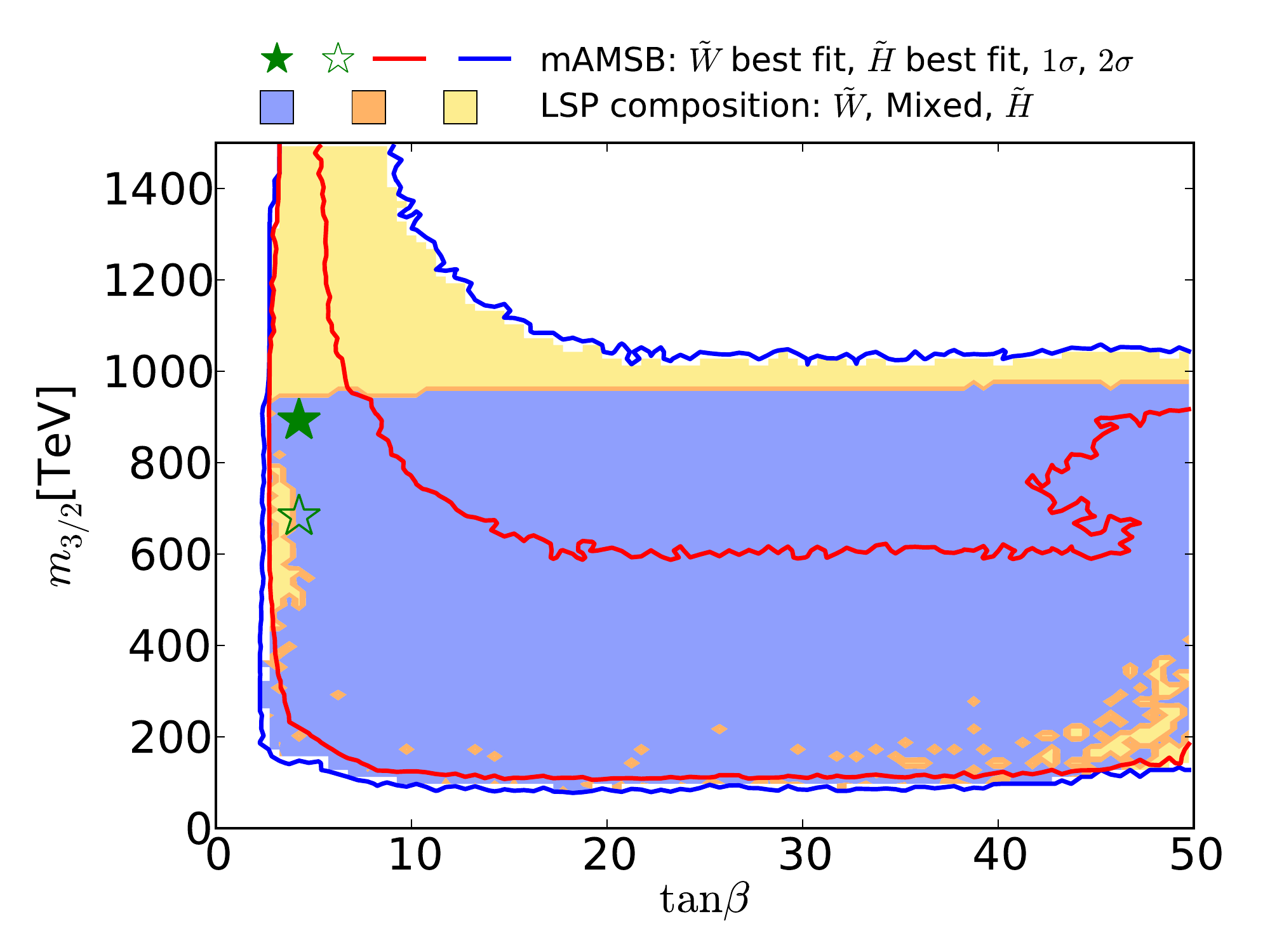}}\put(-95, +123){\footnotesize$\mu<0$, $\Omega_{\neu1}<\Omega_{\rm CDM}$}
\end{center}
\vspace{-1.0cm}
\caption{\it {The $(m_0, \mgrav)$ planes (upper panels) and the $(\tb, \mgrav)$ planes (lower panels)
for $\mu > 0$ (left panels) and for $\mu < 0$ (right panels), allowing the $\neu1$ to contribute only part of the CDM density.
{The shadings are the same as in Fig.~\protect\ref{fig:m0m32}.}}}
\label{fig:m0m32UL}
\end{figure*}

\begin{figure*}[htb!]
\vspace{0.5cm}
\begin{center}
\resizebox{7.5cm}{!}{\includegraphics{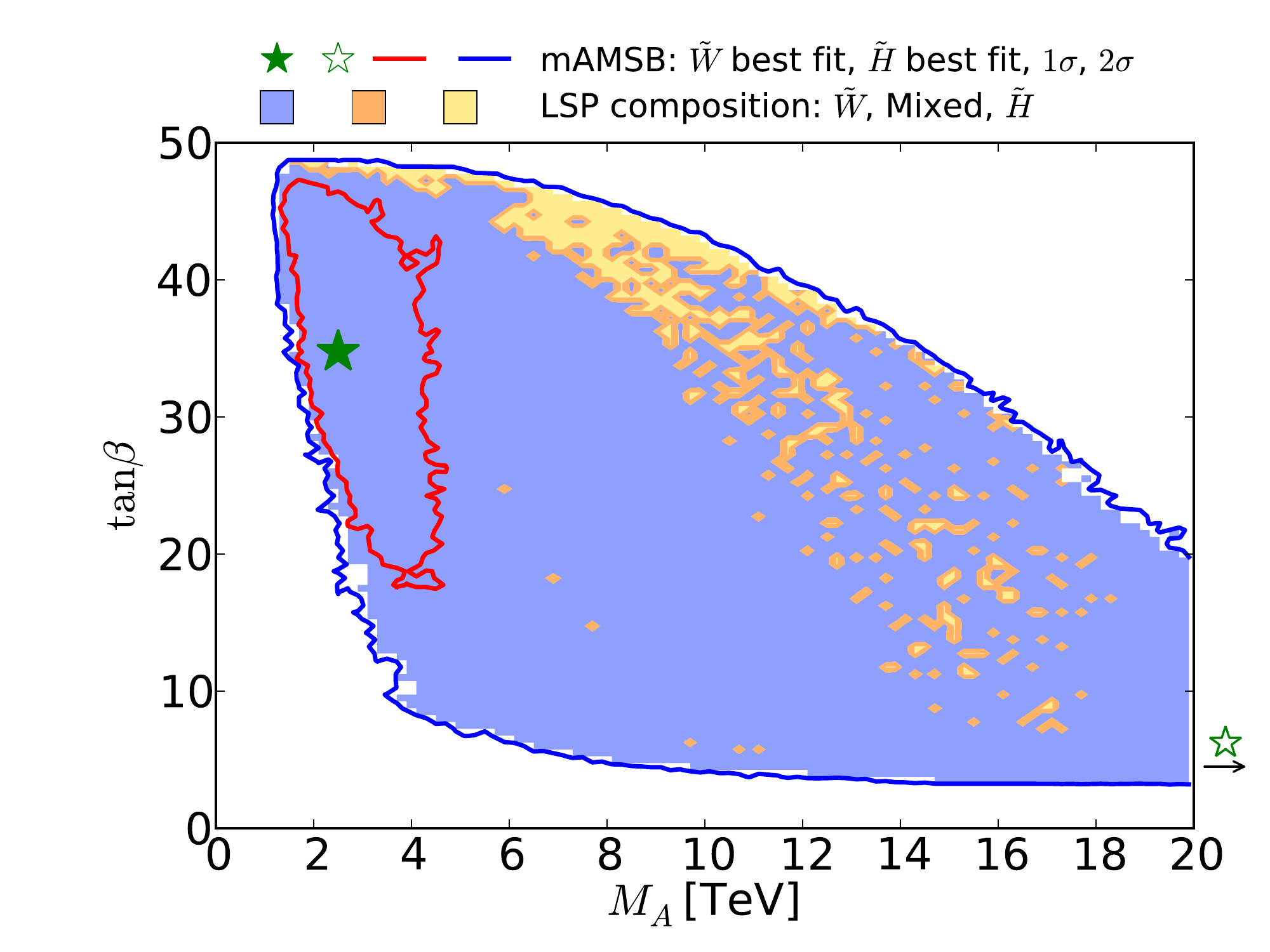}}\put(-95, +123){\footnotesize $\mu>0$, $\Omega_{\neu1}<\Omega_{\rm CDM}$}
\resizebox{7.5cm}{!}{\includegraphics{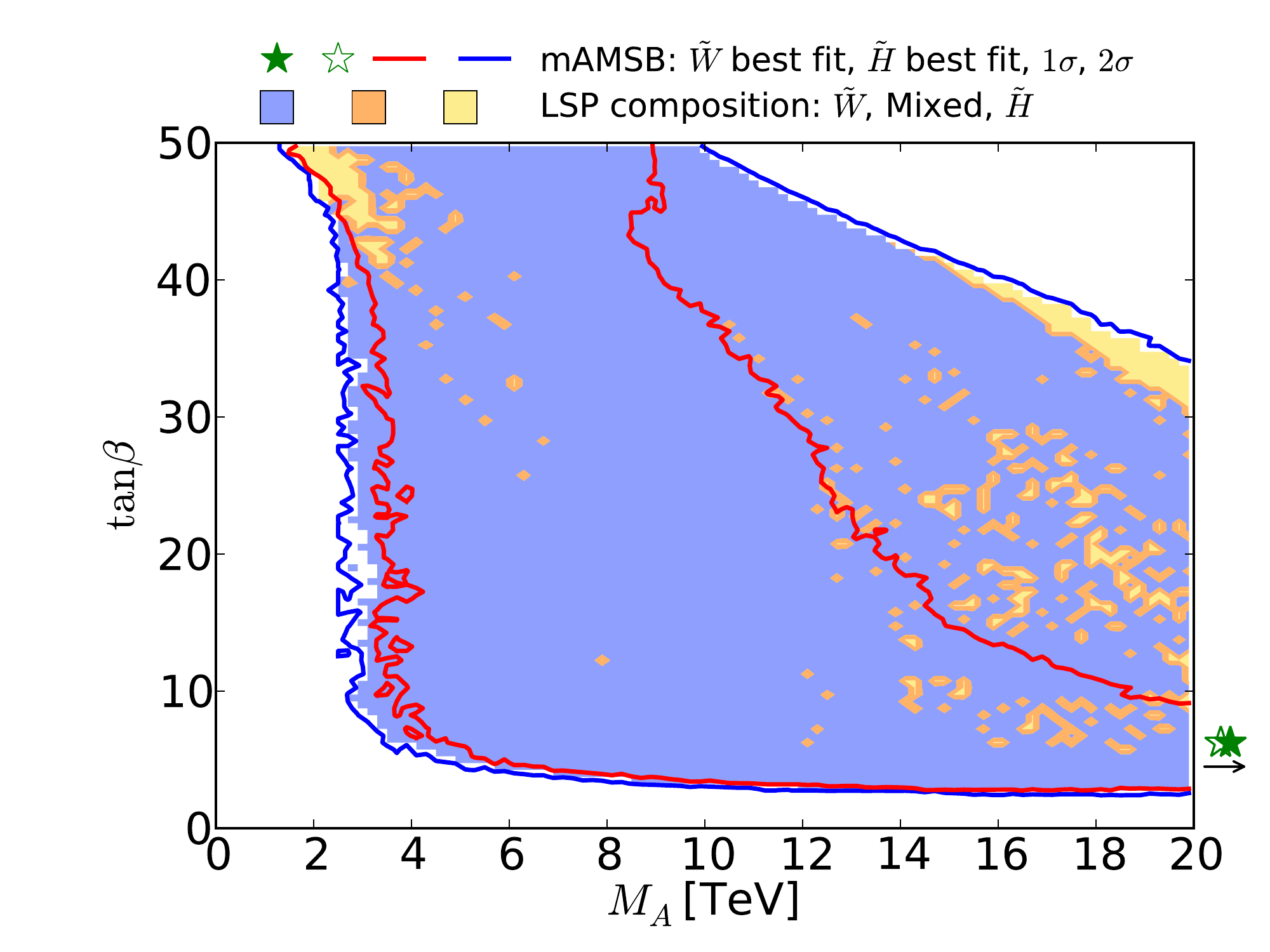}}\put(-95, +123){\footnotesize $\mu<0$, $\Omega_{\neu1}<\Omega_{\rm CDM}$}
\end{center}
\vspace{-1.0cm}
\caption{\it {The $(M_A, \tb)$ planes for $\mu > 0$ (left panel) and for $\mu < 0$ (right panel), allowing the $\neu1$ to contribute only part of the CDM density.
{The shadings are the same as in Fig.~\protect\ref{fig:m0m32}.}}}
\label{fig:MAtbUL}
\end{figure*}

\begin{figure*}[htb!]
\vspace{0.5cm}
\begin{center}
\resizebox{7.5cm}{!}{\includegraphics{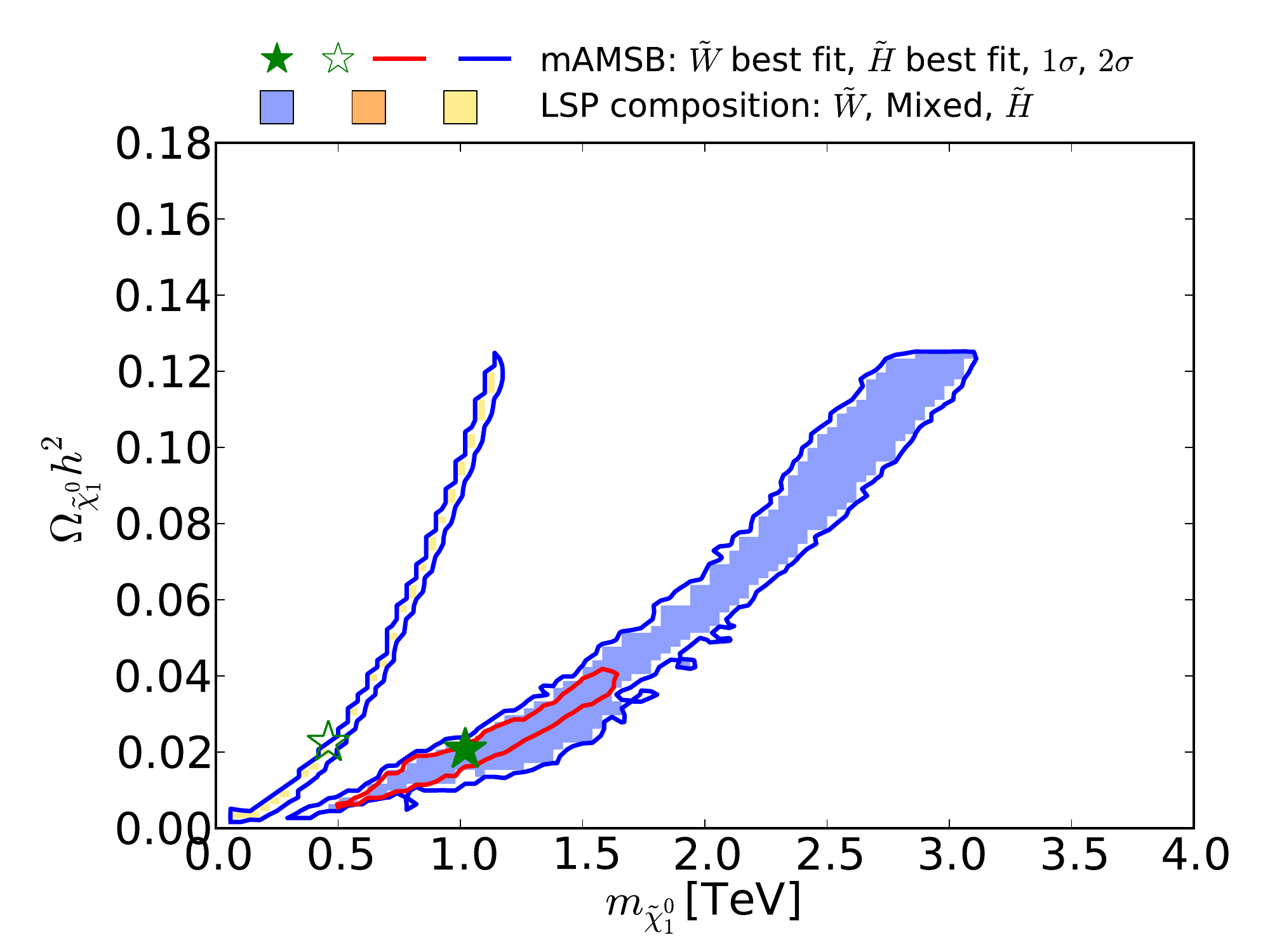}}\put(-169, +123){\footnotesize $\mu>0$, $\Omega_{\neu1}<\Omega_{\rm CDM}$}
\resizebox{7.5cm}{!}{\includegraphics{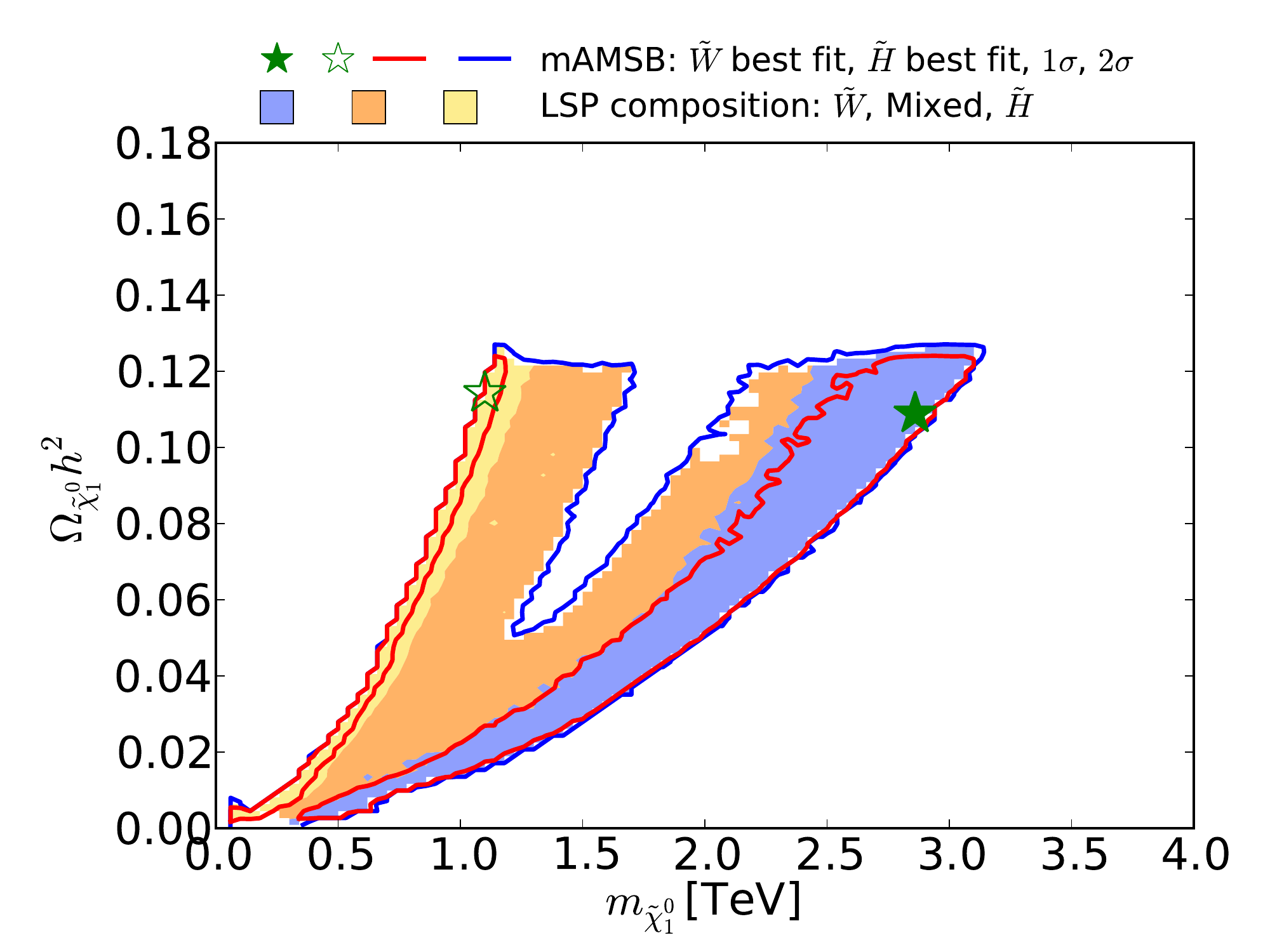}}\put(-169, +123){\footnotesize $\mu<0$, $\Omega_{\neu1}<\Omega_{\rm CDM}$}
\end{center}
\caption{\it The $(\mneu1, \Omega_{\neu1} h^2)$ planes in the mAMSB for $\mu>0$ (left) and $\mu<0$
(right), allowing the $\neu1$ to contribute only part of the CDM density. The red (blue) contours represent the $68\%$ ($95\%$) CL contours. {The shadings are the same as in Fig.~\protect\ref{fig:m0m32}.}}
\label{fig:omegaDM}
\end{figure*}

The lower panels of Fig.~\ref{fig:m0m32UL} show the {$(\tb, \mgrav)$} planes
for $\mu > 0$ (left) and for $\mu < 0$ (right). Comparing with the corresponding planes in 
Fig.~\ref{fig:tbm32} for the case in which the LSP provides all the dark matter, we see a large
expansion of the wino-like region, that the allowed range of $\mgrav$ extends down to $\sim 100 \tev$,
and the 68\% CL region extends to large values of $\tb$.

{We display in Fig.~\ref{fig:MAtbUL} the $(\MA, \tb)$ planes in the partial-CDM case for
$\mu > 0$ (left panel) and $\mu < 0$ (right panel). Comparing with the corresponding Fig.~\ref{fig:MAtball}
for the all-CDM case, we see that a large region of smaller values of $\MA$ and $\tb$ are allowed
in this case. We also note that the best-fit point in the wino-like region for $\mu > 0$ has moved
to a much smaller value of $\MA$ and a much larger value of $\tb$, much closer to the region
currently excluded by LHC searches. In this connection, we note that the fit
including only the LHC 8-TeV $H/A \to \tau^+ \tau^-$ constraint~\cite{HA8}
{is slightly} weaker in this region than that including the 13-TeV constraint~\cite{ICHEP}. 
This gives hope that future improvements in the LHC $H/A$ search may be sensitive
to the preferred region of the mAMSB parameter space in the partial-CDM case.}

Figure~\ref{fig:omegaDM} displays the $(\mneu1, \Omega_{\neu1} h^2)$ planes
for $\mu > 0$ (left panel) and $\mu < 0$ (right panel) in the partial-CDM case.
{We see that the allowed range of $\neu1$ masses decreases with $\Omega_{\neu1} h^2$, as expected.
Pure wino or Higgsino LSP states are slightly preferred over mixed ones 
because the latter are accompanied by larger scattering cross sections 
on protons and are thus in tension with direct DM searches (see Section~\ref{sec:directDetection}).}
{The preferred region in the wino-like LSP $\mu > 0$ case appears at
small values of $\mneu1$ and $\Omega_{\neu1} h^2$, 
pulled down by the possibility of negative interference in
the $B_{s,d} \to \mu^+ \mu^-$ decay amplitudes and the consequent decrease in \bsdmm, as discussed in Section~5.3.
In the Higgsino-LSP $\mu>0$ case and in all $\mu<0$ cases, all $\Omega_{\neu1} h^2$ 
values below the Planck preferred density are equally likely.}

Figure~\ref{fig:mspectrum_noDM} shows the mass spectra allowed in the wino-like LSP case
for $\mu > 0$ (top panel) and $\mu < 0$ (second panel), and also in the Higgsino-like LSP case
for $\mu > 0$ (third panel) and $\mu < 0$ (bottom panel). The one- and two-$\sigma$ ranges are again
shown in dark and light orange respectively, and the best-fit values are represented by blue lines.
We see that the spectra in the wino-like LSP case are quite different for the two signs of $\mu$,
whereas those in the Higgsino-like LSP case resemble each other more. Table~\ref{tab:fit_noDM} provides
numerical values for the 68\% CL ranges for the neutralino masses, the gluino mass, the mass difference between the lightest chargino and neutralino, as well as for the corresponding chargino lifetime.


\begin{figure*}[htb!]
\begin{center}
\resizebox{14cm}{!}{\includegraphics{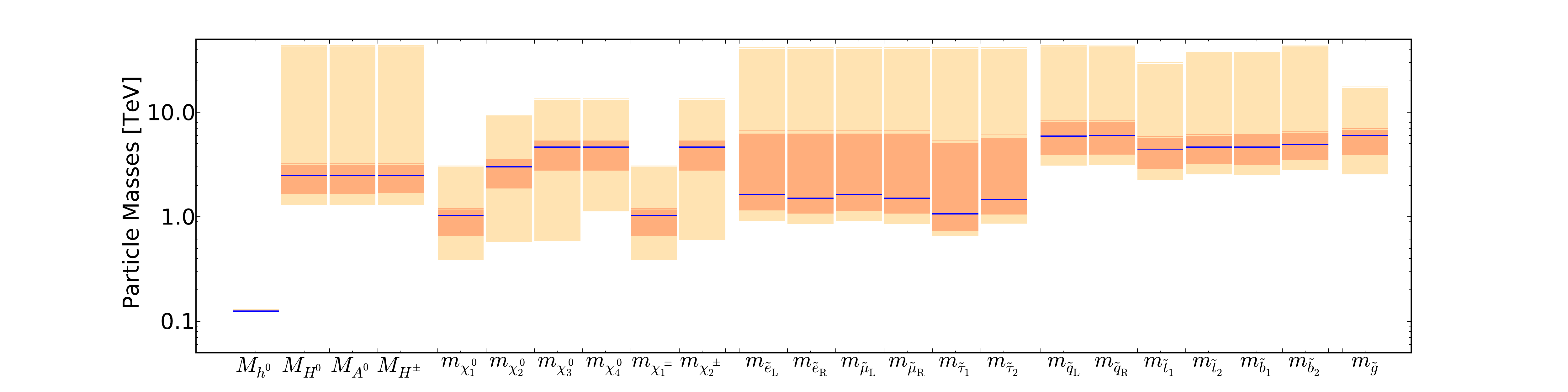}}
\put(-283, +102){\small  $\tilde{W}$-LSP  for $\mu>0$, $\Omega_{\neu1}<\Omega_{\rm CDM}$}
\\\vspace{0.2cm}
\resizebox{14cm}{!}{\includegraphics{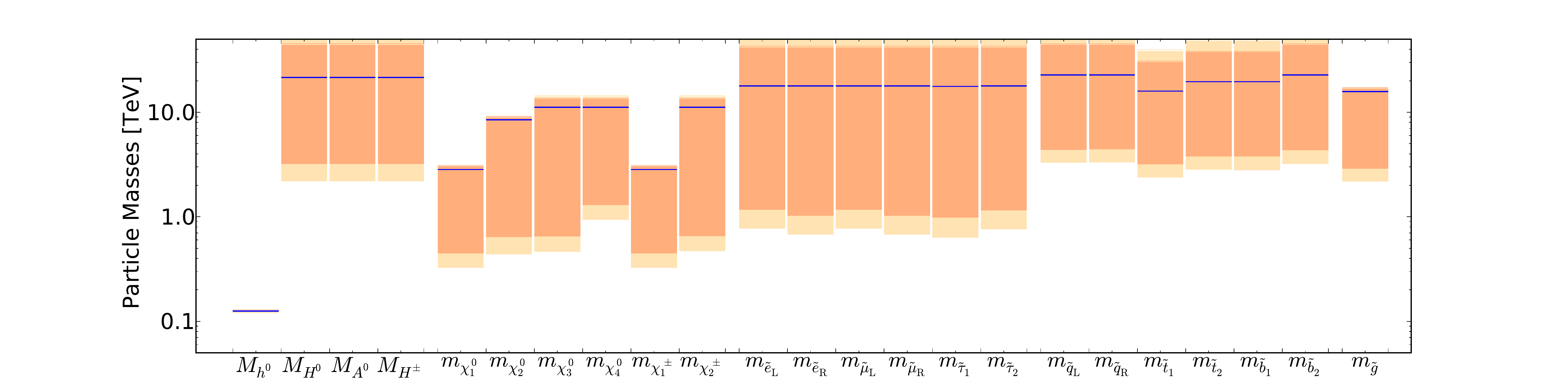}}
\put(-283, +102){\small  $\tilde{W}$-LSP  for $\mu<0$, $\Omega_{\neu1}<\Omega_{\rm CDM}$}
\\\vspace{0.2cm}
\resizebox{14cm}{!}{\includegraphics{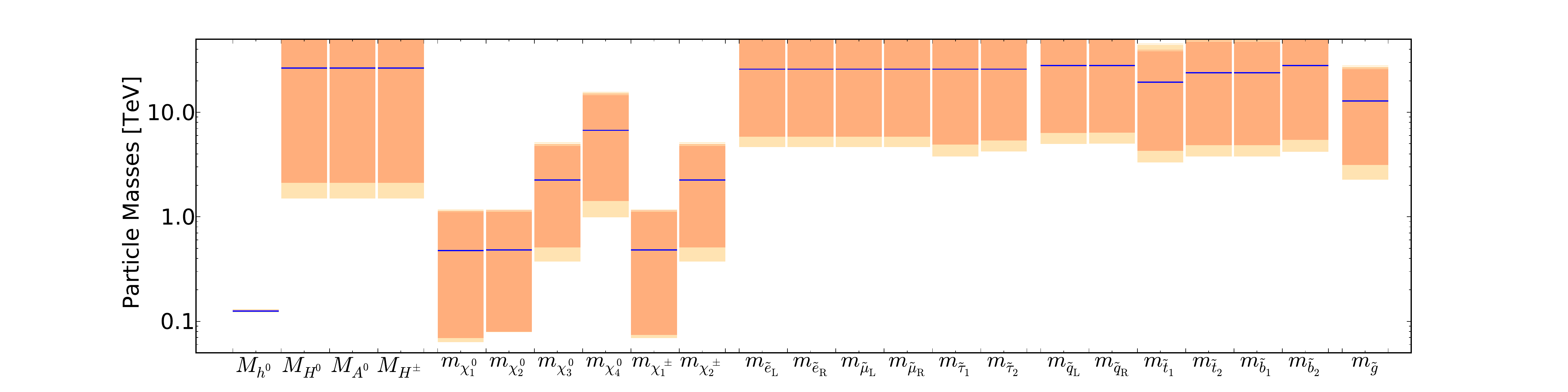}}
\put(-283, +105){\small  $\tilde{H}$-LSP  for $\mu>0$, $\Omega_{\neu1}<\Omega_{\rm CDM}$}
\\\vspace{0.2cm}
\resizebox{14cm}{!}{\includegraphics{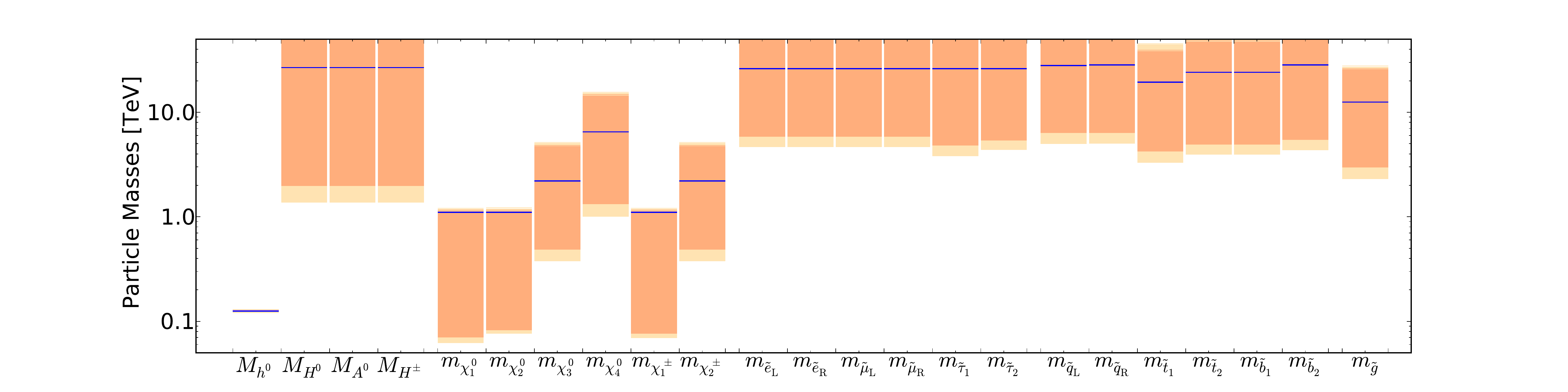}}
\put(-283, +105){\small  $\tilde{H}$-LSP  for $\mu<0$, $\Omega_{\neu1}<\Omega_{\rm CDM}$}
\end{center}
\vspace{-0.5cm}
\caption{\it  The ranges of masses obtained for the wino-like LSP case with $\mu > 0$ (top panel) and 
$\mu < 0$ (second panel), and also for the Higgsino-like LSP case for $\mu > 0$ (third panel)
and $\mu < 0$ (bottom panel), relaxing
  the assumption that the LSP contributes all the cold dark matter density. The one- and two-$\sigma$ CL regions are shown in {dark and light} orange respectively, and the best-fit values are represented by blue lines.
  }
\label{fig:mspectrum_noDM}
\end{figure*}

{Finally, Fig.~\ref{fig:spectrumno} displays the spectra of our best-fit points in the case that
the LSP contributes only a fraction of the cold dark matter density. As previously,
the left panels are for $\mu > 0$ and the right panels are for $\mu < 0$ (note the different scales on the
vertical axes). Both
the wino- (upper) and the Higgsino-like LSP (lower) best-fit points are shown. In each case, we also
indicate all the decay modes with branching ratios above 20\%.}

\begin{figure*}[htb!]
\begin{center}
  \resizebox{7.5cm}{!}{\includegraphics{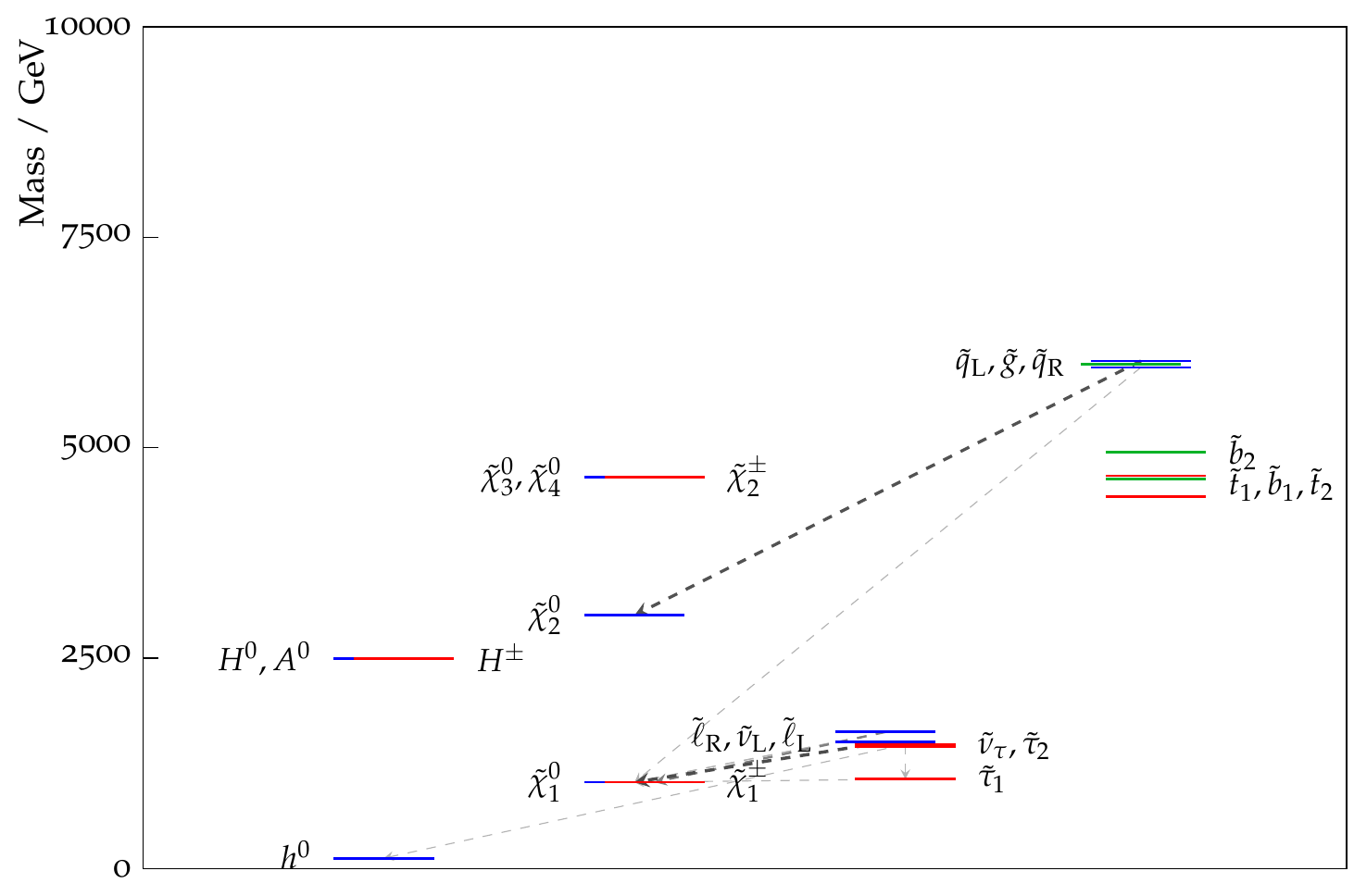}}\put(-174, +144){\small $\tilde{W}$-LSP  for $\mu>0$, $\Omega_{\neu1}<\Omega_{\rm CDM}$}
  \resizebox{7.5cm}{!}{\includegraphics{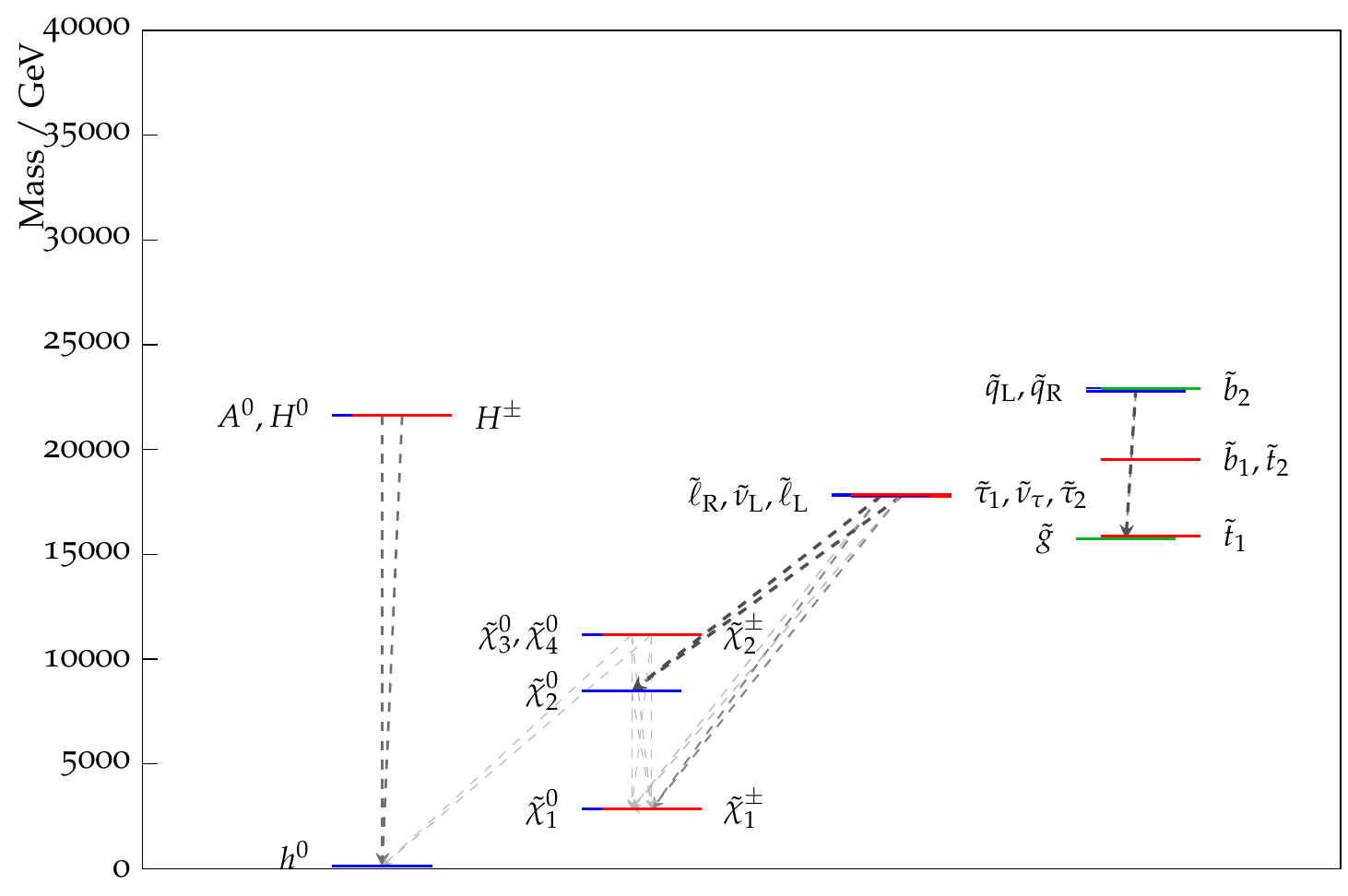}}\put(-174, +144){\small  $\tilde{W}$-LSP  for $\mu<0$, $\Omega_{\neu1}<\Omega_{\rm CDM}$}
\vspace{0.5cm}
\resizebox{7.5cm}{!}{\includegraphics{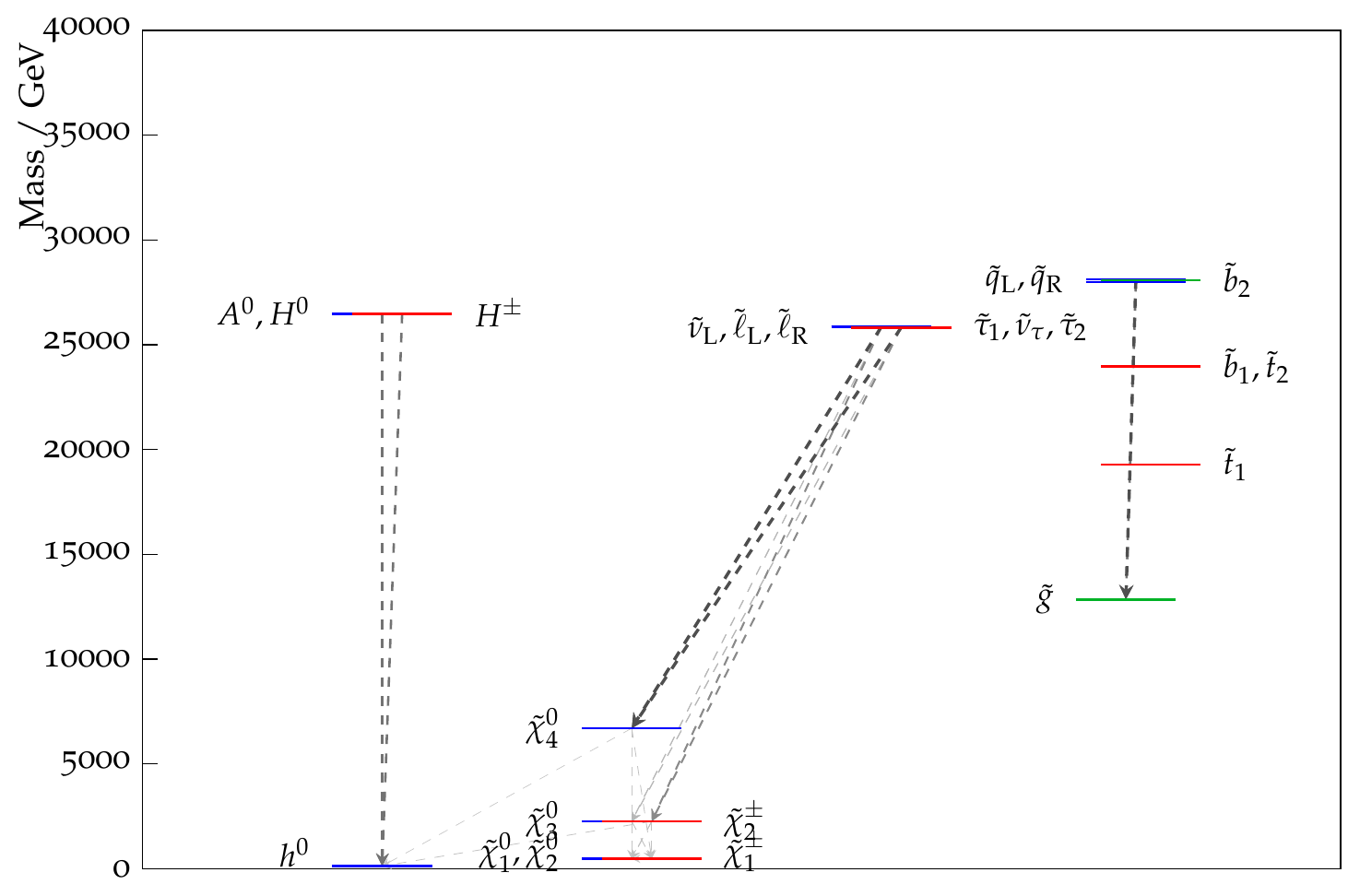}}\put(-174, +144){\small  $\tilde{H}$-LSP  for $\mu>0$, $\Omega_{\neu1}<\Omega_{\rm CDM}$}
\resizebox{7.5cm}{!}{\includegraphics{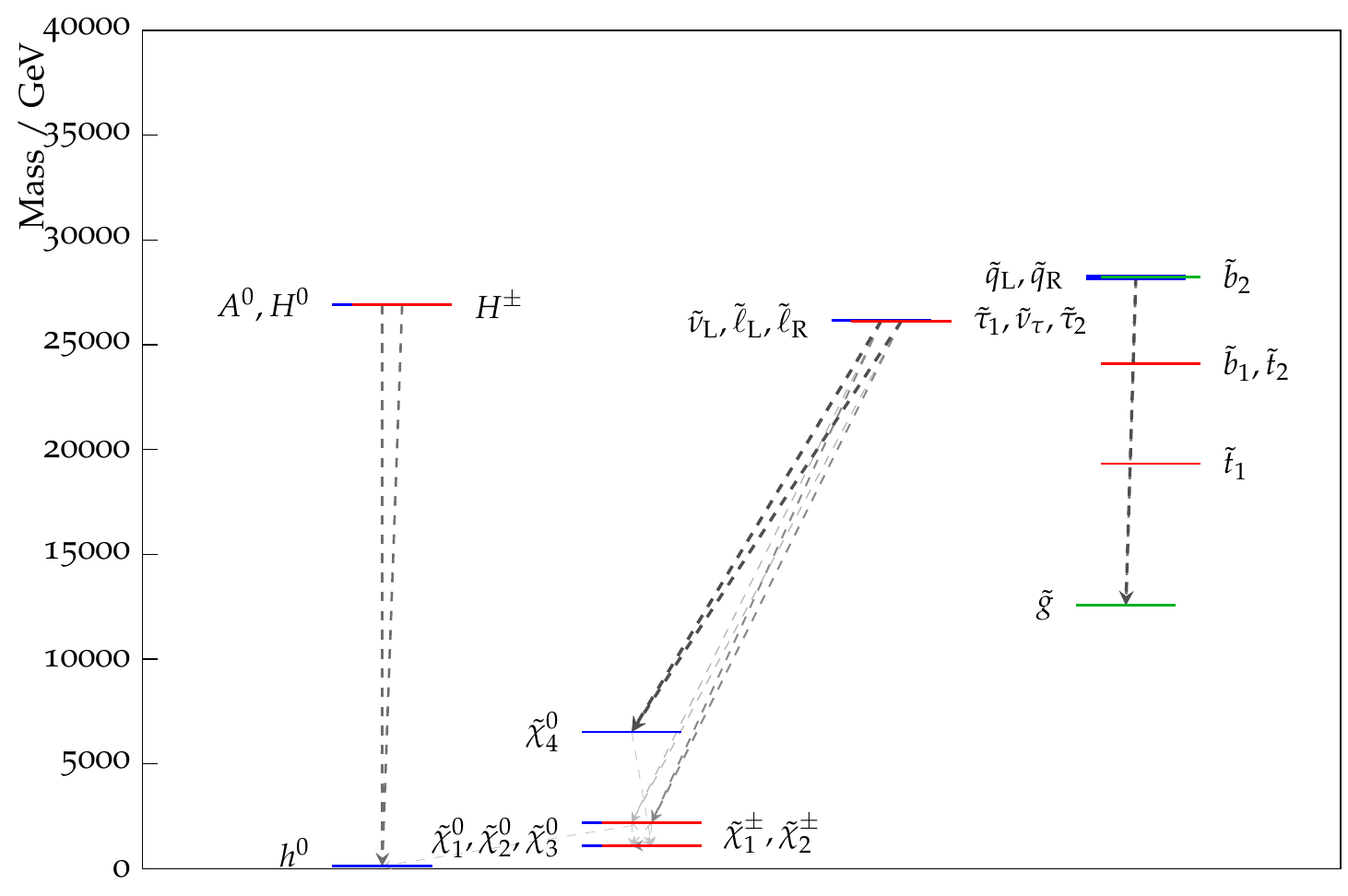}}\put(-174, +144){\small  $\tilde{H}$-LSP  for $\mu<0$, $\Omega_{\neu1}<\Omega_{\rm CDM}$}
\end{center}
\vspace{-0.5cm}
\caption{\it {The spectra of our best-fit points for $\mu > 0$ (left panel) and $\mu < 0$ (right panel), allowing
the LSP to contribute only part of the cold dark matter density. Both the wino- (upper) and the Higgsino-like LSP (lower) best-fit points are shown. In each case, we also
indicate all the decay modes with branching ratios above 20\%. {The range of masses shown for the $\tilde{W}$-LSP $\mu>0$ best fit point (top-left panel) is smaller than the others, since its mass spectrum is considerably lighter.} } }
\label{fig:spectrumno}
\end{figure*}

\subsection{\boldmath{$\chi^2$} Likelihood Functions for Observables}

We show in Fig.~\ref{fig:1dchi2} the one-dimensional likelihoods for several sparticle masses.
In all cases the solid lines correspond to the case in which the LSP accounts for all of the CDM, and the dashed lines for the case in which it may provide only a fraction of the CDM. The blue lines are for $\mu > 0$
and the red lines are for $\mu < 0$. It is apparent that in the all-CDM case the sparticles
in the mAMSB are expected to be too heavy to be produced at the LHC: $\mgl, m_{\tilde t_1} \gtrsim 10 \tev,
\msq \gtrsim 15 \tev, m_{\tilde \tau_1} \gtrsim 3 \tev, \mneu1 \gtrsim 1 \tev$.
However, in the part-CDM case the sparticle masses may be much lighter, with strongly-interacting
sparticles possibly as light as $\sim 2 \tev$ and much lighter $\neu3$ and $\cha2$ also possible,
so that some of them may become accessible at LHC energies. Indeed, as we {discuss below,
future LHC runs} should be able to explore parts of the allowed parameter space.

\begin{figure*}[htb!]
\vspace{0.5cm}
\begin{center}
\resizebox{7.5cm}{!}{\includegraphics{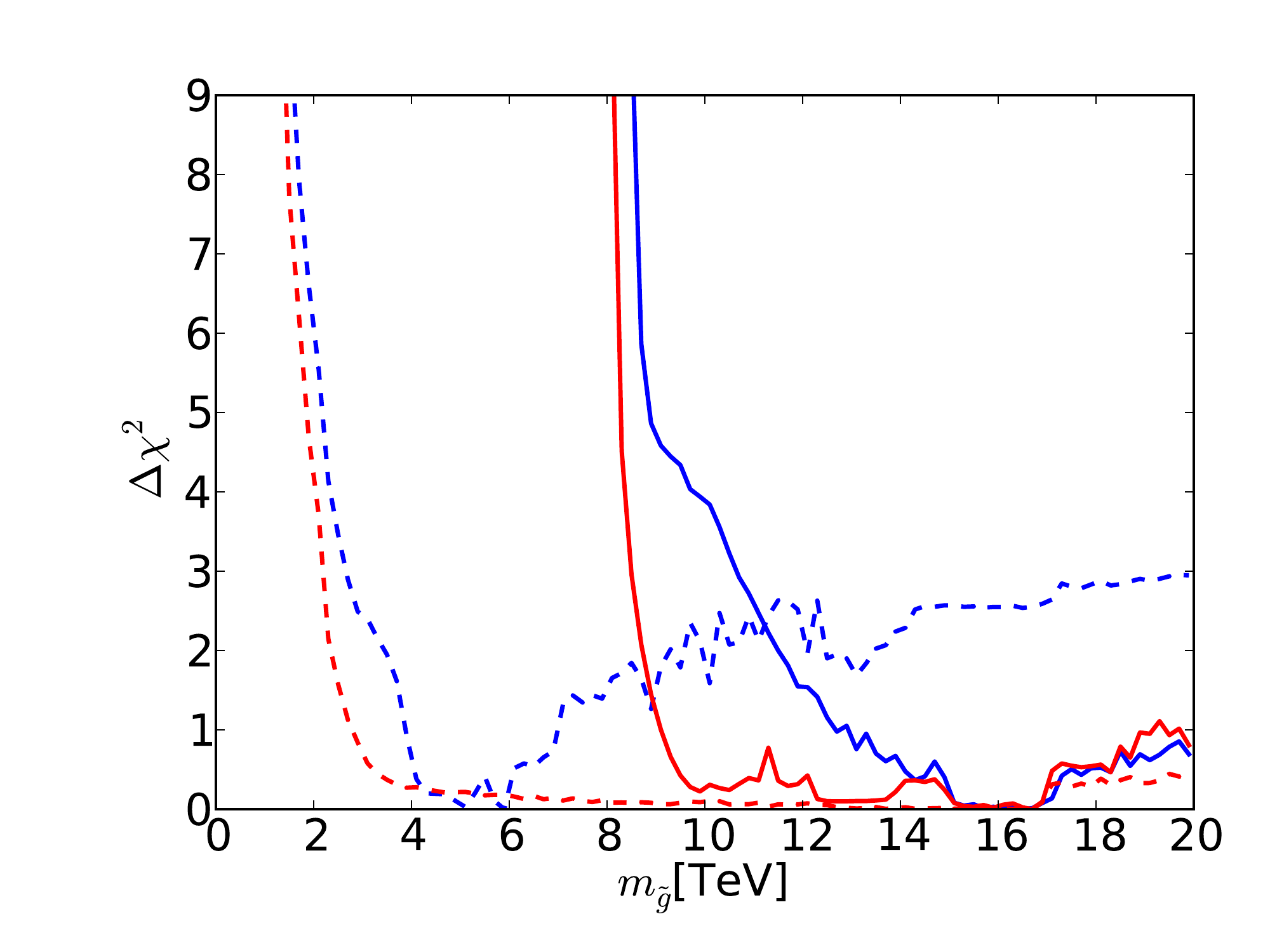}}
\resizebox{7.5cm}{!}{\includegraphics{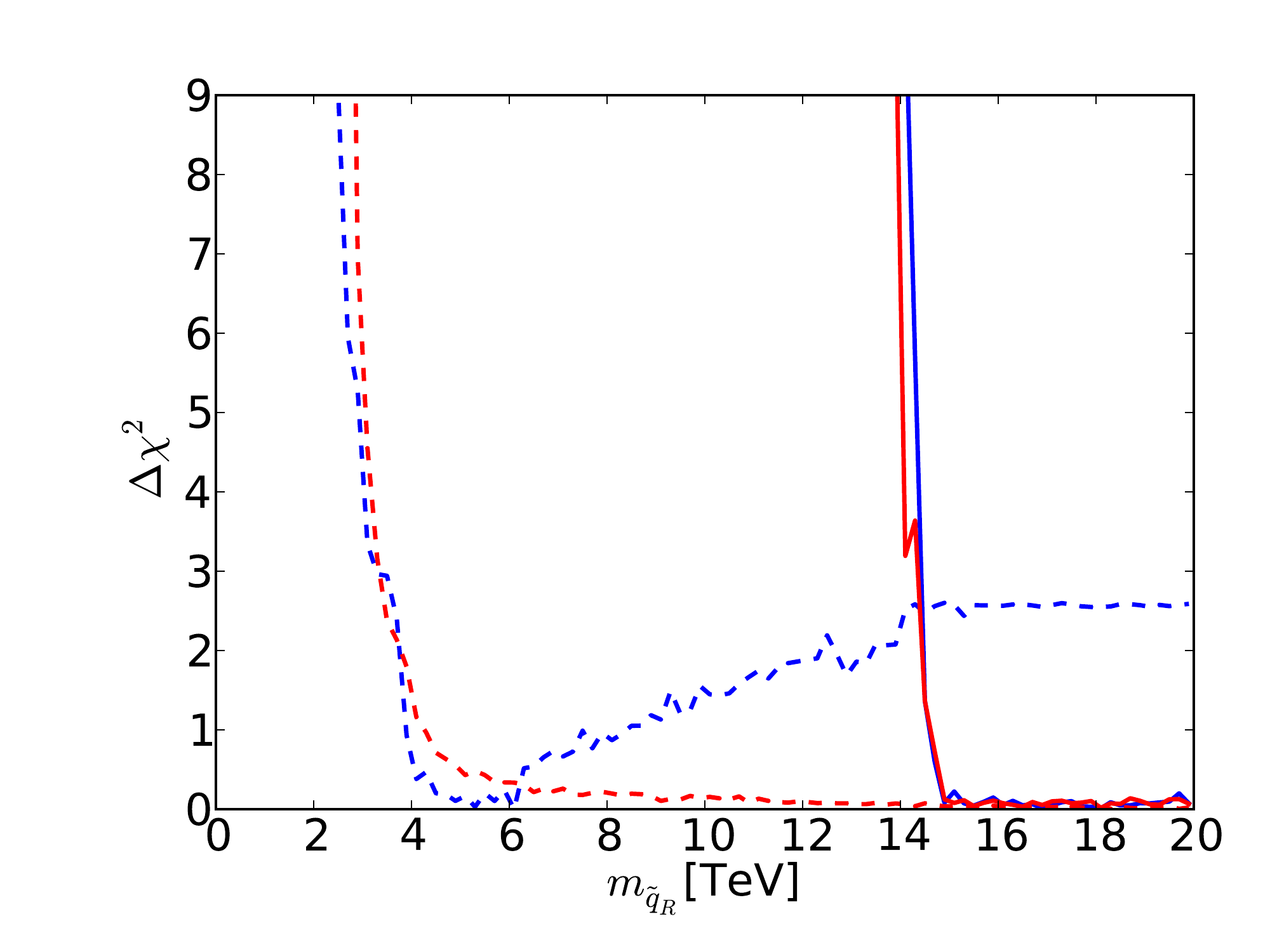}}\\
\resizebox{7.5cm}{!}{\includegraphics{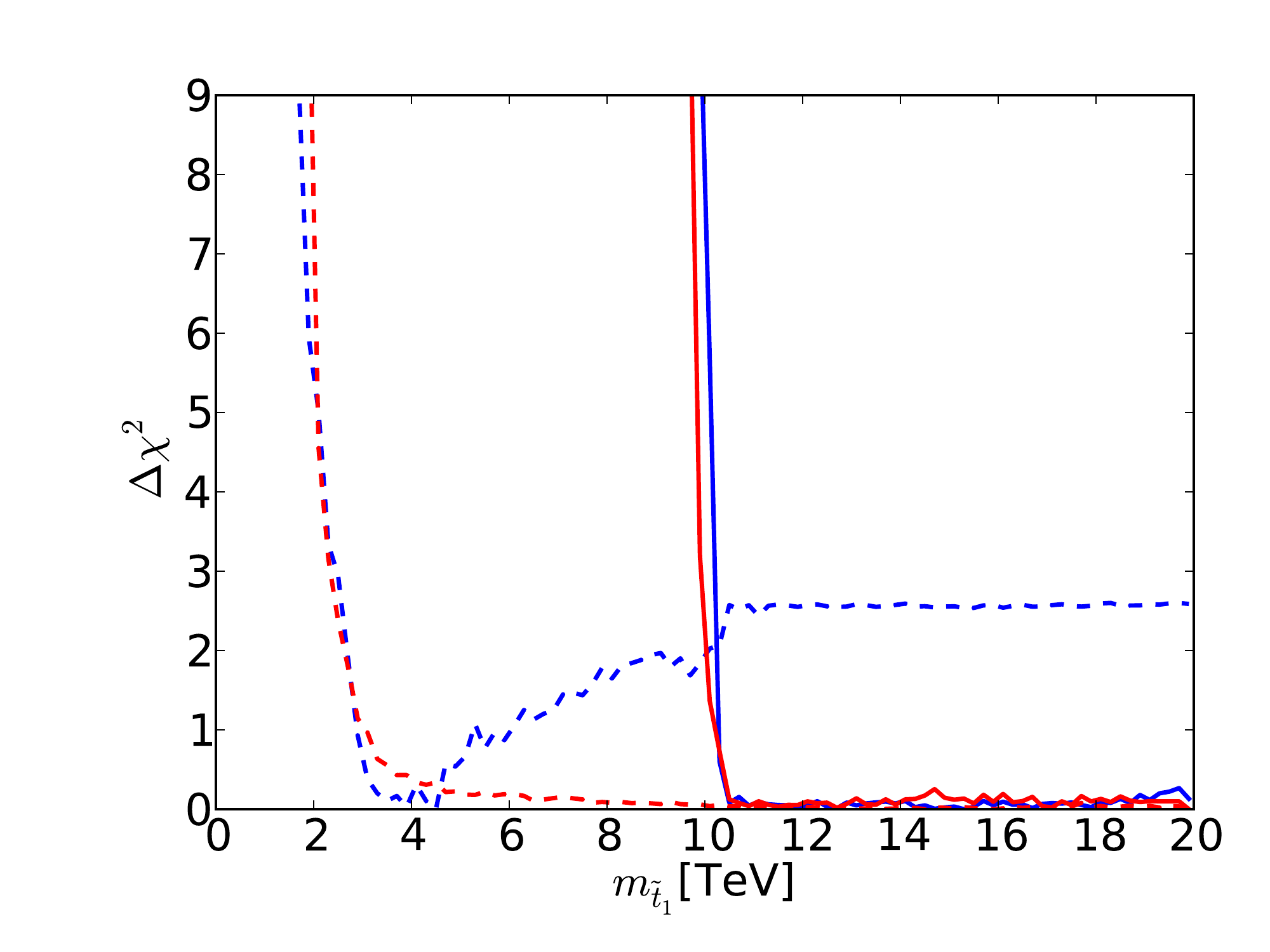}}
\resizebox{7.5cm}{!}{\includegraphics{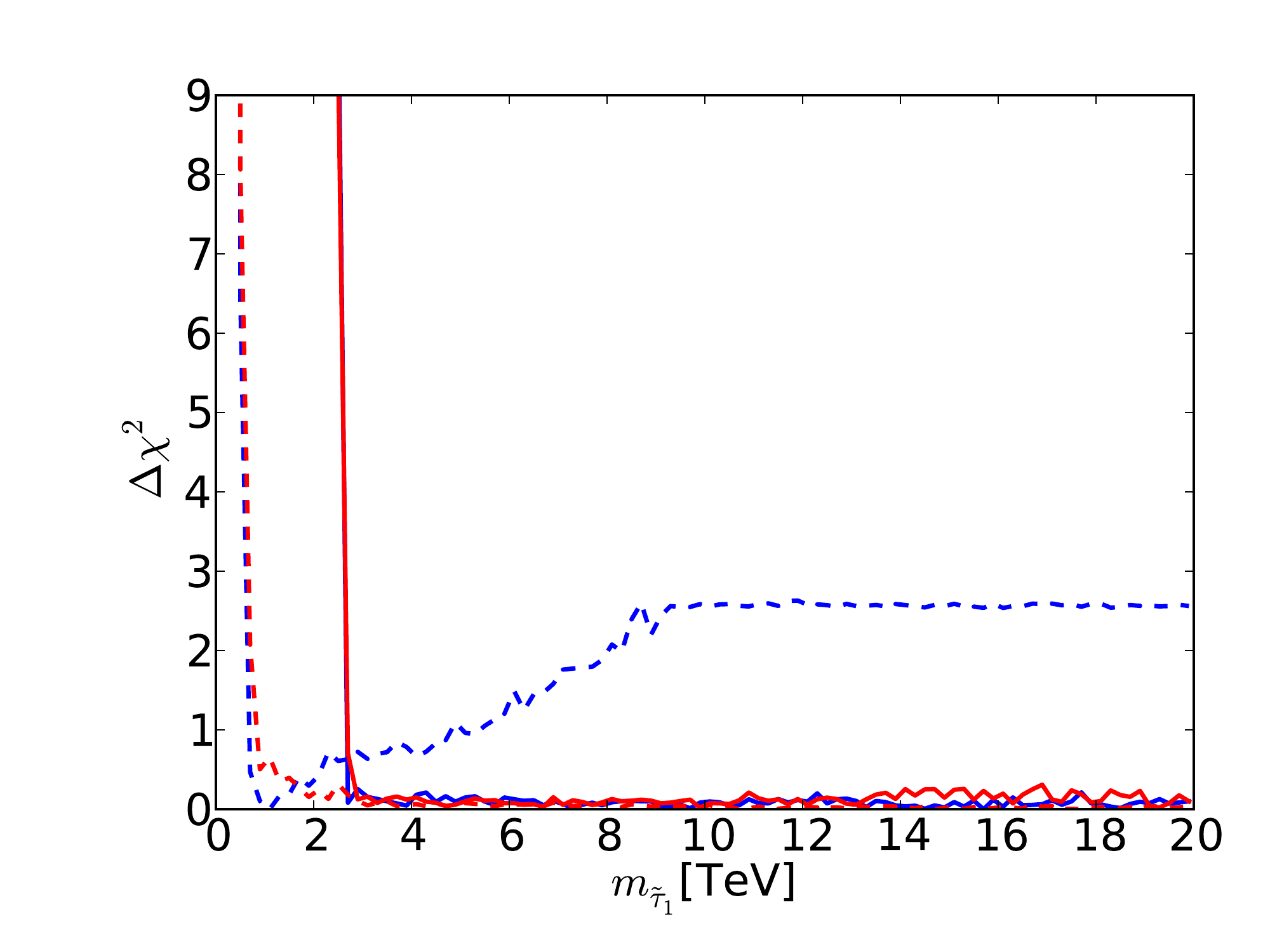}}\\
\resizebox{7.5cm}{!}{\includegraphics{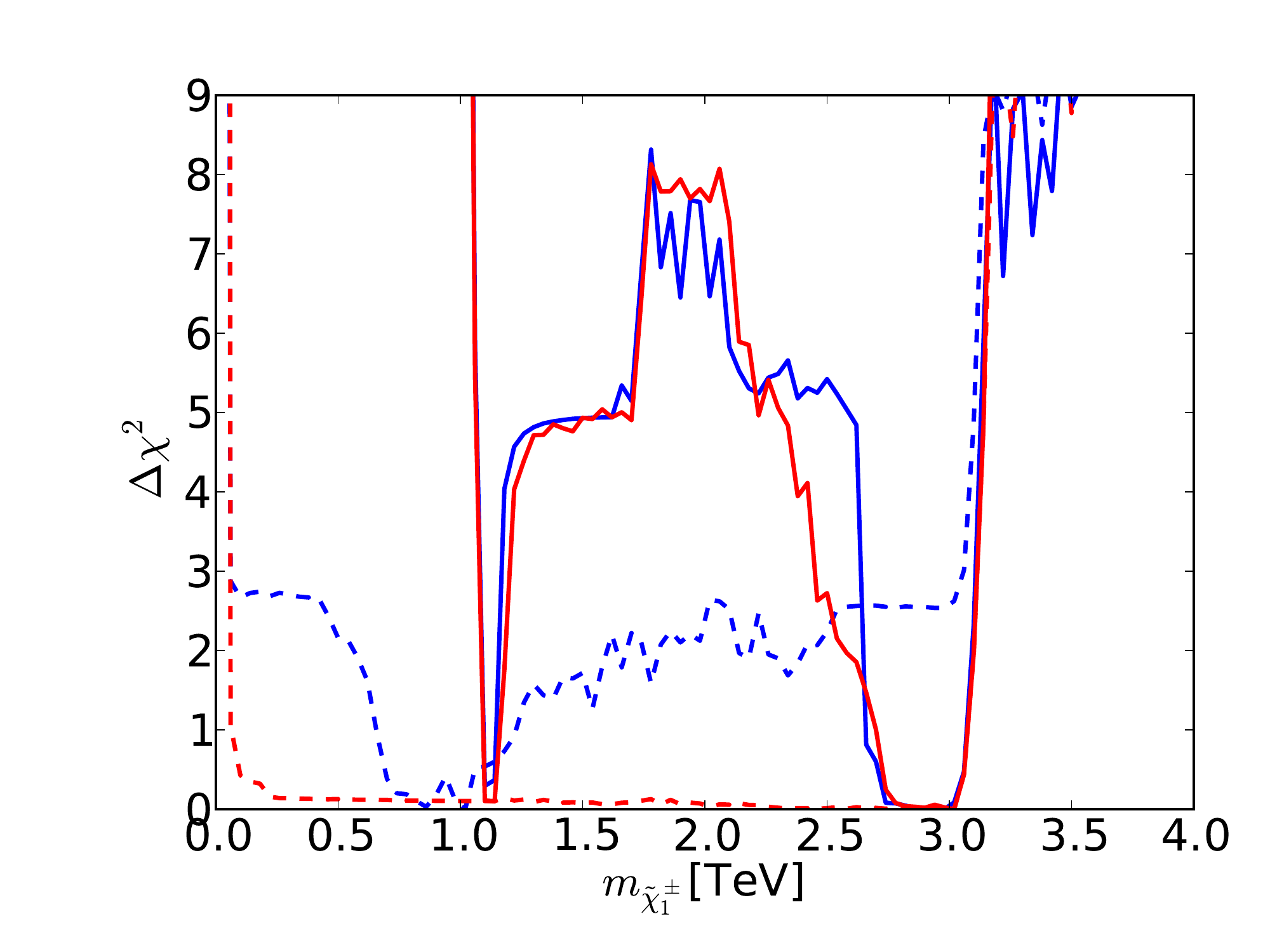}}
\resizebox{7.5cm}{!}{\includegraphics{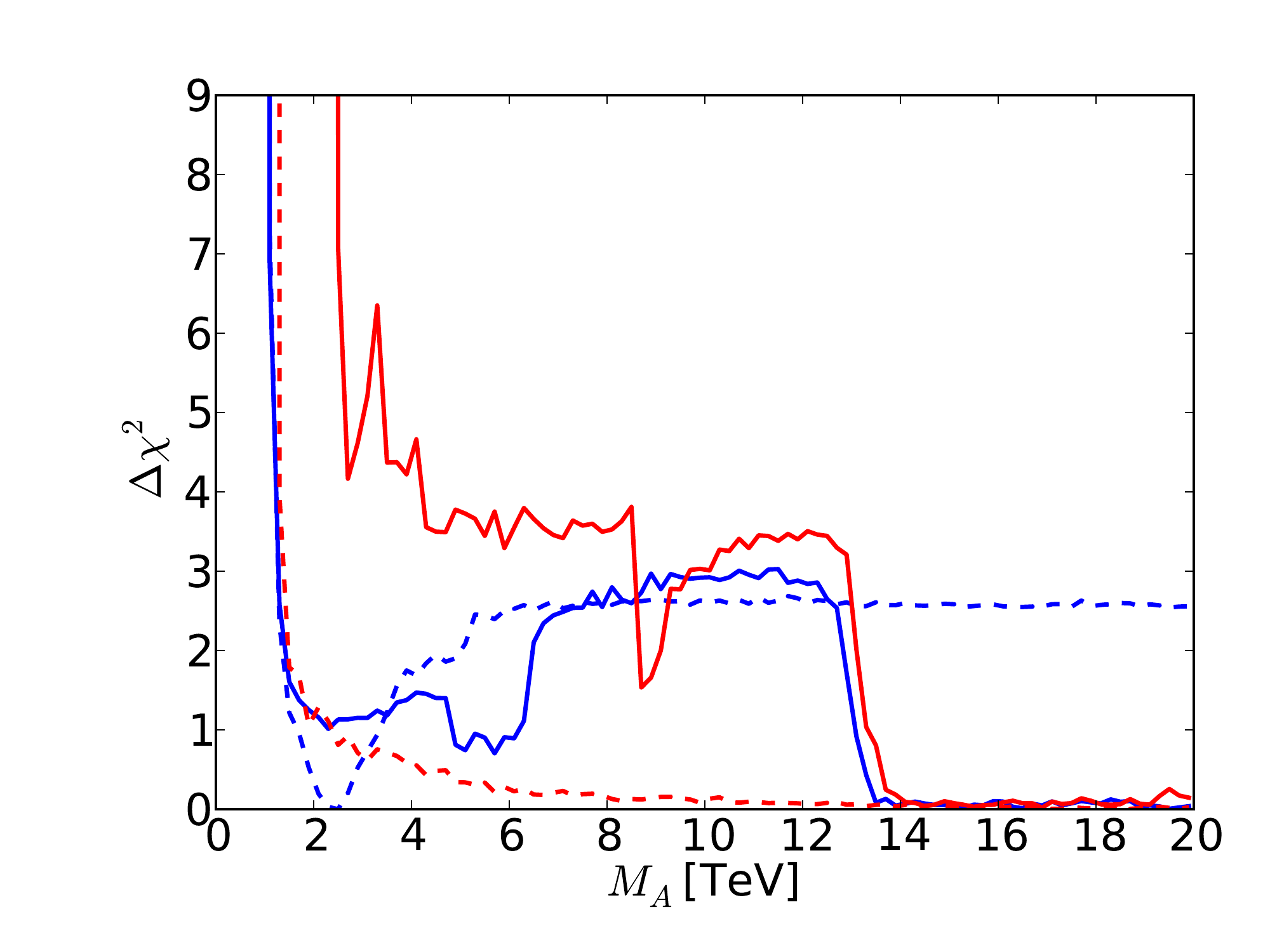}}\\
\end{center}
\vspace{-0.5cm}
\caption{\it The $\chi^2$ likelihood functions for $\mgl, \msq, \mst1, \mstau1, \mcha1$ and $\MA$.  We show curves with both $\Omega_{\neu1} = \Omega_{\rm CDM}$ (solid lines), and with $\Omega_{\neu1} \le \Omega_{\rm CDM}$ (dashed lines), for both the $\mu>0$ and $\mu<0$ cases (blue and red, respectively).} 
\label{fig:1dchi2}
\end{figure*}

As shown in Fig~\ref{fig:indirect_searches}, there are some interesting prospects for indirect searches 
for mAMSB effects. There are in general small departures from the SM if the LSP accounts for all of the CDM,
whereas much more significant effects can arise if the CDM constraint is relaxed. 
In particular, we find that significant destructive interference between mAMSB
effects and the SM may cause a sizeable decrease of the \bsdmm\ branching ratio in the positive $\mu$ case,
which can be significant within the range of model parameters allowed at the 2-$\sigma$ level and improves
the agreement with the experimental measurement shown by the dotted line.
This effect arises from a region of parameter space at large \tb\ where $\MA$ can be below $5 \tev$, 
as seen in the bottom right panel of Fig~\ref{fig:1dchi2}. 
We find that the destructive interference in \bsdmm\ is always 
accompanied by a constructive interference in \bsg. There is also some possibility of positive
interference in \bsdmm\ and negative interference in \bsg\ when $\mu < 0$ and the LSP does not provide
all the dark matter, though this effect is much smaller. Finally, we also note that
only small effects at the level of $10^{-10}$ can appear in \gmt, for either sign of $\mu$.

\begin{figure*}[htb!]
\vspace{0.5cm}
\begin{center}
\resizebox{7.5cm}{!}{\includegraphics{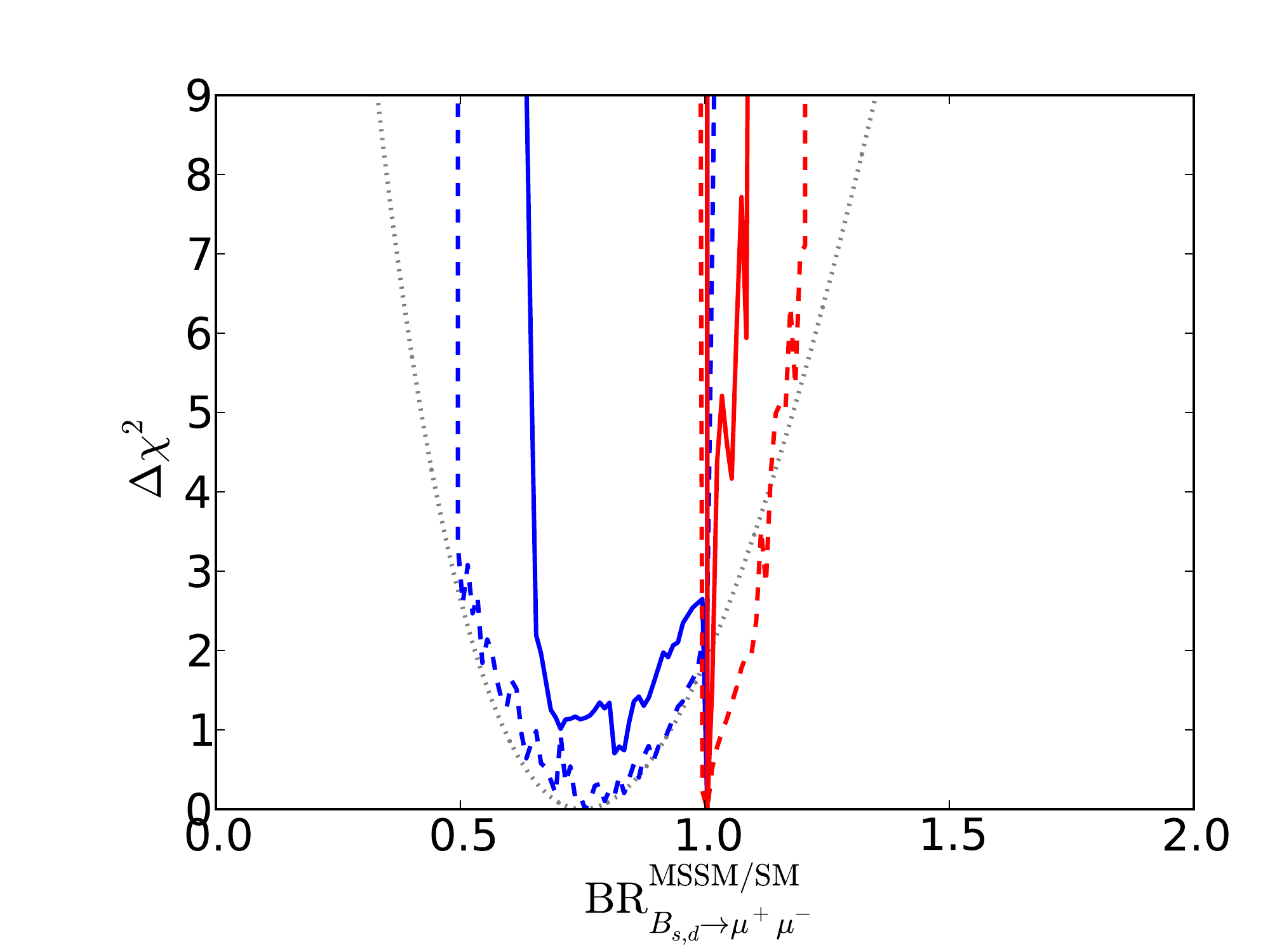}}\\
\resizebox{7.5cm}{!}{\includegraphics{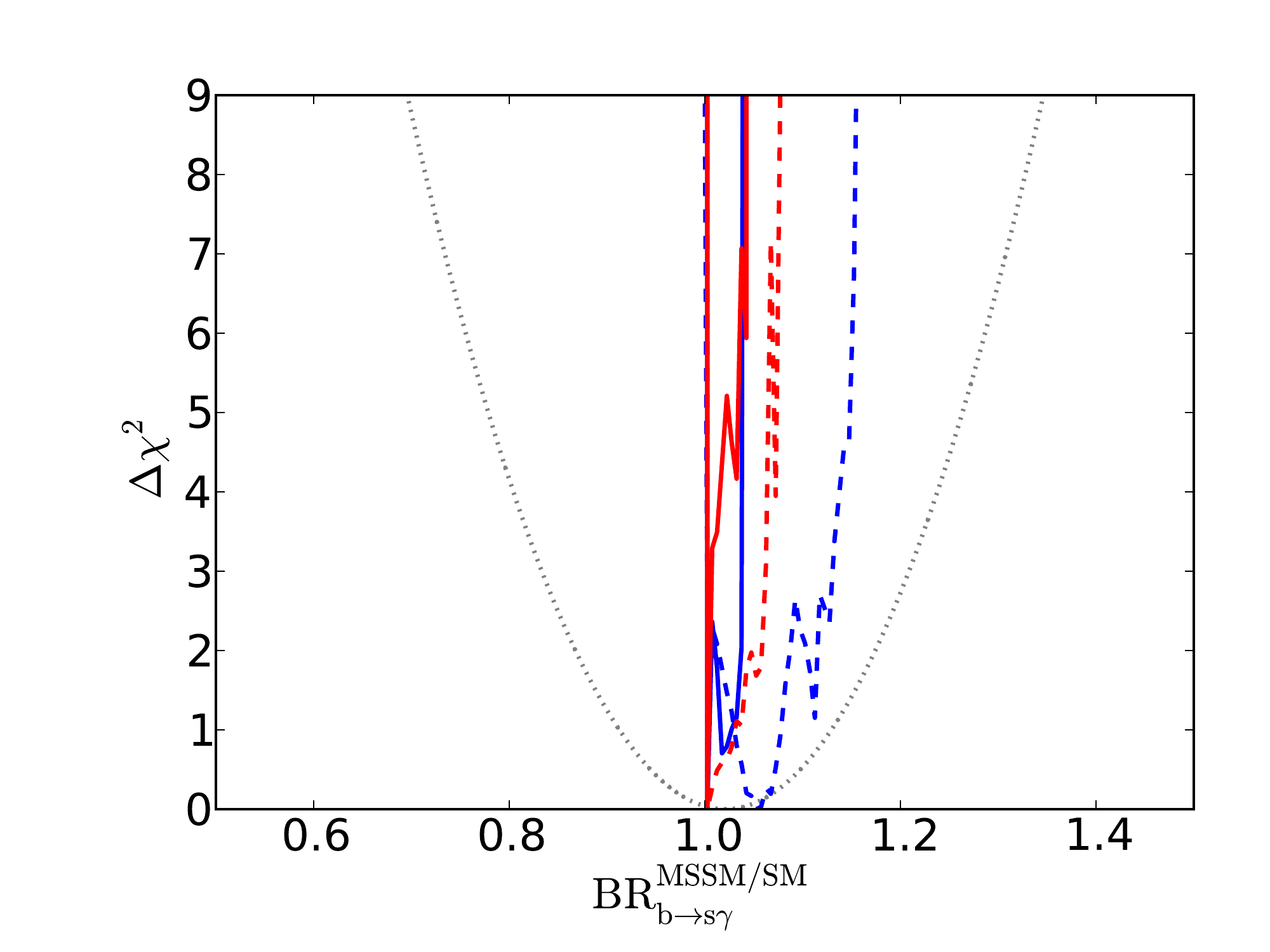}}
\resizebox{7.5cm}{!}{\includegraphics{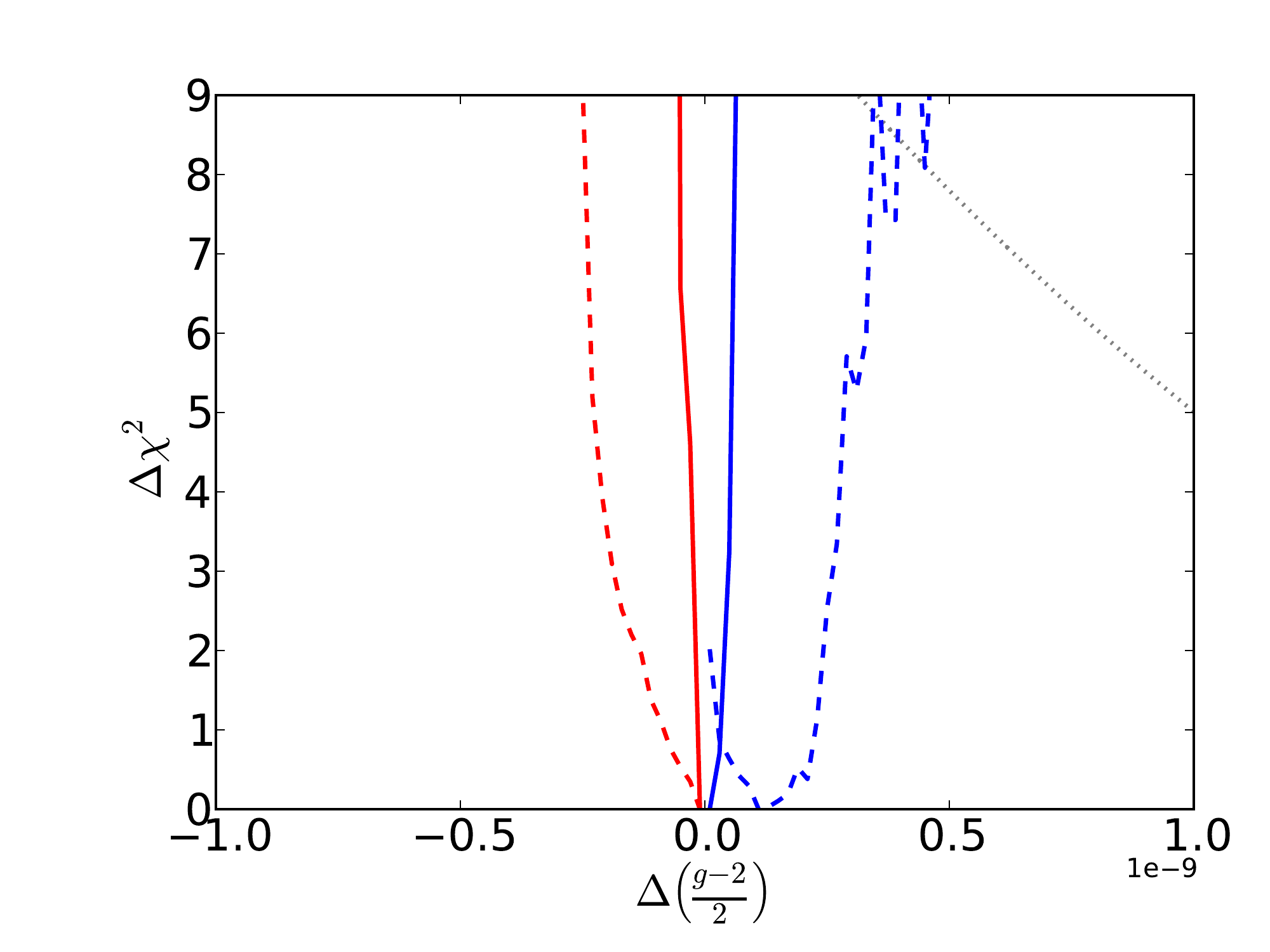}}
\end{center}
\vspace{-0.5cm}
\caption{\it The $\chi^2$ likelihood functions for the ratios of \bsdmm, \bsg\ to their SM values,
and for the contribution to \gmt/2.  We show curves with both
$\Omega_{\neu1} = \Omega_{\rm CDM}$ (solid lines), and with $\Omega_{\neu1} \le \Omega_{\rm CDM}$ (dashed lines), as well as both the $\mu>0$ and $\mu<0$ cases (blue and red, respectively).
The dotted lines represent the current experimental measurements of these observables.
} 
\label{fig:indirect_searches}
\end{figure*}

\subsection{{Discovery Prospects at the LHC and FCC-hh}}

\begin{figure*}[htb!]
\vspace{0.5cm}
\begin{center}
  \resizebox{7.5cm}{!}{\includegraphics{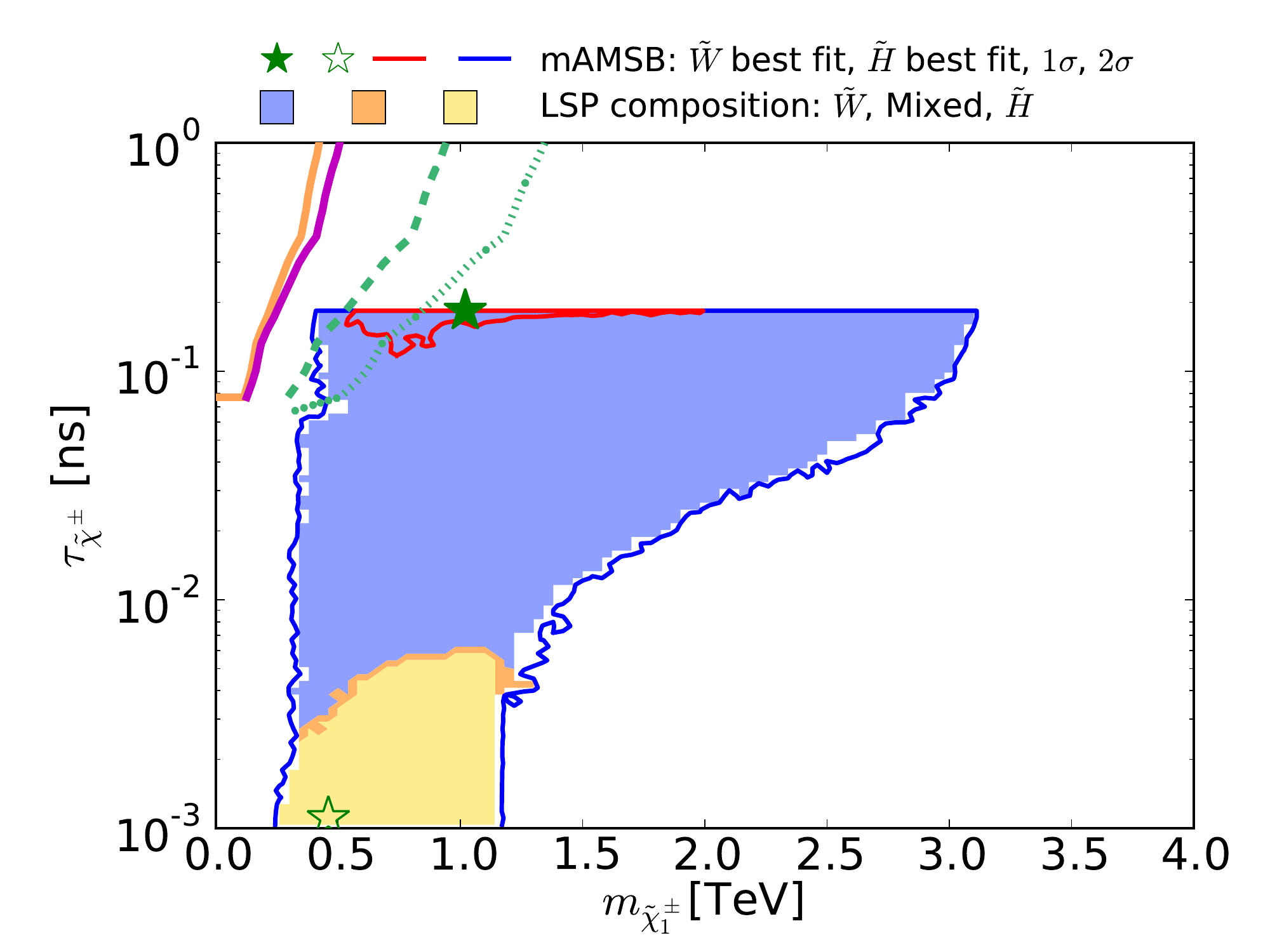}}\put(-95, +123){\footnotesize $\mu>0$, $\Omega_{\neu1}<\Omega_{\rm CDM}$}
\resizebox{7.5cm}{!}{\includegraphics{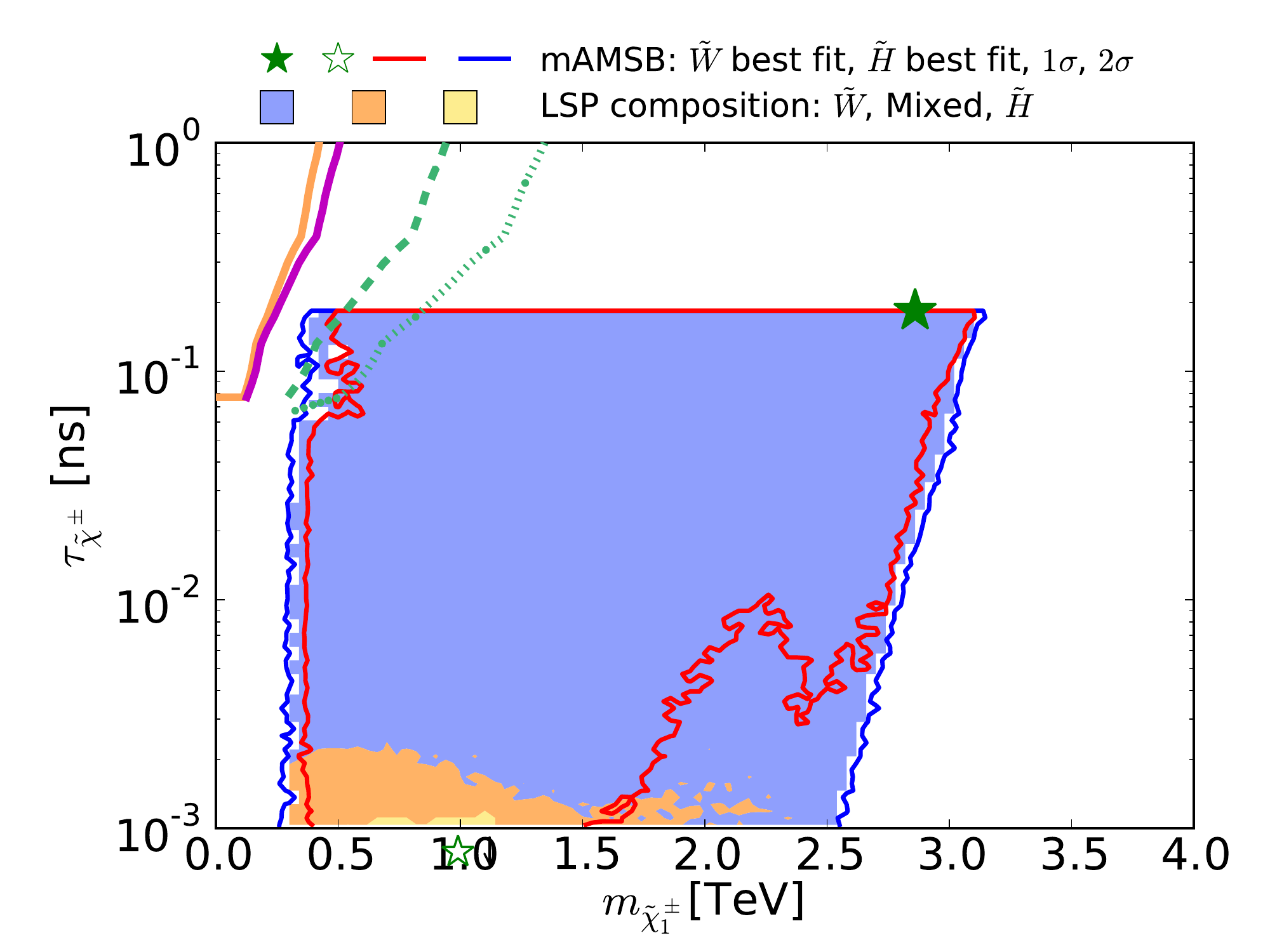}}\put(-95, +123){\footnotesize $\mu<0$, $\Omega_{\neu1}<\Omega_{\rm CDM}$}
\end{center}
\caption{\it {The region of the $(\mcha1, \tau_{\cha1})$ plane allowed
in the  $\Omega_{\neu1} \le \Omega_{\rm CDM}$ case for $\mu>0$ (left) and $\mu<0$ (right). The orange solid line represents the limit from the ATLAS 8-TeV search for disappearing tracks~\cite{disappearing-ATLAS}. The magenta solid, green dashed and green dotted lines represents the projection of this limit to 13-$\tev$ data with 13, 300 and 3000~$fb^{-1}$, respectively. 
{The shadings are the same as in Fig.~\protect\ref{fig:m0m32}.}
}
}
\label{fig:winoLifetime}
\end{figure*}

\begin{figure*}[htb!]
\vspace{1.0cm}
\begin{center}
\resizebox{7.5cm}{!}{\includegraphics{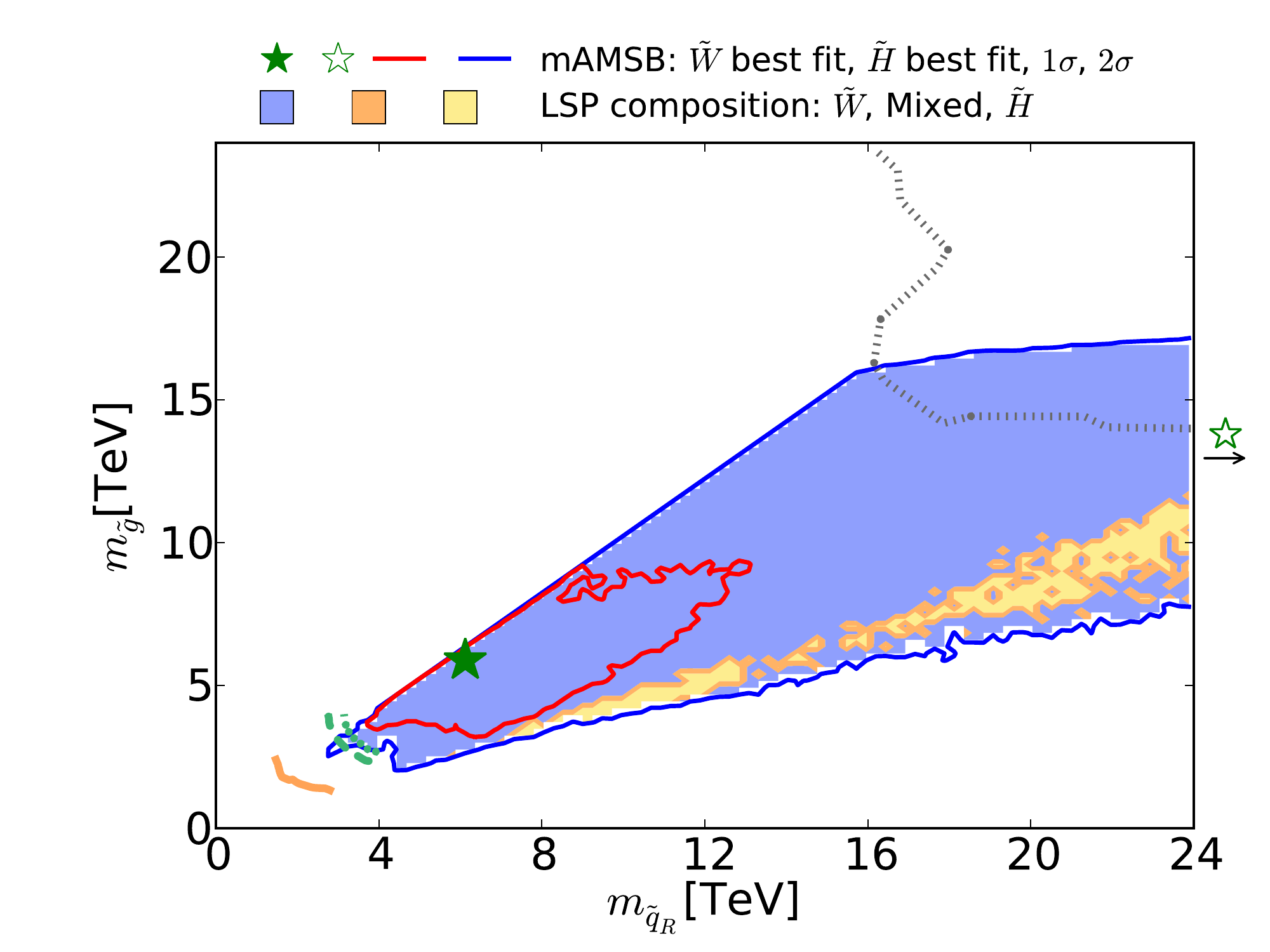}}\put(-169, +123){\footnotesize $\mu>0$, $\Omega_{\neu1}<\Omega_{\rm CDM}$}
\resizebox{7.5cm}{!}{\includegraphics{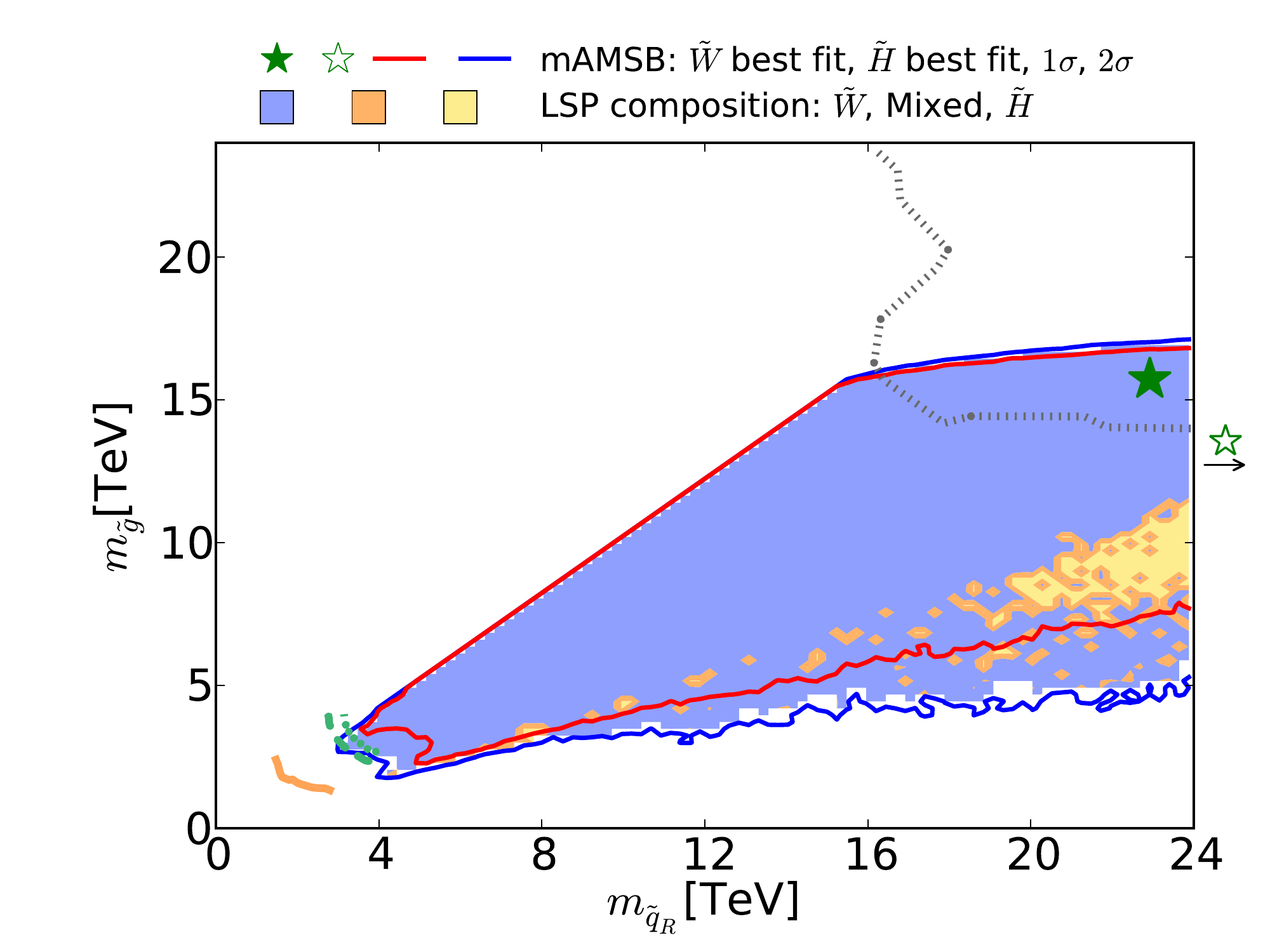}}\put(-169, +123){\footnotesize $\mu<0$, $\Omega_{\neu1}<\Omega_{\rm CDM}$}
\end{center}
\caption{\it The region of the $(\msqr, \mgl)$ plane allowed in the  $\Omega_{\neu1} \le \Omega_{\rm CDM}$ case for $\mu>0$ (left) and $\mu<0$ (right). The orange solid line represents the LHC 8-TeV $
95\%$ CL exclusion~\cite{Aad:2014wea}. The {green} dashed and dotted lines show the projection estimated by ATLAS \cite{ATLAS-Collaboration:2012jwa} for 14-TeV data with 300 and 3000~$fb^{-1}$, respectively.
The grey dotted line is the 95\% CL sensitivity expected at a 100 TeV $pp$ collider with a 3000~$fb^{-1}$~integrated luminosity 
obtained from \cite{Cohen:2013xda}.
All contours assume massless $\neu1$.
{The shadings are the same as in Fig.~\protect\ref{fig:m0m32}.}
}
\label{fig:mQ-mG}
\end{figure*}

\begin{figure*}[htb!]
\vspace{0.5cm}
\begin{center}
\resizebox{7.5cm}{!}{\includegraphics{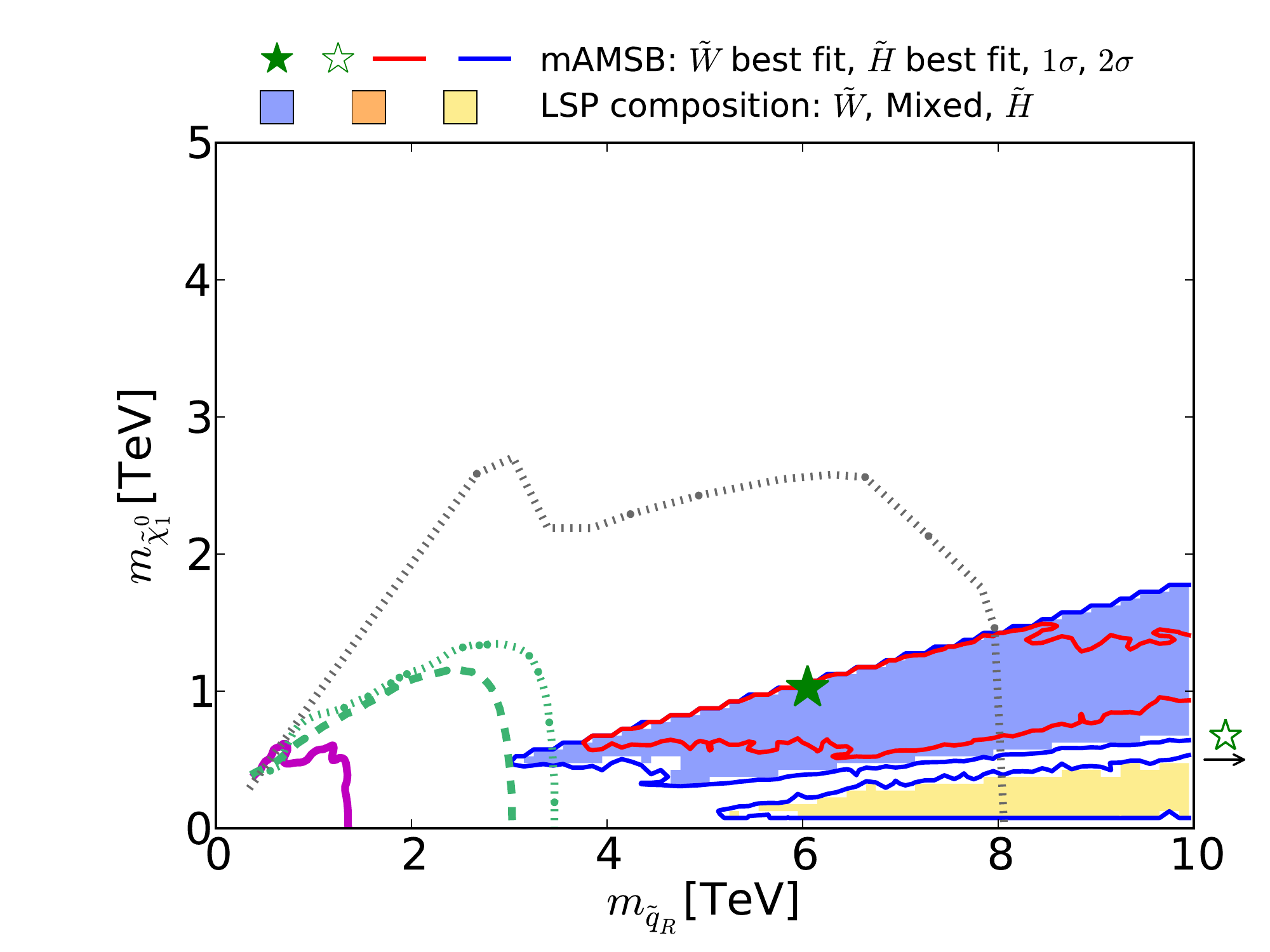}}\put(-169, +123){\footnotesize $\mu>0$, $\Omega_{\neu1}<\Omega_{\rm CDM}$}
\resizebox{7.5cm}{!}{\includegraphics{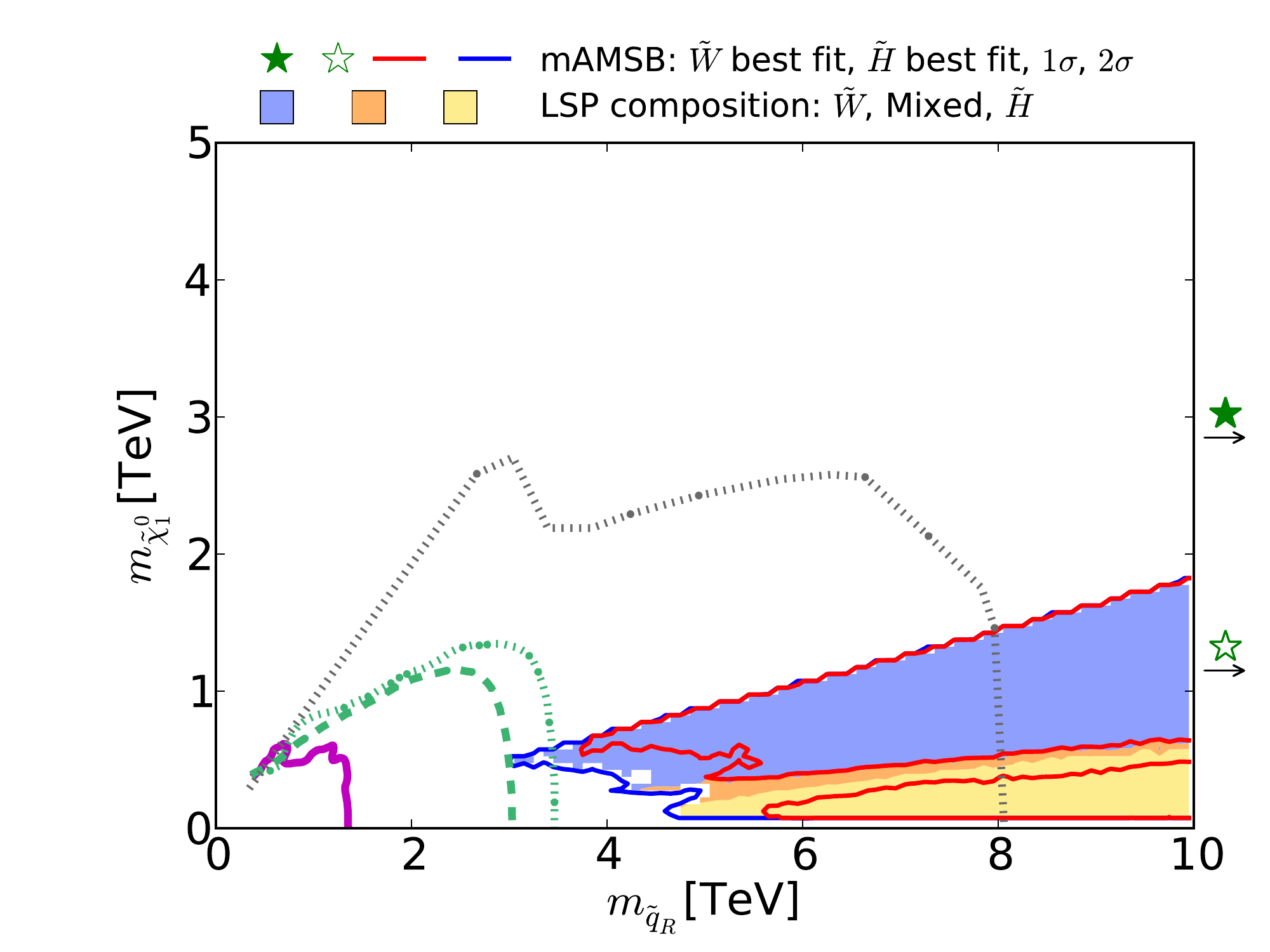}}\put(-169, +123){\footnotesize $\mu<0$, $\Omega_{\neu1}<\Omega_{\rm CDM}$}
\end{center}
\vspace{-0.5cm}
\caption{{\it The region of the $(\msqr, \mneu1)$ plane allowed in the  $\Omega_{\neu1} \le \Omega_{\rm CDM}$ case for $\mu>0$ (left) and $\mu<0$ (right). The purple solid line represents the ATLAS 13-TeV $95\%$ CL exclusion using 13~$fb^{-1}$ of data ~\cite{ATLAS:2016kts}.
The green dashed and dotted lines show the projected 95\% CL sensitivity estimated by ATLAS \cite{ATL-PHYS-PUB-2014-010} for 14-TeV data with integrated luminosities of 300 and 3000~$fb^{-1}$, respectively. 
The grey dotted line is the 95\% CL sensitivity expected at a 100 TeV $pp$ collider with a 3000~$fb^{-1}$~integrated luminosity 
obtained from \cite{Cohen:2013xda}.
All contours assume a simplified model with $100\%$ BR for $\tilde{q} \to q \neu1$.
The current limit and 100 TeV projection assumes decoupled gluino, while the projection to the higher luminosity LHC
assumes a 4.5-TeV gluino. {The shadings are the same as in Fig.~\protect\ref{fig:m0m32}.}
}}
\label{fig:mQ-mN1}
\end{figure*}

\begin{figure*}[htb!]
\vspace{0.5cm}
\begin{center}
\resizebox{7.5cm}{!}{\includegraphics{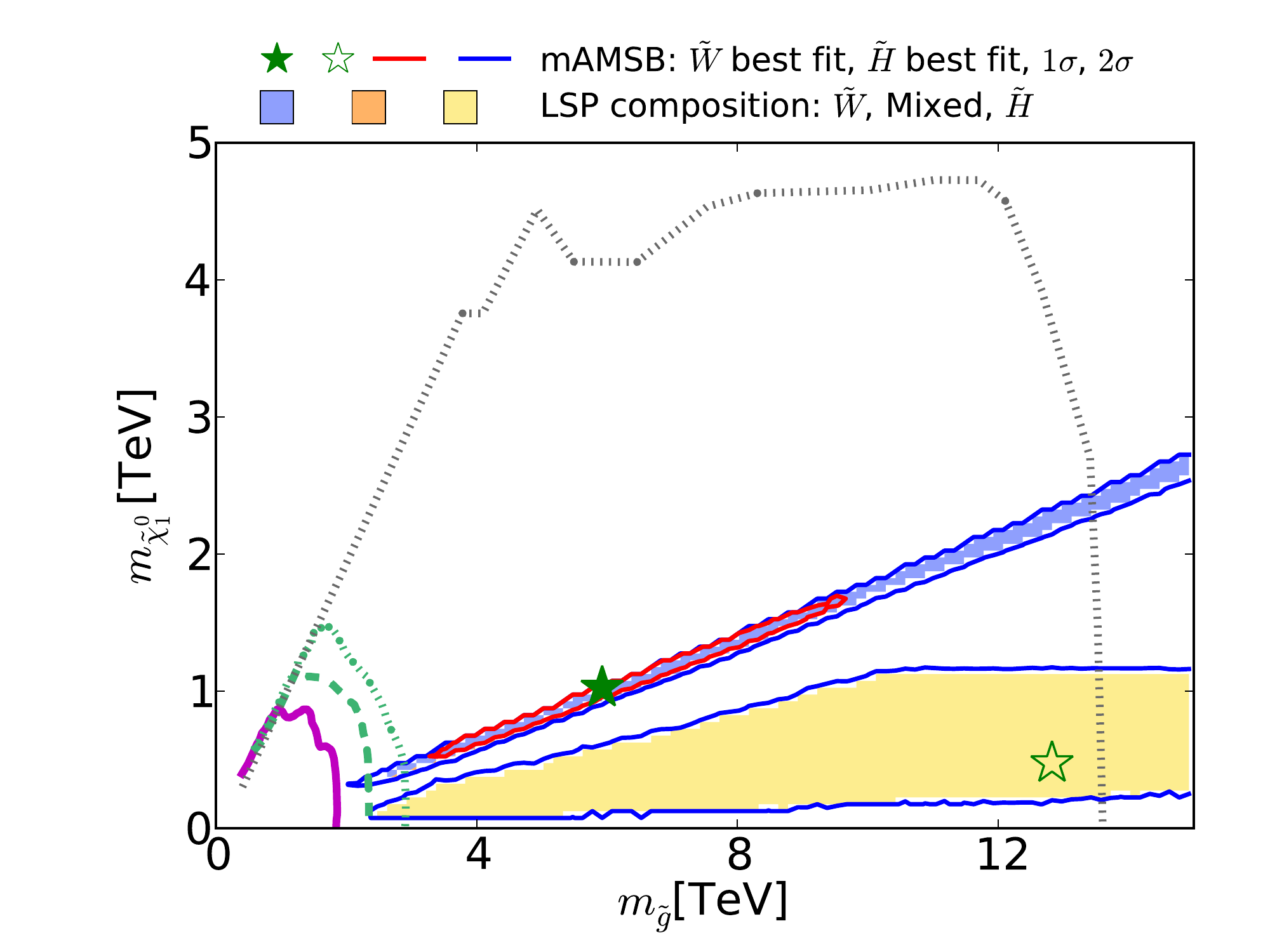}}\put(-169, +123){\footnotesize $\mu>0$, $\Omega_{\neu1}<\Omega_{\rm CDM}$}
\resizebox{7.5cm}{!}{\includegraphics{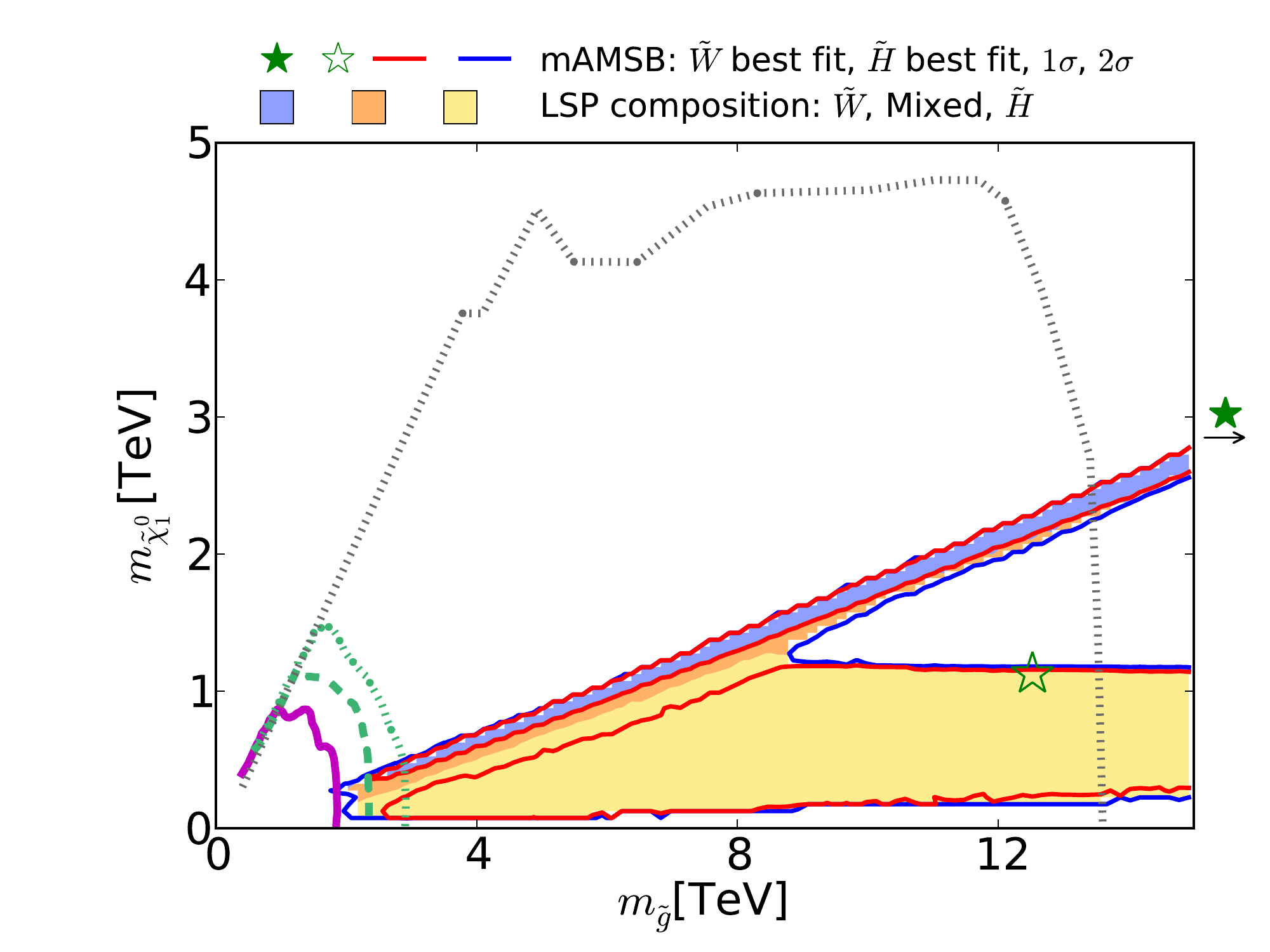}}\put(-169, +123){\footnotesize $\mu<0$, $\Omega_{\neu1}<\Omega_{\rm CDM}$}
\end{center}
\vspace{-0.5cm}
\caption{{\it The region of the $(\mgl, \mneu1)$ plane allowed in the  $\Omega_{\neu1} \le \Omega_{\rm CDM}$ case for $\mu>0$ (left) and $\mu<0$ (right). 
The purple solid line represents the ATLAS 13-TeV $95\%$ CL exclusion with the data with 13~$fb^{-1}$ ~\cite{ATLAS:2016kts}.
The green dashed and dotted lines show the projected 95\% CL sensitivity estimated by ATLAS \cite{ATL-PHYS-PUB-2014-010} for 14-TeV data with integrated luminosities of 300 and 3000~$fb^{-1}$, respectively. 
The grey dotted line is the 95\% CL sensitivity expected at a 100 TeV $pp$ collider with a 3000~$fb^{-1}$~integrated luminosity 
obtained from \cite{Cohen:2013xda}.
All contours assume a simplified model with $100\%$ BR for $\tilde{g} \to q q \neu1$. 
{The shadings are the same as in Fig.~\protect\ref{fig:m0m32}.}}
}
\label{fig:mG-mN1}
\end{figure*}

\begin{figure*}[htb!]
\vspace{0.5cm}
\begin{center}
\resizebox{7.5cm}{!}{\includegraphics{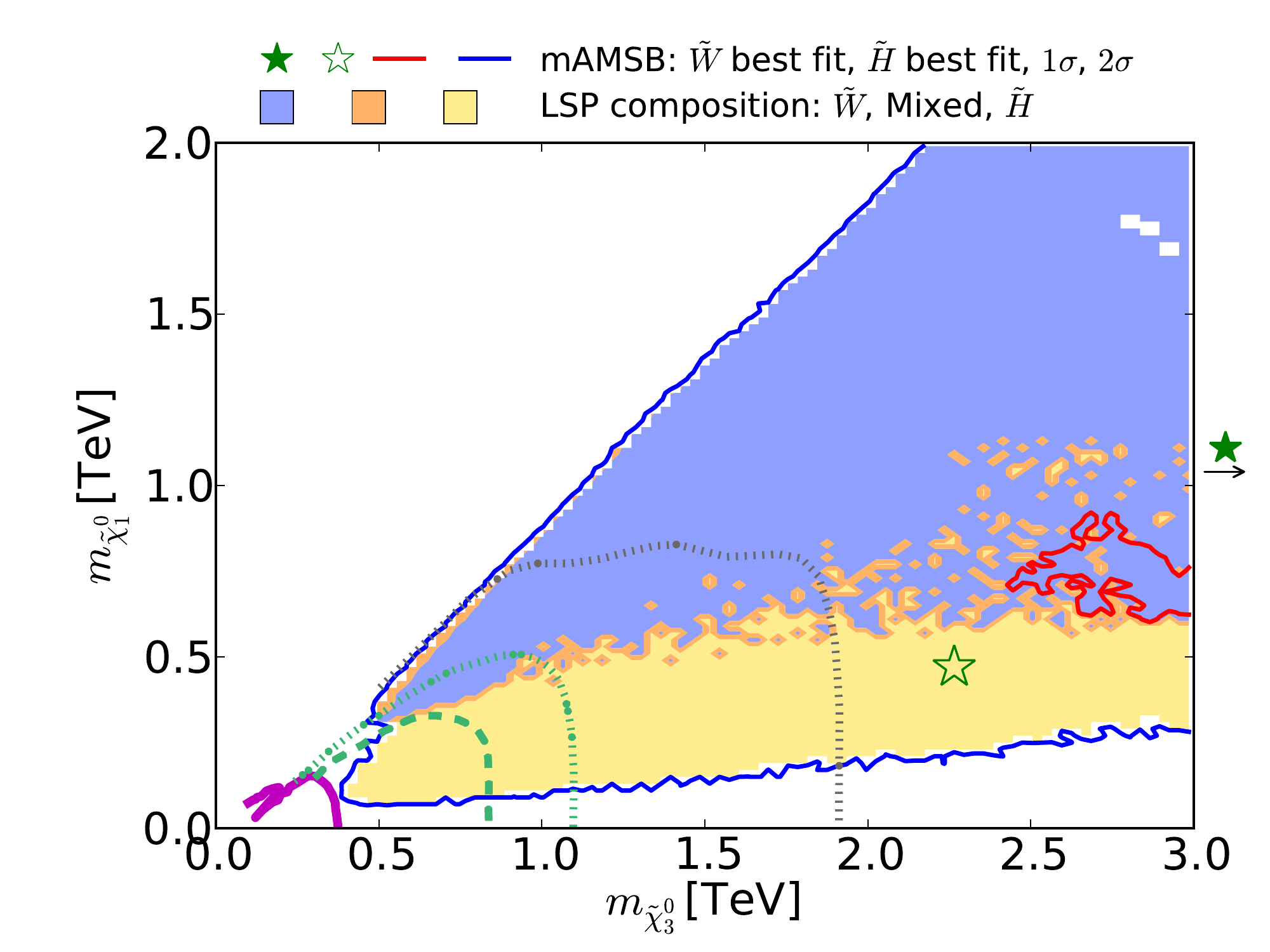}}\put(-169, +123){\footnotesize $\mu>0$, $\Omega_{\neu1}<\Omega_{\rm CDM}$}
\resizebox{7.5cm}{!}{\includegraphics{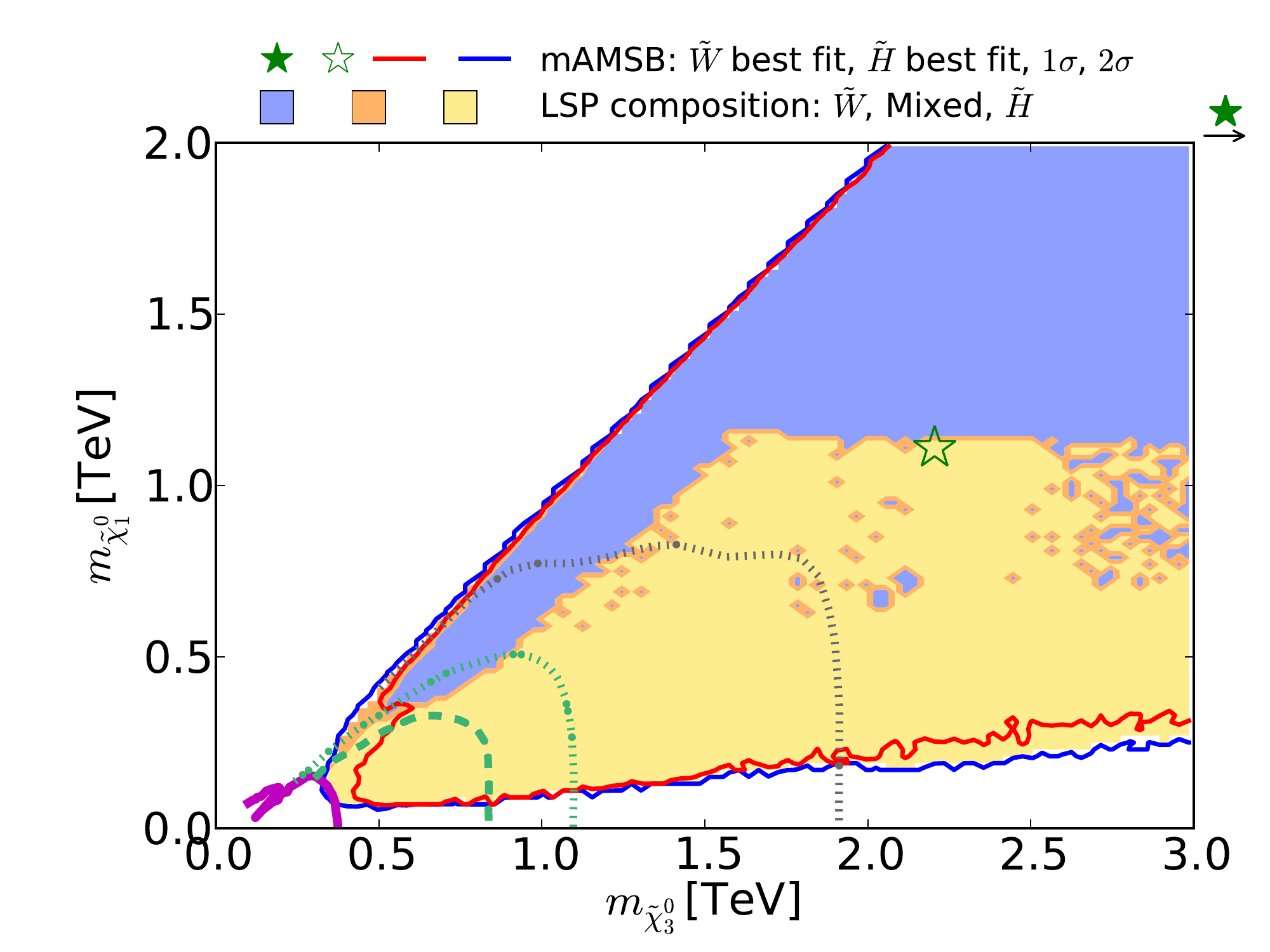}}\put(-169, +123){\footnotesize $\mu<0$, $\Omega_{\neu1}<\Omega_{\rm CDM}$}
\end{center}
\vspace{-0.5cm}
\caption{{\it The region of the $(\mneu3, \mneu1)$ plane allowed in the $\Omega_{\neu1} \le \Omega_{\rm CDM}$ case for $\mu>0$ (left) and $\mu<0$ (right). The purple solid line represents the CMS 13-TeV $95\%$ CL exclusion~\cite{EWKino_cms13} assuming a simplified model with wino-like chargino and neutralino production and $100\%$ BR for the $W^\pm Z + \ETslash$ final state. The green dashed (dotted) line shows the projected sensitivity for 14-TeV data with an integrated luminosity of 300~$fb^{-1}$ (3000~$fb^{-1}$) estimated by ATLAS \cite{ATL-PHYS-PUB-2014-010}.
The grey dotted line is the 95\% CL sensitivity expected at a 100 TeV $pp$ collider with a 3000~$fb^{-1}$~integrated luminosity 
obtained from \cite{Acharya:2014pua}.
{The shadings are the same as in Fig.~\protect\ref{fig:m0m32}.}
} }
\label{fig:mN3-mN1}
\end{figure*}

\begin{figure*}[htb!]
\vspace{0.5cm}
\begin{center}
\resizebox{7.5cm}{!}{\includegraphics{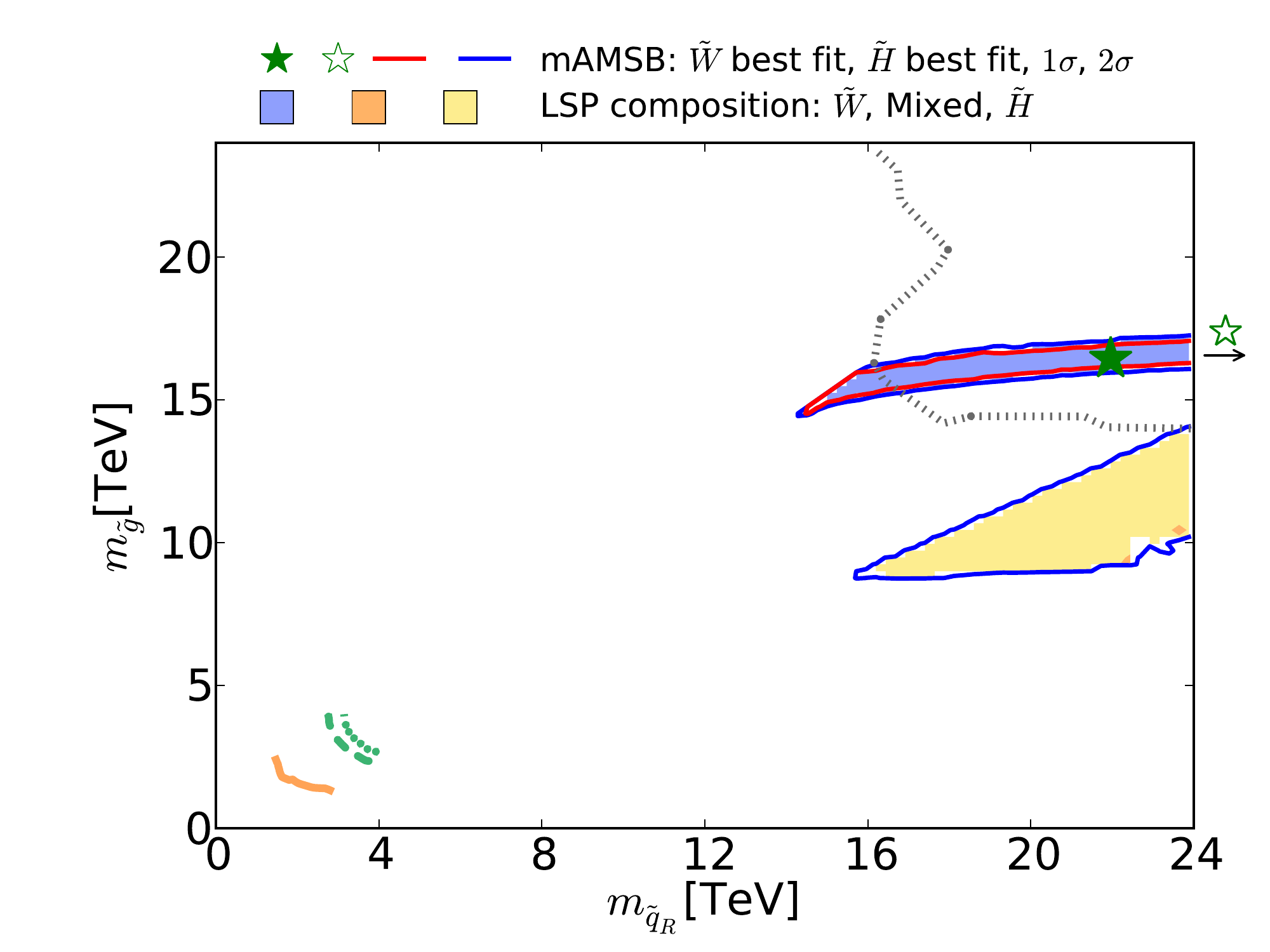}}\put(-169, +123){\footnotesize $\mu>0$, $\Omega_{\neu1}=\Omega_{\rm CDM}$}
\resizebox{7.5cm}{!}{\includegraphics{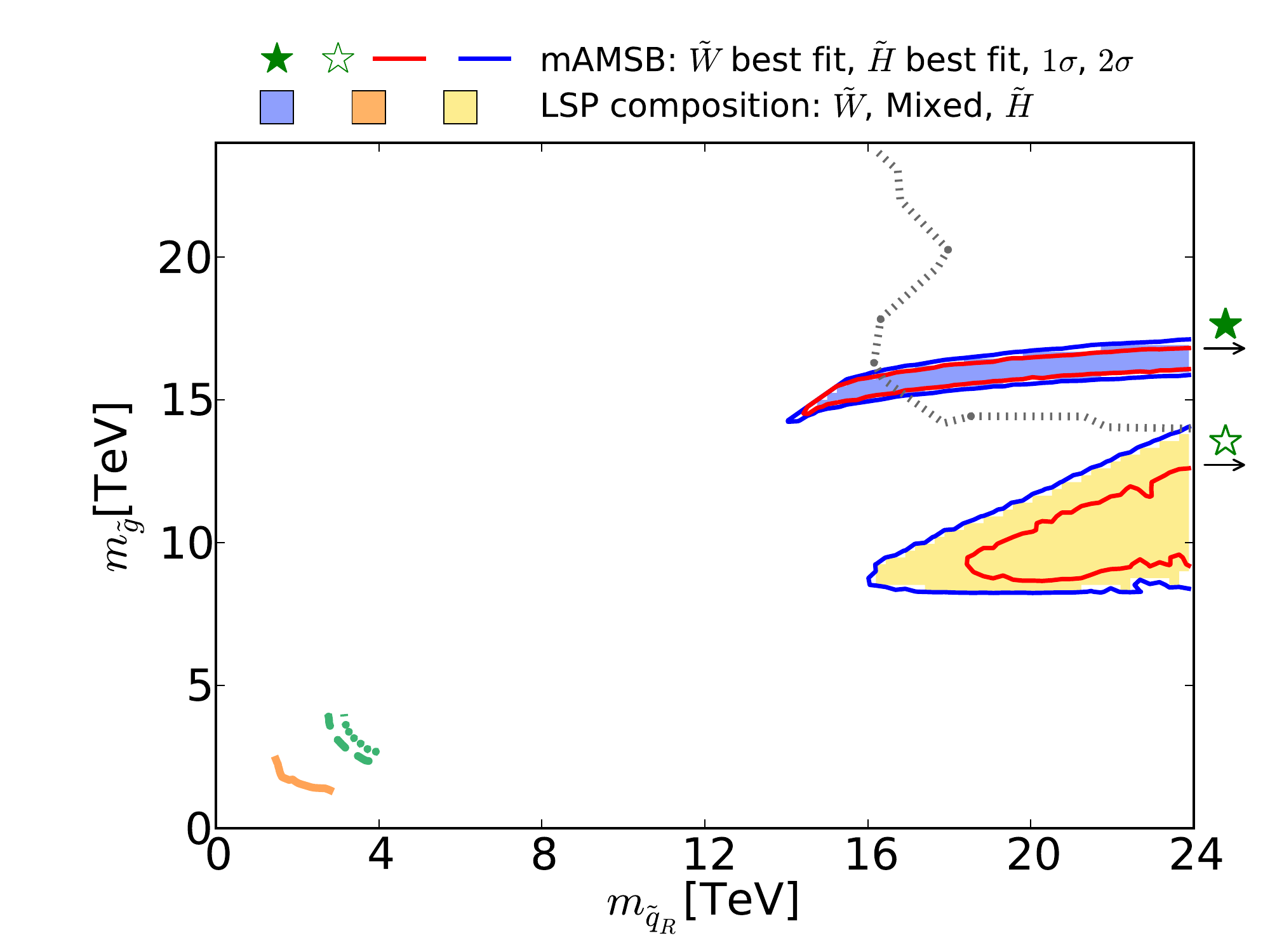}}\put(-169, +123){\footnotesize $\mu<0$, $\Omega_{\neu1}=\Omega_{\rm CDM}$}
\end{center}
\vspace{-0.5cm}
\caption{\it The region of the $(\msqr, \mgl)$ plane allowed in the  $\Omega_{\neu1} = \Omega_{\rm CDM}$ case for $\mu>0$ (left) and $\mu<0$ (right). The orange solid line represents the LHC 8-TeV $
95\%$ CL exclusion~\cite{Aad:2014wea}. The {green} dashed and dotted lines show the projection estimated by ATLAS \cite{ATLAS-Collaboration:2012jwa} for 14-TeV data with 300 and 3000~$fb^{-1}$, respectively.
The grey dotted line is the 95\% CL sensitivity expected at a 100 TeV $pp$ collider with a 3000~$fb^{-1}$~integrated luminosity 
obtained from \cite{Cohen:2013xda}.
All contours assume massless $\neu1$.
{The shadings are the same as in Fig.~\protect\ref{fig:m0m32}.}
} 
\label{fig:mQ-mG_wDM}
\end{figure*}

As mentioned above, current LHC searches are not sensitive to the high-mass spectrum of mAMSB, even if the LSP is not the only component of CDM. However, simple extrapolation indicates that
there are better prospects for future LHC searches with 300 or 3000~\ifb. 
Searches for chargino tracks disappearing in the tracker, such as that performed by 
ATLAS~\cite{disappearing-ATLAS} start to be sensitive with 300~\ifb of data,
and become much more sensitive with 3000~\ifb, as shown in Figure~\ref{fig:winoLifetime}.
We have obtained the projected contours for the 13 TeV LHC with 13, 300 and 3000~\ifb
by rescaling the Run-1 sensitivity presented in~\cite{disappearing-ATLAS}.
In doing so, at a given lifetime we shift the Run-1 value of the 95\% CL reach for the 
wino mass to the higher value at which
the wino cross-section times luminosity (13, 300 and 3000~\ifb) at the 13-TeV 
LHC coincides with the reach achieved during Run-1.    
This method is often used and is known to give a reasonable estimate \cite{colliderreach}.
We find that the disappearing-track search is more sensitive in the $\mu>0$ case than in the $\mu<0$
case because, in order to accommodate the reduced \bsmm, the wino-LSP solution is 
preferred over the Higgsino-LSP one.

{The large mass reach of 
a 100 TeV $pp$ collider would extend these sensitivities greatly.
Ref.~\cite{Low:2014cba} studied the capability of exploring the wino LSP scenario at a 100-TeV collider
and found that the sensitivity reaches around $m_{\cha1} \sim 3$ TeV at 3000~\ifb.
We therefore expect that at a 100-TeV collider with 3000~\ifb ~almost the entire 68\% CL region 
and a part of the 95\% CL with $\tau_{\cha1} > 0.1$\,ns will be explored for $\mu > 0$ and $< 0$, respectively.
If improvements on the detector and the analysis allow the sensitivity to be extended
to $m_{\cha1} \gsim 3$ TeV,
the wino-like dark matter region in the scenario with $\Omega_{\neu1} = \Omega_{\rm CDM}$ 
can also be probed.
}

{Coloured sparticle} searches that will become sensitive in future LHC runs are shown in 
Figs.~\ref{fig:mQ-mG}, \ref{fig:mQ-mN1}, and \ref{fig:mG-mN1} 
in the ($m_{\tilde g}, m_{\tilde q_R}$), ($m_{\tilde q_R}, m_{\tilde \chi_1^0}$) and ($m_{\tilde g}, m_{\tilde \chi_1^0}$) planes, respectively,
the latter being the most promising one. 
In these figures, exclusion contours from the Run-1 (orange) and the 13~TeV with 13~\ifb ~(purple) data 
are taken from ATLAS analyses~\cite{Aad:2014wea} and~\cite{ATLAS:2016kts}, respectively.
Also superimposed are the projected 95\% CL contours at the 14 TeV LHC with 
300 (green dashed) and 3000 (green dotted)~\ifb estimated by ATLAS \cite{ATLAS-Collaboration:2012jwa, ATL-PHYS-PUB-2014-010}.
We also show, by grey dotted contours, the sensitivity at a 100 TeV collider with 3000~\ifb ~taken from \cite{Cohen:2013xda}. 
These contours assume simplified models with $\mathrm{BR}(\tilde q \to q \tilde \chi_1^0) = 
\mathrm{BR}(\tilde g \to qq \tilde \chi_1^0) = 100\%$ for Figs.~\ref{fig:mQ-mN1} and \ref{fig:mG-mN1}.
For Fig. \ref{fig:mQ-mG}, in addition to these decays the heavier of the gluino and squark 
can also decay into the lighter one with an appropriate branching ratio.
In Fig.~\ref{fig:mQ-mN1} the projected LHC contours are estimated postulating $m_{\tilde g} = 4.5$ TeV, 
which is the right ball-park in our scenario.
{We see that with 3000~\ifb ~the LHC could nip the tip of the 95\% CL region in these planes,
whereas with the same luminosity a 100-TeV collider would explore a sizeable region of the parameter space.
In particular, the best-fit point {and the 68\% CL region are}
within this sensitivity for the $\mu > 0$ case.}

{Allowing $\Omega_{\neu1} < \Omega_{\rm CDM}$, we found in our sample very light winos as well as Higgsinos.
If both of them are light but  with a sufficiently large mass hierarchy between them,
the LHC and a 100-TeV collider may be able to detect 
the production of a heavier state decaying subsequently into the lightest state by emitting the heavy bosons
$W^\pm$, $Z$ and $h$.
In Fig.~\ref{fig:mN3-mN1} we plot the current and future LHC
reaches as well as the sensitivity expected at a 100-TeV collider
with the same luminosity assumptions as in the previous figures.
The current limit (purple) and projected sensitivity (green) at the LHC are estimated by
CMS \cite{EWKino_cms13} and ATLAS \cite{ATL-PHYS-PUB-2014-010}
and assume wino-like chargino and neutralino production and a 100\% rate for 
decay into the $W^\pm Z + \ETslash$ final state.
As can be seen in Fig.~\ref{fig:mN3-mN1}, the region that can be explored is mainly the Higgsino-like LSP region, 
whereas we are interested in the wino-like chargino and neutralino production.
However, unlike the simplified model assumption employed by ATLAS and CMS, 
the charged wino decays into neutral or charged Higgsinos
emitting $W^\pm$, $Z$ and $h$ with 50, 25 and 25\% branching ratio, 
respectively~\cite{Jung:2014bda,Acharya:2014pua, Gori:2014oua}.  
Similarly, the branching ratios of the neutral wino are 50, 25 and 25\%
for decays into $W^\pm$, $Z$ and $h$, respectively.
In total, only 25\% of the associated charged and neutral wino production 
events contribute to the $W^\pm Z + \ETslash$ channel. 
The LHC contours shown in Fig.~\ref{fig:mN3-mN1} should be considered with this caveat.
Also shown by the grey dotted line is the sensitivity expected at a 100-TeV
collider with 3000~\ifb ~luminosity studied in \cite{Acharya:2014pua}
(see also \cite{Gori:2014oua}) assuming a Higgsino-like LSP and wino-like chargino and neutralino production,
taking into account the correct branching ratios mentioned above.
As can be seen, a 100-TeV collider is sensitive up to $m_{\tilde \chi_3^0} \sim 2$ TeV,
and a large part of the 95\% CL region would be within reach, {and also a substantial portion of
the 68\% CL region if $\mu < 0$, though not the best-fit point for either sign of $\mu$}.

Finally, in Fig.~\ref{fig:mQ-mG_wDM} we show the ($m_{\tilde g}, m_{\tilde q_R}$) plane 
for the scenario with $\Omega_{\neu1} = \Omega_{\rm CDM}$.
We found that a small part of the wino-like dark matter region and a good amount
of the Higgsino-like dark matter region
are within the 95\% CL sensitivity region at a 100-TeV collider with 3000~\ifb.
In particular, the sensitivity contour reaches the best-fit point for the $\mu > 0$ case.}

\begin{table*}[htb!]
\begin{center}
\begin{tabular}{|c|c|c|c|c|} 
\hline
 					   & \multicolumn{2}{|c|}{Wino-LSP}  					& \multicolumn{2}{|c|}{Higgsino-LSP}       \\
Parameter   	       &      $\mu>0$      			& $\mu<0$ 	            & $\mu>0$            & $\mu<0$            \\ 
\hline         
$m_0$: best-fit value       &      $2.0 \tev$     	&  $18 \tev$ 	       & $ 26 \tev$         & $26 \tev$    \\
  	      ~~68\% range   &      $(1, 8) \tev$     	&  $(1, 40) \tev$ 	& $(6,50) \tev$     & $(6, 50) \tev$    \\
	      \hline
$\mgrav$: best-fit value    &      $320 \tev$    	&  $880 \tev$ 	& $700 \tev$         & $700 \tev$ \\
~~68\% range                &   $(200, 400) \tev$       &  $(150, 950) \tev$ & $(150, 1500) \tev$ & $(150, 1500) \tev$ \\
\hline
$\tb$: best-fit value       &      $35 $ 		       &  $4.4$               & $4.4$              & $4.2$            \\
  ~~68\% range              &      $(28, 45) $ 	       &  $(3, 50)$         & $(3,50)$            & $(3,50)$            \\
\hline \hline
$\chi^2 / {\rm d.o.f}$ & 33.7 / 27	       		& 36.4 / 27             & 36.4 / 27          & 36.4 / 27          \\
$\chi^2$ probability        & $17.5\%$               	& $10.6\%$                  & $10.6\%$               & $10.7\%$               \\
\hline
\end{tabular}
\caption{\it Fit results for the mAMSB assuming that the LSP accounts for just a fraction of the cold dark matter density. The 68\% CL ranges correspond to $\Delta\chi^2 = 1$. We also display the values of the global $\chi^2$ function {omitting the contributions from {\tt HiggsSignals}}, and the {corresponding $\chi^2$ probability} values. Each mass range is shown for both the wino- and Higgsino-LSP scenarios and both signs of $\mu$.
} 
\label{tab:parameters_noDM}
\end{center}
\end{table*}

\begin{table*}[htb!]
\begin{center}
\begin{tabular}{|c|c|c|c|c|} 
\hline
 & \multicolumn{2}{|c|}{Wino-LSP}  & \multicolumn{2}{|c|}{Higgsino} \\
Parameter   			&  $\mu>0$      & $\mu<0$ & $\mu>0$      & $\mu<0$ \\ 
\hline                 
$\mneu1$        	       & $(0.7, 1.2) \tev$ 	       & $(0.5, 3.1) \tev$         & $(0.07, 1.15) \tev$       & $(0.07, 1.15) \tev$\\
$\mneu2$        		& $(1.9, 3.5) \tev$ 		& $(0.6, 9.2) \tev$         & $(0.08, 1.15) \tev$       & $(0.08, 1.15) \tev$\\
$\mneu3$        		& $(2.8, 5.4) \tev$ 	       & $(0.6, 13.7) \tev$        & $(0.5, 4.9)\tev$          & $(0.5, 4.8)\tev$ \\
$\mneu4$        		& $(2.8, 5.4) \tev$ 		& $(1.3, 13.7) \tev$        & $(1.4, 15.0)\tev$         & $(1.3, 14.8)\tev$\\
$m_{\tilde{g}}$     	       & $(3.9, 6.9) \tev$ 		& $(2.9, 17.2) \tev$,       & $(3.1, 27)\tev$            & $(3.0, 26) \tev$ \\
$\mcha1-\mneu1$        	& $(0.16, 0.17) \gev$ 	& $(0.16, 4.5) \gev$ & $(0.7, 6.0) \gev$ & $(1.3, 7.0) \gev$ \\
$\tau_{\cha1}$        	& $(0.15, 0.17) ~{\rm ns} $ & $(0.02, 0.17) ~{\rm ns} $  & $< 5.0 \times 10^{-3} ~{\rm ns}$ &$< 1.0 \times 10^{-3} ~{\rm ns}$ \\
\hline
\end{tabular}
\caption{\it Ranges for the masses of the LSP $\neu1$, the next-to-lightest neutralino $\neu2$
and the mass splitting between the lighter chargino $\cha1$ and
the LSP and the corresponding lifetime of $\cha1$ for the case in which the $\neu1$ may accounts for only a fraction of the CDM density. Each parameter is shown for both the wino- and Higgsino-LSP scenarios as well as for the two signs of $\mu$.}
\label{tab:fit_noDM}
\end{center}
\end{table*}


\subsection{Prospects for Direct Detection of Dark Matter}
\label{sec:directDetection}

While the heavy spectra of mAMSB models may lie beyond the reach of the LHC constraints, future
direct DM search experiments may be capable of detecting the interaction of a mAMSB neutralino, 
even if it does not provide all the CDM density \cite{Cheung:2012qy}. 
Fig~\ref{fig:DMdirect} displays the cross section for 
spin-independent scattering on a proton, \ssi, versus the neutralino mass. 
As previously, {the left plane is for $\mu > 0$, the right plane is for $\mu < 0$,
the} 1 and 2\,$\sigma$ CL contours are shown as red and blue lines,
and the wino- and Higgsino-LSP regions are shaded in pale blue and yellow. The pale-green-shaded 
region represents the range of \ssi\ excluded at the 95\% CL by {our} combination of the latest PandaX 
and LUX results~\cite{pandax,lux16}, while the purple and {blue} line{s} show the prospective sensitivities of the LUX-Zeplin (LZ), XENON1T {and XENONnT} experiment{s}~\cite{LZ,XENON1T}. 
Also shown, as a dashed orange line, is the neutrino `floor', below which astrophysical neutrino 
backgrounds would dominate any DM signal~\cite{Snowmass} {(grey region)}.
The mAMSB region allowed at the $2\,\sigma$ level includes points where \ssi\ is nominally larger than that
excluded by LUX and PandaX at the $95\%$ CL, which become allowed when the
large theoretical uncertainty in \ssi\ is taken into account.

{This uncertainty stems largely from the 
uncertainty in the strangeness contribution to the nucleon, which receives contributions from two sources.
The strange scalar density can be written as $y = 1 - \sigma_0 / \Sigma_{\pi N}$ 
where $\sigma_0$ is the change in the nucleon mass due to the 
non-zero $u, d$ quark masses, and is estimated from octet 
baryon mass differences to be $\sigma_0 = 36 
\pm 7$ MeV~\cite{oldsnp}. This is the dominant source of error in the computed cross section.
In addition, the $\pi$-nucleon $\Sigma$ term is taken as $50 \pm 8$ MeV. Another non-negligible source of 
error comes from the uncertainty in quark masses. The resulting 68\% CL uncertainty in the calculated 
value of \ssi\ is more than 50\%.}

However, the current data already put pressure on the mAMSB {when $\mu > 0$}, 
in both the wino- and Higgsino-like LSP cases, corresponding to the left and right vertical
strips in Fig~\ref{fig:DMdirect}~\footnote{{The arches between them,
where the LSP has a mixed wino/Higgsino
composition, are under severe pressure from LUX and PandaX.}}. The Higgsino-like $\neu1$ 
{with this sign of $\mu$} could be explored completely with
the prospective LZ sensitivity, while the wino-like LSP may have \ssi\ below the neutrino `floor'. 
The wino-like LSP region lying below the LZ sensitivity could be {partly}
accessible to a 20-tonne DM experiment such as Darwin~\cite{Darwin}. 
{When $\mu < 0$, \ssi\ may be lower than for positive $\mu$, possibly lying below the LZ sensitivity
in the Higgsino case and far below the neutrino `floor' in the wino case.}

Fig~\ref{fig:DMdirect_ul} {extends the analysis to} the case in which the LSP 
{is allowed to contribute} only a fraction of the CDM density. {In this case
we weight the model value \ssi\ by the ratio
$\Omega_{\neu1}/\Omega_{\rm CDM}$}, {since this would be the fraction of the galactic halo
provided by the LSP in this case.}
There are {still reasonably} good prospects for future DM direct detection experiments
when $\mu > 0$, with only a small fraction of the parameter space lying below the neutrino `floor'. 
{However, when $\mu < 0$ \ssi\ may fall considerably below the `floor', because of
cancellations~\cite{cancellations} in the scattering matrix element.}

\begin{figure*}[htb!]
\vspace{0.5cm}
\begin{center}
\resizebox{7.5cm}{!}{\includegraphics{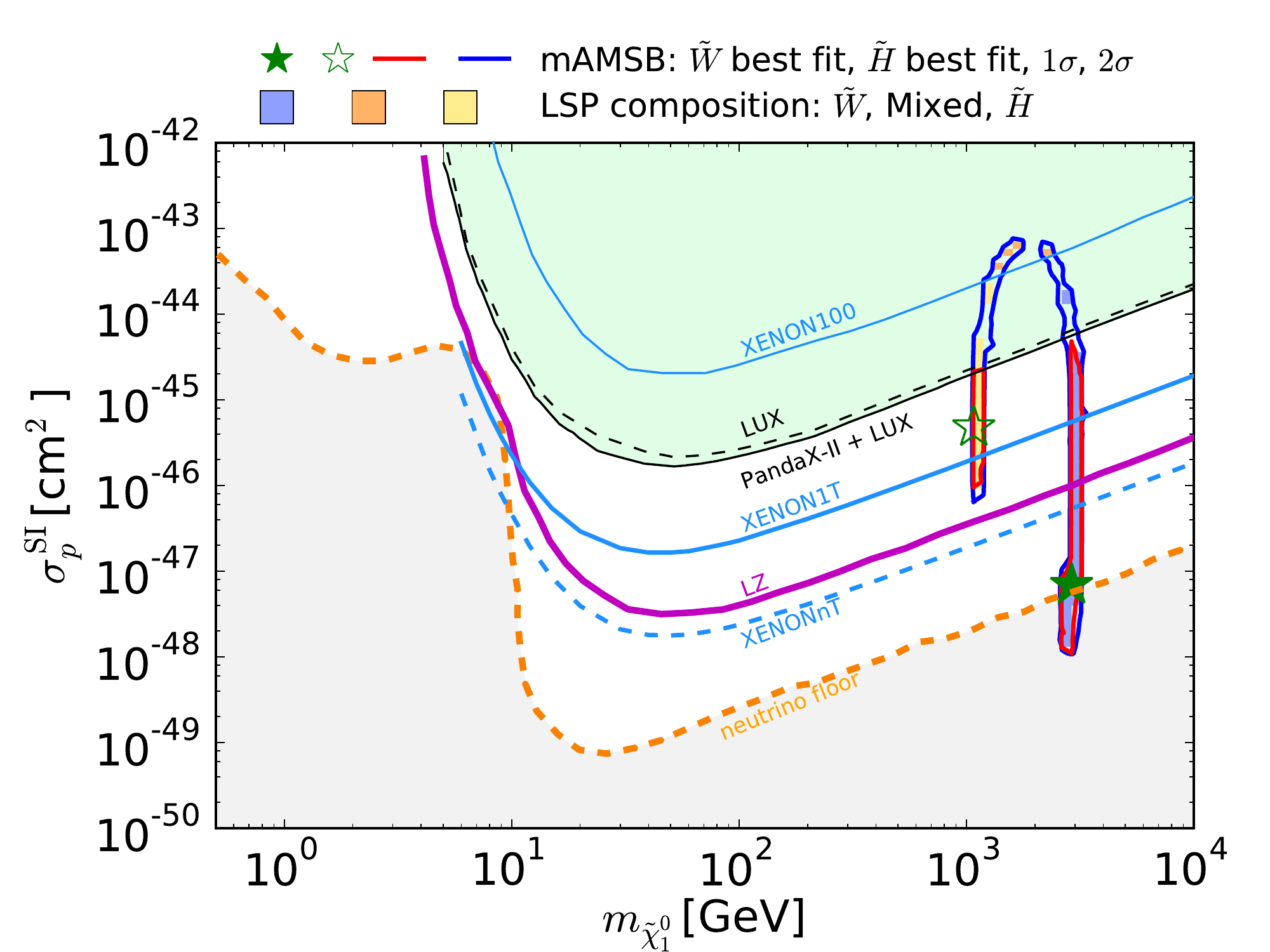}}\put(-95, +123){\footnotesize $\mu>0$, $\Omega_{\neu1}=\Omega_{\rm CDM}$}
\resizebox{7.5cm}{!}{\includegraphics{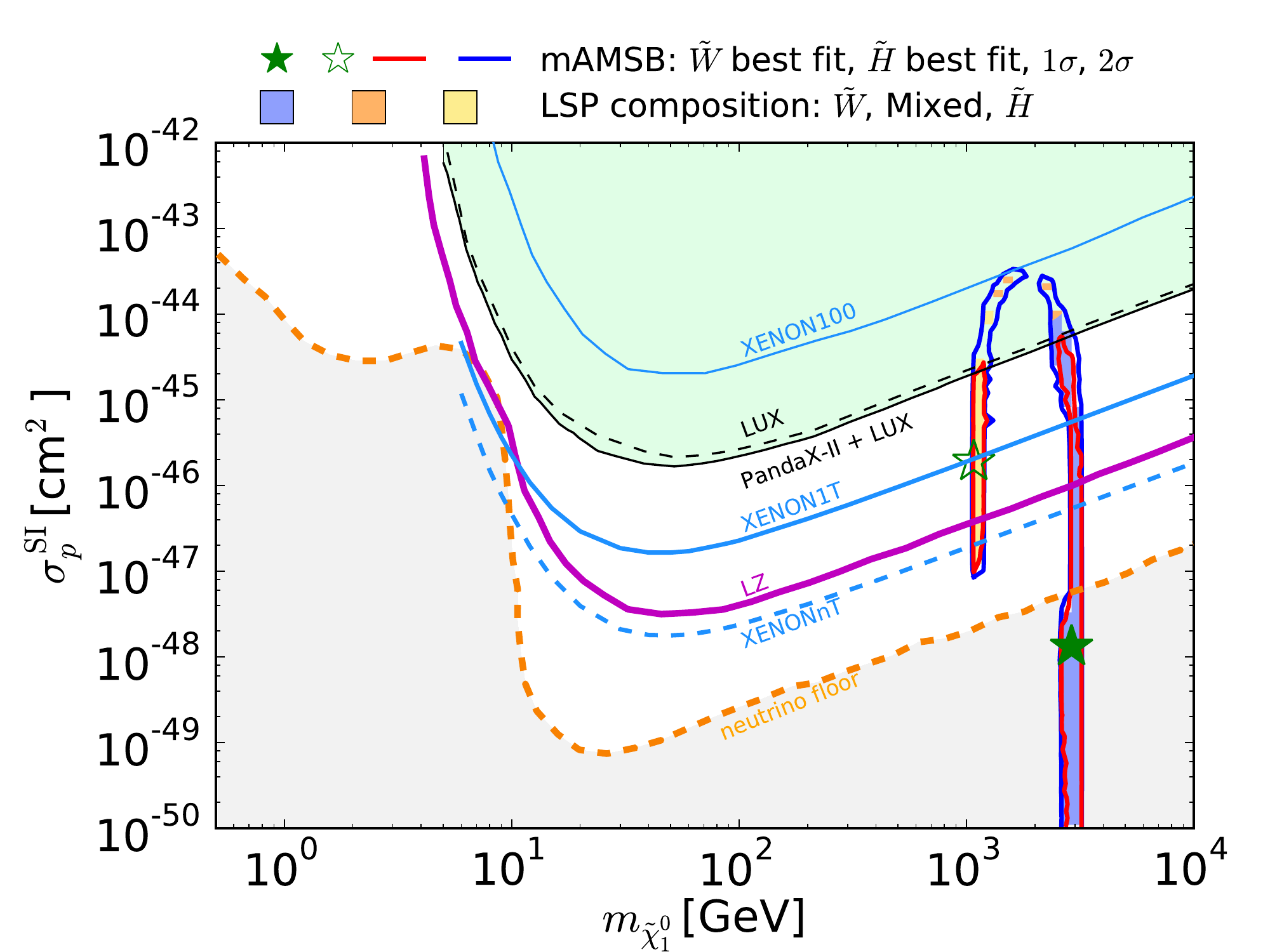}}\put(-95, +123){\footnotesize $\mu<0$, $\Omega_{\neu1}=\Omega_{\rm CDM}$}
\end{center}
\caption{\it The $(\mneu1, \ssi)$ planes in the mAMSB for $\mu>0$ (left) and $\mu<0$
(right) in the case when the LSP accounts for the whole DM density.
The red and blue solid lines are  the 1 and 2\,$\sigma$ CL contours,
and the solid purple and {blue} lines show the projected 95\% exclusion sensitivities of the LUX-Zeplin
{(LZ)~\protect\cite{LZ} and XENON1T/nT experiments~\protect\cite{XENON1T},} respectively. The green line and shaded region show the combined limit from the LUX and PandaX experiments~\protect\cite{pandax,lux16}, and the dashed orange line shows the astrophysical neutrino `floor'~\protect\cite{Snowmass}, below which astrophysical neutrino backgrounds dominate (grey region). {The blue, orange and yellow shadings are the same as in Fig.~\protect\ref{fig:m0m32}.}}
\label{fig:DMdirect}
\end{figure*}

\begin{figure*}[htb!]
\vspace{0.5cm}
\begin{center}
\resizebox{7.5cm}{!}{\includegraphics{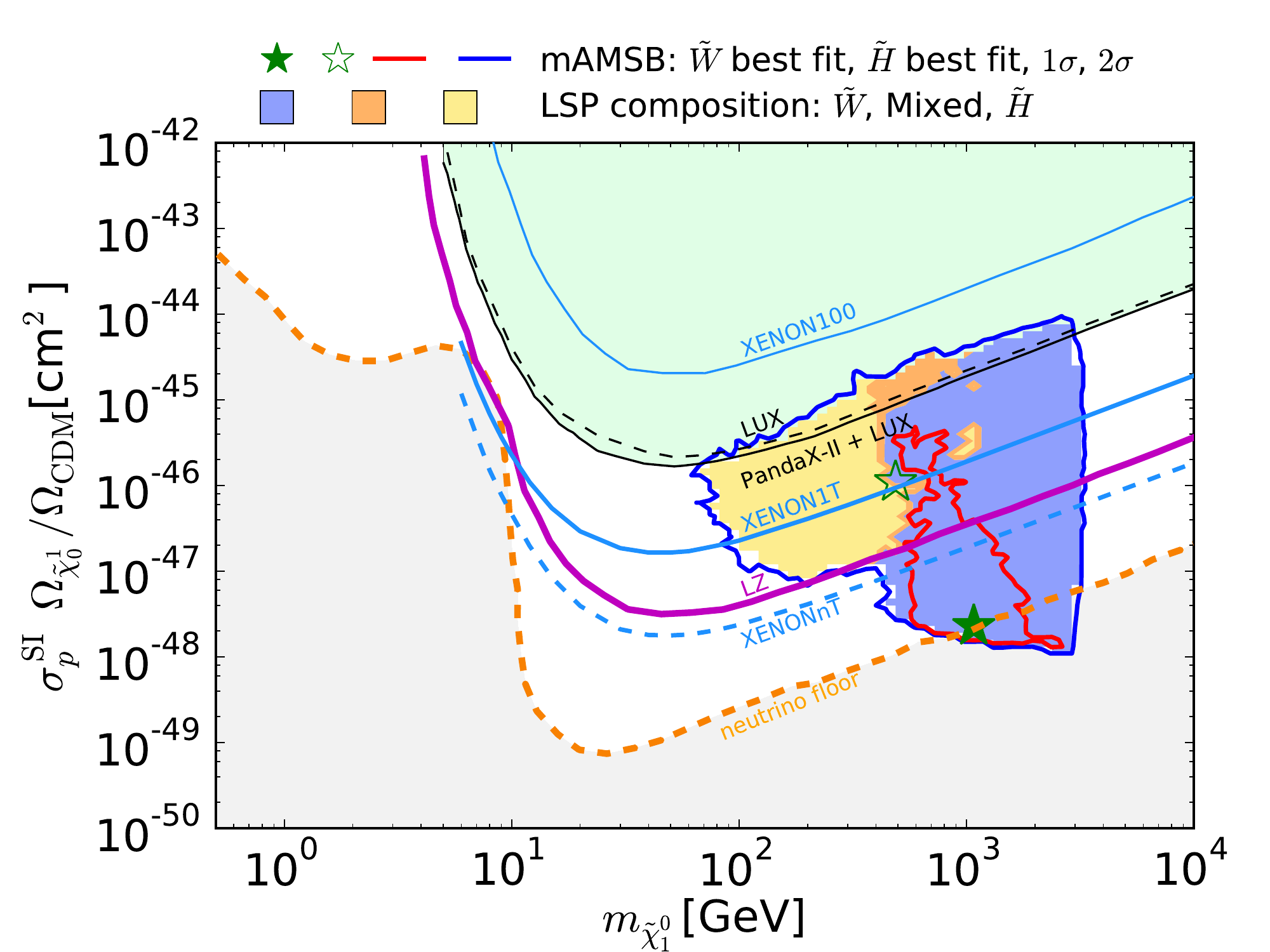}}\put(-95, +123){\footnotesize $\mu>0$, $\Omega_{\neu1}<\Omega_{\rm CDM}$}
\resizebox{7.5cm}{!}{\includegraphics{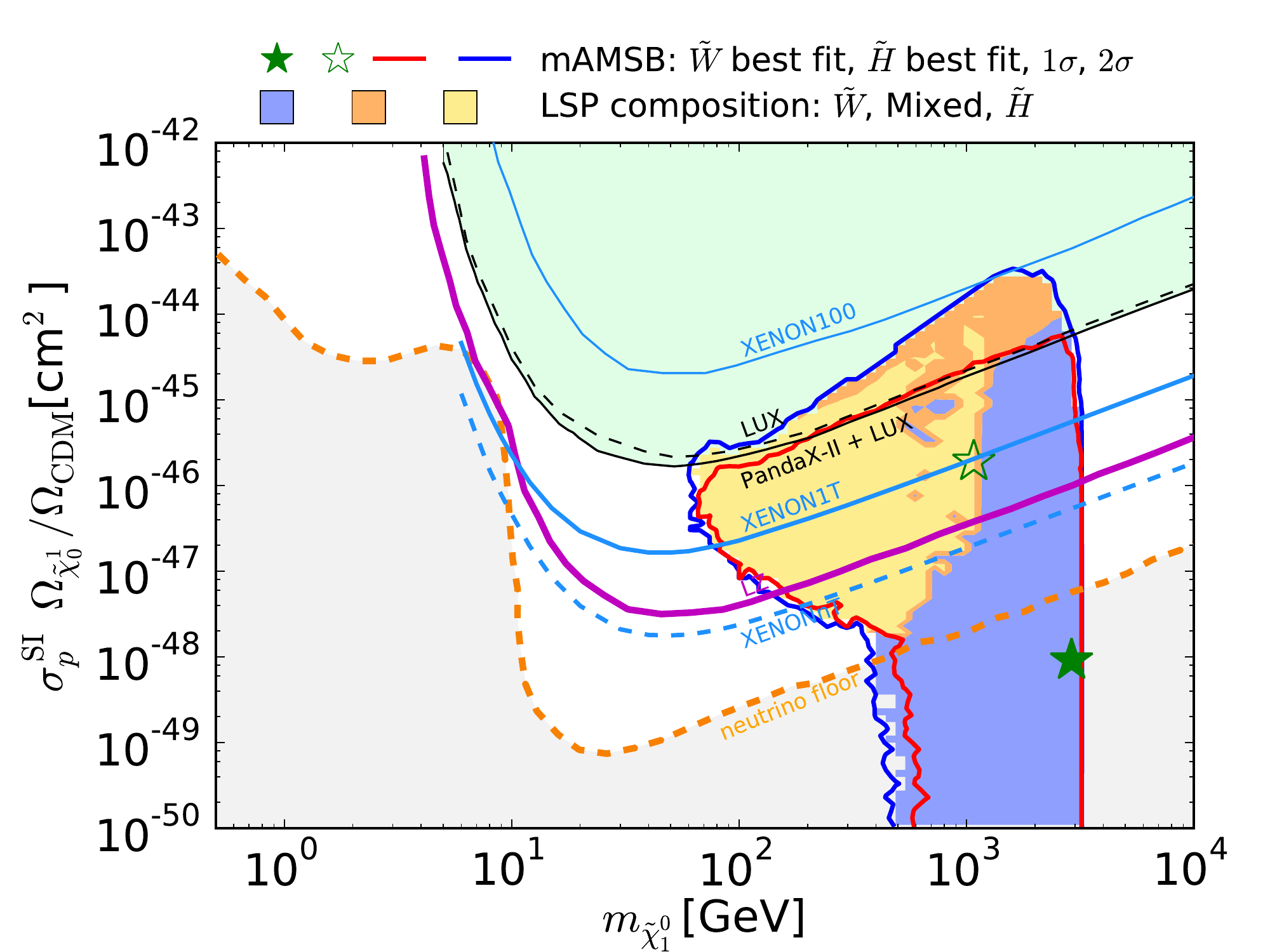}}\put(-95, +123){\footnotesize $\mu<0$, $\Omega_{\neu1}<\Omega_{\rm CDM}$}
\end{center}
\caption{\it The $(\mneu1, \ssi)$ planes in the mAMSB for $\mu>0$ (left)
and $\mu<0$ (right) in the case when the LSP only accounts for a fraction of the CDM density. 
The legends, {line styles and shadings are} the same as in Fig~\protect\ref{fig:DMdirect}.
}
\label{fig:DMdirect_ul}
\end{figure*}

\section{Summary}
\label{sec:conx}

{Using the {\tt MasterCode} framework},
we have constructed  in this paper a global
likelihood function for the minimal AMSB model and explored  the constraints imposed
by the available data on flavour, electroweak and Higgs observables, as well as by LHC searches
for gluinos via $\ETslash$ signatures. We have also included
the constraint imposed by the cosmological cold dark matter density, which we interpret as
either a measurement or an upper limit on the relic LSP density, and searches for dark matter scattering. 

In the all-CDM case, we find that the spectrum is relatively heavy, with strongly-interacting sparticles
weighing $\gtrsim 10 \tev$, but much smaller masses are possible if the LSP contributes only a
fraction of the overall CDM density. In the all-CDM case, the LSP composition may be either wino-
or Higgsino-like with almost equal likelihood, weighing $\sim 3 \tev$ and $\sim 1 \tev$, respectively.
On the other hand, in the part-CDM case much lighter LSP masses are allowed at the 68\% CL,
as are intermediate LSP masses.

Because of the high masses in the all-CDM case, the prospects for discovering sparticles at the
LHC are small, and there are limited prospects for observing significant deviations for the SM
predictions for flavour observables. However, in the part-CDM case some sparticles may well be
within reach of the LHC, and there are more interesting possibilities for observing mAMSB effects
on flavour observables, e.g., \bsdmm. In both cases, wide ranges of dark matter scattering cross-sections, \ssi\,
are allowed: \ssi\ may be very close to the upper limits established recently by the PandaX and LUX
experiments, or it may be within reach of the planned LUX-Zeplin experiment, or it may even be
far below the neutrino `floor'.

The mAMSB scenario discussed in this paper clearly presents different challenges from the
models with GUT-scale unification of (at least some of) the soft SUSY-breaking
parameters that we have studied previously~\cite{oldmc,mc9,mc10,mc12,mc14-SU5}, and
does not share the flexibility of pMSSM models~\cite{mc11,mc15-pMSSM11}.
As such, the mAMSB serves as a useful reminder that SUSY phenomenology
may differ significantly from what is preferred in these other scenarios.
}
\clearpage
\subsubsection*{Acknowledgements}

The work of E.B. and G.W. is supported in 
part by the Collaborative Research Center SFB676 of the DFG, ``Particles, Strings and the early Universe", 
and by the European Commission through the ``HiggsTools" Initial Training Network PITN-GA-2012-316704.The work of R.C. is supported in part by the National Science Foundation under 
Grant No. PHY-1151640 at the University of Illinois Chicago and in part by Fermilab, 
operated by Fermi Research Alliance, LLC under Contract No. De-AC02-07CH11359 
with the United States Department of Energy.
This work of M.J.D.
is supported in part by the Australian Research Council. The work of J.E. is supported in part by STFC
(UK) via the research grant ST/L000326/1, and
the work of H.F. is also supported in part by STFC (UK).
The work of S.H. is supported 
in part by CICYT (grant FPA 2013-40715-P) and by the
Spanish MICINN's Consolider-Ingenio 2010 Program under grant MultiDark
CSD2009-00064. The work of D.M.-S. is supported by the European Research Council 
via Grant BSMFLEET 639068.
The work of F.L. is supported by World Premier International Research Center Initiative (WPI), MEXT, Japan.
The work of K.A.O. is supported in part by DOE grant
DE-SC0011842 at the University of Minnesota. KS is supported by STFC through the IPPP grant.  
The work of K.S. is partially supported by the National Science Centre, Poland, under research grants
DEC-2014/15/B/ST2/02157 and DEC-2015/18/M/ST2/00054.

  \end{document}